\documentclass[smallcondensed]{svjour3}    
\smartqed  
\usepackage{graphicx}

\begin{document}

\title{On the Detection of (Habitable) Super-Earths Around Low-Mass Stars 
Using {\em Kepler} And Transit Timing Variation Method}

\titlerunning{Detecting super-Earths around low-mass stars using TTV with {\it Kepler}} 

\author{Nader Haghighipour \\
        Sabrina Kirste}

\authorrunning{Haghighipour \& Kirste} 

\institute{N. Haghighipour \at
           Institute for Astronomy and NASA Astrobiology Institute,
           University of Hawaii, Honolulu, HI 96825, USA,
              Tel: +1-808-956-6098,
              Fax: +1-808-956-4532,
              \email{nader@ifa.hawaii.edu} \\
          S. Kirste \at
          Zentrum f\"ur Astronomie und Astrophysik (ZAA), 
          Technische Universit\"at Berlin, 10623 Berlin, Germany.}  
\maketitle

\begin{abstract}

We present the results of an extensive study of the detectability of Earth-sized
planets and super-Earths in the habitable zones of cool and low-mass stars using transit timing
variation method. We have considered a system consisting of a star, a transiting 
giant planet, and a terrestrial-class perturber, and calculated 
TTVs for different values of the parameters of the system. 
To identify ranges of the parameters for which these variations would be detectable
by {\it Kepler}, we considered the analysis presented by Ford et al. (2011, ArXiv:1102.0544)
and assumed that a peak-to-peak variation of 20 seconds
would be within the range of the photometric sensitivity of this telescope.
We carried out simulations for resonant and non-resonant orbits, and identified
ranges of the semimajor axes and eccentricities of the transiting and perturbing bodies
for which an Earth-sized planet or a super-Earth in the habitable zone of a low-mass star 
would produce such 
TTVs. Results of our simulations indicate that in general, outer perturbers near first-
and second-order resonances show a higher prospect for detection. 
Inner perturbers are potentially detectable only when near 1:2 
and 1:3 mean-motion resonances. For a typical M star with a Jupiter-mass transiting 
planet, for instance, an Earth-mass perturber in the habitable zone can 
produce detectable TTVs when the orbit of the transiting planet is between 15 and 80 days. 
We present the details
of our simulations and discuss the implication of the results for the detection
of terrestrial planets around different low-mass stars.


\end{abstract}

\section{Introduction}
\label{intro}

Cool and low-mass stars (such as M dwarfs) present the most promising candidates 
for searching for Earth-like planets and super-Earths. These stars show large reflex 
accelerations in response to an orbiting body, and their low surface temperatures, 
which place their liquid water habitable zones (Kasting et al 1993) at close 
distances of 0.1--0.2 AU (corresponding to orbital periods of 20--50 days), cause 
their light-curves to show large decrease when they are transited by a close-in planet. 
In the past few years,
these advantages have resulted in the detection of several super-Earths around low-mass 
stars including a 6.0--7.5 Earth-mass planet around the star GJ 876 
(Rivera et al. 2005, 2010; Correia 2010), 4--6 planets with masses ranging
from 1.7 to 7.0 Earth-masses around star GL 581 (Bonfils et al 2005; Udry et al 2007;
Mayor et al 2009; Vogt et al 2010),
and two transiting planets around M stars GJ 436 (Butler et al 2004; Gillon 2007) 
and GJ 1214 (Charbonneau 2009).

The existence of systems such as GJ 876 and GL 581, where an M star is host to 
multiple planets in short-period orbits, suggests that other single-planet low-mass 
stars may also host additional objects. In systems where one (or more) of these bodies 
transits, the interaction between this object and other planets of the system results
in the exchange of energy and angular momentum, and creates small, 
short-term oscillations in their semimajor axes and orbital eccentricities. 
On the transiting planet, these oscillations manifest themselves as variations 
in the time and duration of the planet's transit.  
In this paper, we discuss the possibility
of the detection of these transit timing variations when they are generated by
an Earth-like planet or a super-Earth in the habitable zone of a low-mass star.

Measurements of transit timing variations can be used in detecting small, 
non-transiting planets (Miralda-Escud\'e 2002, Schneider 2003, 
Holman and Murray 2005, Agol et al 2005, Heyl and Gladman 2007).
Despite the complication in extracting the mass and orbital elements of 
the perturbing body from the values of TTVs 
(Nesvorn\'y and Morbidelli 2008, Nesvorn\'y 2009, Nesvorn\'y and Beaug\'e 2010,
Meschiari and Laughlin 2010, Veras et al 2011), this method has also been recognized as a potential
mechanism for detecting Trojan planets (Ford and Gaudi 2006, Ford and Holman 2007,
Haghighipour et al, in preparation), as well as
the moons of transiting gas-giants (Sartoretti and Schneider 1999, 
Simon et al 2007, Kipping 2009a\&b). 
TTVs are also sensitive to the precession (Miralda-Escud\'e 2002) and 
inclination of the orbit of the transiting object (Payne et al 2010), and 
in close, eclipsing binary systems, they can be used
to detect circumbinary planets
(Schneider and Chevreton 1990, Schneider and Doyle 1995, 
Doyle et al 1998, Doyle and Deeg 2003, Deeg et al 2008, 
Sybilski et al 2010, Schwarz et al 2011).

In this paper, we focus our study on the detection of TTVs using {\it Kepler} 
space telescope. {\it Kepler} monitors more than 150,000 stars with spectral 
types of A to M, and with apparent magnitudes brighter than 14
(Oshagh et al 2010; Batalha et al 2010, Coughlin et al 2011).
A survey of the discovery-space of this telescope indicates that low-mass stars 
such as M dwarfs, constitute a significant number of its 
targets. The large number of the targets of this telescope, 
combined with its unprecedented photometric precision,
and the long duration of its operation, makes {\it Kepler} ideal for 
searching for transit timing variations. A recent success, for instance, 
has come in the form 
of detecting variations in the orbital periods of transiting planets around stars 
Kepler-9 (Holman et al 2010) and Kepler-11 (Lissauer et al 2011). Further analysis
of the currently available data from this telescope also points to the existence
of many additional candidates that have 
potentially detectable variations in the times of their transits
(Steffen et al 2010,  Ford et al 2011).

As shown by Agol et al (2005), Steffen and Agol (2005), and  Agol and Steffen (2007), 
the amplitude of TTV is greatly enhanced when the transiting
and perturbing planets are near a mean-motion resonance (MMR). In this case, transit 
times vary as a function of the ratio of the masses of the two planets which 
enhances the sensitivity of the TTV method by one to two orders of magnitude. 
For instance, as shown by Agol et al (2005), 
around a solar-mass star, an Earth-mass 
planet can create TTVs with amplitudes of up to 180 s 
when near an exterior 2:1 MMR with a transiting Jupiter-size body in a 3-day orbit. 
In the recently detected Kepler-9 system where two Saturn-sized planets are near 
a 2:1 resonance,
the rate of the transit timing variation on the inner planet was even
higher and reached to approximately 50 minutes. 
This is two orders of magnitude higher than
the $\sim$10 s transit timing precisions that was achieved by a 1.2  m ground-based 
telescope in the measurements of the transit timing of HD 209458 (Brown et al 2001).
The fact that a perturber near a mean-motion resonance can produce large TTVs 
implies that such resonant systems will have a higher chance of being detected 
by the TTV method. In this paper, we will pay especial attention to 
such resonant orbits.

The outline of this paper is as follows. In section 2, we present the details of 
our methodology for calculating TTVs. Since we are interested in the detectability 
of an Earth-like planet or a super-Earth in the habitable zone of a star, we will analyze 
the dependence of the magnitudes of TTVs on different parameters of a system and present 
the results in section 3. Section 4 has to do with the implications of this
analysis for the detection of planets in the habitable zone, and section 5
concludes this study by presenting a summary and reviewing the results.

\section{Calculation of TTV}

The system of our interest consists of a star, a transiting planet,
and a perturbing body. Since the focus of our study is on low-mass stars, we choose
the star to have a mass between 0.25${M_\odot}$  and 0.7${M_\odot}$. 
The transiting planet is taken to be initially on a circular orbit with a period 
$(P_{\rm tr})$ between 3 and 10 days. The mass of this planet $(m_{\rm tr})$ is 
varied between 0.1 and 3 Jupiter-masses $(M_J)$. The perturber is considered to be smaller 
with a mass $(m_{\rm pr})$ ranging from 0.3 to 10 Earth-masses $(M_\oplus)$.
 The initial orbital
eccentricity of this planet is varied from 0 to 0.3 in increments of
0.05, and its orbital period $(P_{\rm pr})$ is chosen such that
$0.3 \leq P_{\rm pr}/P_{\rm tr} \leq 10$. We consider the system to be coplanar,
and unless stated otherwise, we set the initial values of the mean-anomaly,
argument of pericenter, and the longitude of the ascending node of the transiting planet
equal to zero.

We integrated our systems using the Bulirsch-Stoer algorithm of the N-body 
integration package Mercury (Chambers 1999). This algorithm was modified to
include general relativistic effects (Saha and Tremaine 1992). We carried 
out our integrations for three time spans of 1, 3, and 30 years. The 3-year time 
span was chosen to correspond to the duration of {\it Kepler}'s primary mission.
The 1-year and 30-year integrations were meant to show the short-term and long-term effects.
The 30-year time span was chosen to ensure that long-term variations in the amplitude of 
TTVs will also be modeled. 

To calculate TTVs, we obtained the difference between the time of the mid-transit
in the unperturbed system (star and transiting planet), $t_1$, and its corresponding
value obtained from the numerical integrations of the perturbed system
(star, transiting and perturbing planet), $t_2$. We assumed that at $t=0$, 
the centers of the star and the transiting planet were on the $x$-axis and calculated 
the time of transit by interpolating between the
times before and after the center of transiting planet crossed the center of the
star. Figure 1 shows a sample
of our results. The system in this figure consists of a 0.367$M_\odot$ star,
a Jovian-type transiting planet with a mass of $1.05 {M_J}$ on a 10-day orbit, 
and an Earth-mass perturber. Table 1 shows the orbital elements of this object.
 
\begin{table}
\centering
 \caption{Initial orbital elements of the perturber in figure1.}
  \begin{tabular}{lccc}
  \hline
 Orbital Elements       &  Top & Middle  & Bottom\\
  \hline
Semimajor Axis (AU) & 0.089  & 0.089   &  0.042  \\
Eccentricity & 0  & 0   &  0 \\
Inclination         & 0  & 0   &  0 \\
Argument of Periastron  & 352.8$^\circ$  & 352.8$^\circ$   & 0 \\
Mean-Anomaly  & 7.192$^\circ$  & 7.192$^\circ$ & 0 \\
Longitude of Periastron  & 352.8$^\circ$ & 352.8$^\circ$  &  0 \\
Longitude of Ascending Node  & 0 & 0   &  0 \\
\hline
\end{tabular}
\label{tab5}
\end{table}

The top two panels in figure 1 show the TTVs for a non-resonant case where the
perturber is in an exterior orbit with a period of ${P_{\rm pr}}=16.11$ days.
As expected, the amplitude of TTVs in this system is small. The results shown here are
consistent with the analytical prediction of TTVs as given by Agol et al (2005). 
These authors indicated that in a non-resonant system, when the orbital period
of the perturber is half-way between two resonances, the timing deviation
is approximately given by 
$0.7{P_{\rm tr}}({m_{\rm pr}}/M)[{a_{\rm tr}}/({a_{\rm pr}}-{a_{\rm tr}})]^2$
where $M$ is the mass of the central star.
For the top two panels of figure 1, the orbital period of the perturbing planet
(16.11 days) is between the 4:3 and 2:1 exterior resonances. In this case, the
formula above predicts a timing deviation of approximately 37 s which is consistent
with the RMS in these two panels. The bottom panel of figure 1 corresponds to a case of near 
interior 1:2 mean-motion resonance. 
As expected, the amplitude of TTVs in this case vary in a large range with a maximum of 
$\sim$800 s.

As shown by Agol et al (2005), high-order resonances result in smaller magnifications 
in the amplitude of TTV and as such they are more suitable for demonstrating the 
contrast between the short- and long-term variations in transit timing.
To show this effect, we placed the perturber of the system of figure 1 
near an exterior 3:1 resonance and integrated the system for a longer time.
Table 2 shows the orbital elements of the perturber at the beginning of the integrations.
The results are shown in figure 2. The top panel in this figure shows TTVs for 250 days.
As shown here, although the planets are near a resonance, the amplitudes of TTVs are much 
smaller than those of the 1:2 MMR (bottom panel of figure 1). This is an expected result as the 
3:1 MMR is one order higher than the 1:2 MMR. As shown by Agol et al (2005), the analytically
predicted value of the amplitude of TTVs in this case is given by 
$({P_{\rm tr}}/{2^{3/2}}\pi)({{m_{\rm pr}}/M})({a_{\rm pr}}/{a_{\rm tr}})\sim 1.67$ s
which agrees with the amplitude of TTVs as shown in this graph. 

The continuation of integrations 
to 1000 days (middle panel) reveals the onset of long-term secular variations.
As the integrations continue to $10^4$ days, the secular long-term variations are fully
developed and reach an amplitude as high as 253 s. The increase in the amplitude of
TTVs in this case can be attributed to the increase in the eccentricity of the
perturber as a result of its long-term interaction with the massive transiting planet.
Figure 3 show the eccentricities of these objects for the duration of integration.

\begin{table}
\centering
 \caption{Initial orbital elements of the perturber in figure 2.}
  \begin{tabular}{lc}
  \hline
 Orbital Elements       & All Panels \\
  \hline
Semimajor Axis (AU) & 0.137  \\
Eccentricity & 0 \\
Inclination  & 0 \\
Argument of Periastron  & 354.81$^\circ$\\
Mean-Anomaly  & 5.194$^\circ$\\
Longitude of Periastron  & 354.81$^\circ$\\
Longitude of Ascending Node & 0 \\
\hline
\end{tabular}
\label{tab5}
\end{table}

Figure 4 shows the maximum values of the amplitudes of TTVs for the system of figure 1
(a 0.367$M_\odot$ star, a Jovian-type transiting planet with a mass of $1.05 {M_J}$ on a 10-day orbit,
and an Earth-mass perturber) when the perturber is in different resonant and non-resonant 
orbits\footnote{We did not show 
the case of 1:1 MMR in figure 4 because our integrations indicated that in 
a large area around this resonance, the orbit of the perturbing planet was unstable. 
We carried out a more detailed study of this resonance and were able to identify ranges 
of orbital elements for which a perturber near a 1:1 MMR would have a stable orbit. We will
present the analysis of transit timing variations for this system, and the detectability
of their TTVs, in an up-coming article (Haghighipour et al. in preparation).}.
As shown here, low-order resonances show high variations in transit 
timing that are well within the range of the photometric precision of 
{\it Kepler} space telescope. These results are also supported by the analytical
formulae given by Agol et al (2005). These authors have shown that when the two planets
are in a first order $j:j+1$ MMR, the maximum value of the TTVs is approximately given by 
$(P/4.5j)({m_{\rm pert}}/{{m_{\rm pert}}+{m_{\rm tr}}})$. For the system of figure 1
when the two planets are in an exterior 2:1 resonance, the maximum TTV as predicted by this
formula is approximately 600 s which agrees with the value shown in figure 4.
More results can be found in the appendix where similar simulations 
have been carried out for resonant and non-resonant orbits in a system with a Sun-like
star, and for a perturber with a mass of 1 $M_\oplus$.

\section {Dependence on the parameters of the system}

The magnitude of TTV changes with the mass and orbital elements of the
planets, and the type of the central star. To identify ranges of these parameters that
produce detectable TTVs, we carried out simulations for different values of the 
mass and orbital eccentricity of the perturber, period and mass of the transiting 
planet, and the mass of the host star. Calculations were carried out for 3 years and for
both near resonance and non-resonant orbits. Except when studying the effect of the orbital
eccentricity of the perturber, calculations began with  both planets in circular orbits.
We present the results of our simulations 
in the following. Except for the case where the mass of the 
central star is varied, we use the system of figure 1 $({m_{\rm star}}=0.367 {M_\odot}\,,\,
{m_{\rm tr}}=1 {M_J}\,,\,{m_{\rm pr}}= 1 {M_\oplus}\,,\,{P_{\rm tr}}=10\, {\rm days})$
as a sample system for our analysis.

\subsection{Mass of the perturbing planet ($m_{\rm pr}$)}

Since the interaction between the two planets is gravitational, the mass of the perturber
plays an important role in varying the time of transits. We used the system mentioned above
and calculated TTVs for different values of this parameter. Since we are interested
in Earth-like planets and super-Earths, we considered different values for $m_{\rm pr}$ 
between 0.1$M_\oplus$ to 10$M_\oplus$. Figure 5 shows the results for two cases of
near exterior resonances (2:1 and 5:2 MMRs), and for a non-resonant system. As shown here, near a resonance,
variations in the time of transit scale with the mass of the perturber.
This is an expected result that has also been reported by Agol et al (2005). 
Our simulations suggest that such a scaling seems to exist for non-resonant orbits as well,
which is a natural consequence of the conservation of energy and momentum (see equation 30
of Agol et al 2005).

\subsection{Eccentricity of the perturbing planet ($e_{\rm pr}$)}

As explained in the beginning of this section, 
we considered the initial orbits of the perturbing and
transiting planets in all our simulations to be circular. During the course of an
integration, the mutual interactions of the two planets cause their orbital eccentricities
to increase (e.g., figure 4). In cases where the planet-planet interaction does not make the system unstable,
the increase in the orbital eccentricity of the perturbing body results in a decrease in its
periastron distance which in turn causes rapid and fast changes in the time of transit.
To examine the enhancing effect of $e_{\rm pr}$ on TTVs and its role in their possible
detection, we carried out simulations where the initial orbital eccentricity of the
perturbing planet was chosen to be between 0 and 0.3. We considered a system with a Jupiter-mass
transiting planet in a 10-day orbit, an Earth-mass perturber, and a 0.367 $M_\odot$ central star.
We chose the system to be co-planar, and set all orbital angles for both planets equal to zero.
Results indicated that
in general, an increase in perturber's orbital eccentricity increases the amplitude
of TTV. However the rate of this increase depends on the resonant state of the system
(figures 6 and 7), and the time of integration (figure 8). 

For large values of the perturber's eccentricity, high-order MMRs play an important role
in the variations of transit timing (Pan and Sari 2004; Agol et al. 2005). 
This can be seen in figure 6 where the amplitudes of TTVs are shown for different values 
of the ratios of the orbital periods of the two planets. As shown here, among the resonant 
cases, the rate of the increase in TTVs is larger for a near 5:2 MMR. Figure 6 also
shows that when the two planets are near a MMR, a range of the perturber's eccentricity 
exists for which the increase in the value of TTV is negligibly small. This can be seen in
figure 7 where the magnitudes of TTV and the period-ratio are shown for a near
exterior 2:1 MMR. As shown here, the amplitude of TTV maintains an almost constant 
maximum value around 600 s for ${e_{\rm pr}} \leq 0.08$ (for the system considered
in these simulations, integration for larger values of eccentricity resulted
in instability). Figures 6 and 7 are also consistent with figure 5 of Agol et al (2005)
where these authors show the dispersion in timing variations for different values of the
outer planet's orbital eccentricity and for different resonances. Calculations by Agol
et al (2005) indicate that when the perturber is in a low-eccentricity resonant orbit,
the variations of its eccentricity that are caused because of its interactions with 
the massive transiting planet, are of the order $({m_{\rm pert}}/{m_{\rm tr}})^{1/3}$. 
For the system of figure 7, this quantity is approximately 0.14. Therefore, for initial
eccentricities up to half of this value (i.e., $\sim 0.07$, as in figure 7), the TTVs
are primarily dominated by the period variations of the perturber and not strongly
affected by its initial eccentricity.

To explore how the amplitude of TTVs vary as the time of integration increases in a system  
where the perturber is in a resonant and eccentric orbit, we calculated TTVs for different 
values of the orbital eccentricity of an Earth-mass
planet in a 1:2 MMR with a Jupiter-mass transiting body. 
The mass of the central star is 0.367 $M_\odot$. Figure 8 shows the results
for different integration times (in terms of the number of transits). As shown here, 
the strongest increase in TTVs is reached when the time of integration varies between 
20 and 30 transits. This can be explained noting that for a system in a $j:j+1$ MMR, 
the libration period scales as $0.5\,{j^{-4/3}}\,({m_{\rm pert}}/{m_{\rm tr}})^{-2/3}$ 
(Agol et al 2005). For the system of figure 8 where $j=1$, this results in approximately 
24 transits. For
longer times of integration, the magnitude of TTV increases almost linearly with 
the perturber's eccentricity.

\subsection{Period of the transiting planet ($P_{\rm tr}$)}

We calculated TTVs for different values of the period of the transiting
planet. Figure 9 shows a sample of results for a non-resonant (top) and a resonant (bottom)
system. Simulations were carried out for a sub-Earth perturber and several super-Earths. 
As shown here, the magnitude of TTV increases with increasing the period of the 
transiting planet. This result agrees with the analytical analysis by 
Holman and Murray (2005)
and Agol et al (2005) where these authors have shown that near a MMR, the magnitude of
TTVs scale as $P_{\rm tr}$. We will explain the implications of these results for
the detection of habitable planets in the next section.

\subsection{Mass of the transiting planet ($m_{\rm tr}$)}

Simulations were also carried out for transiting planets with masses ranging from 
0.1$M_J$ to 3$M_J$. Results point to different $m_{\rm tr}$-dependence based on the 
resonant state of the system and the order of the resonance. Integrations indicated that 
for the system considered here (i.e. the system of figure 1),
except for the interior 1:2 and exterior 2:1 MMRs, other first-order resonances were unstable.
Figure 10 shows the results for when the system is near these two resonances. 
As shown here, the amplitude of
TTV is larger for smaller values of the mass of the transiting planet. As the mass of
this object increases, the corresponding value of its TTV signal becomes smaller.
This is a result that agrees with the theoretical analysis of Agol et al (2005).
As shown by these authors, the maximum variation of transit timing in a resonant 
case scales as ${m_{\rm pr}}/({m_{\rm pr}}+{m_{\rm tr}})$. 

The top panel of figure 11 shows the values of TTV when the system is near
two high-order resonances of 3:1 and 5:2, and for
some of their neighboring non-resonant orbits. As shown here,
the values of TTVs are slightly increased for large transiting  planets. This can be 
attributed to the fact that as the mass of the transiting planet increase, its interaction
with the perturbing body excites the eccentricity of this object
which in turn causes higher order resonances to become important. The increase in the
eccentricity of the perturbing planet can be seen
in the bottom panel of figure 11 where the median of the
eccentricity of this object has been plotted against the mass of the transiting planet
for a sample non-resonant system and also when the system is in a 5:2 MMR.
A shown here, the increase in the median eccentricity is larger in the non-resonant
case indicating that, as shown in the top panel, TTVs would be larger outside
resonances. The decreasing trend of TTV in the case
of 5:2 MMR may also be due to similar effects as in the case of 2:1 MMR (figure 10).

\subsection{Mass of the central star}

Although the focus of our study has been on the possibility of the detection of
TTVs around low-mass stars, we carried out simulations for different values of the 
mass of the central star as well. Figure 12 shows the results for a perturber with
different masses. The graph on the top is for an 
exterior 2:1 MMR. The graph on the bottom corresponds to a non-resonant system  
where the perturber is in a 28.3-day orbit. As shown here, in the non-resonant system,
the magnitude of TTV seems to be inversely proportional to the mass of the central star.
This result has also been reported by Agol et al (2005). In the resonant
system, on the other hand, the mass of the central star does not seem to
play an important role in changing the magnitude of TTV. This characteristic of resonant systems
enables us to examine the
effect of the mass of the star on the possibility of the detection of the perturber
in its habitable zone. We explain this in more detail in the next section.

\section {Implications for the detection of a perturber in the habitable zone}

Using the results of the simulations of section 3, we assessed the detectability
of TTVs that were produced by a perturber in the habitable zone. We determined the 
locations of the boundaries of the habitable zone $(r)$ using the equation 

\begin{equation}
{\bigg ({r \over {r_s}}\bigg)} = {\bigg ({L\over {L_\odot}}\bigg)^{1/2}},
\end{equation} 
\vskip 10pt
\noindent
where $L$ is the luminosity of the central star. The quantity ${r_s}$ in this equation,
corresponds to the (inner,outer) boundaries of the Sun's habitable zone. In general,
this quantity is model dependent and varies with different
atmospheric circulation models. For instance, as suggested by Kasting et al (1993), 
${r_s}$=(0.95 , 1.37) (AU), whereas models by Menou and Tabachnik (2003) and Jones et al 
(2005, 2006) suggest an inner boundary as close as 0.7 AU and an outer boundary 
beyond 2 AU. In this study, we considered some intermediate values, and assumed that
${r_s}$=(0.78 , 1.8) (AU). 

To calculate the luminosity of the star, we used two different mass-luminosity 
relations. First we considered $L\,\sim\, {M^{3.5}}$, where both the
luminosity and mass are in solar units. We also used the mass-luminosity relation
suggested by Duric (2004):

\begin{itemize}
\item $L ({L_\odot})\,\sim \, 0.23 \, {M^{2.3}}\,({M_\odot})\qquad , \qquad
M\, < 0.4 \,{M_\odot}$ 
\vskip 5pt
\item $L ({L_\odot})\,\sim\, {M^4}\,({M_\odot}) \>\>\>\qquad\qquad , \qquad 
M\, >  0.4 \,{M_\odot}$ . 
\end{itemize}
\noindent
Calculations indicate that the locations of the boundaries of HZ, 
as obtained from these two relations, are very close to one another. 
We, therefore, used the
boundaries obtained from the $L\,\sim\, {M^{3.5}}$ for the purpose of our analysis.

Figures 13 and 14 show the HZ for three stars with masses 0.252, 0.367,
and 0.486$M_\odot$. Since we are interested in identifying detectable TTVs
in systems where the transiting planet is at different orbital distances, and also 
because resonant orbits create larger TTVs, we chose the horizontal axis
to represent the period of the transiting body and identified different mean-motion 
resonances. The shaded area in each graph corresponds to systems where the 
variations of TTVs due to a perturber in an initially circular orbit
are larger than 20 s. 
This value is consistent with the photometric precision of {\it Kepler} as reported
by Ford et al (2011). 

An interesting result shown in figures 13 and 14 is the large contrast between the sizes of the 
regions where inner and outer perturbers produce detectable TTVs.
This can be seen for an Earth-mass perturber (figure 13 and the top 
panel of figure 14) as well as a 10$M_\oplus$ super-Earth (bottom panel of figure 14).
As shown in these figures, outer perturbers seem to have higher potential for detection.
An inner perturber also produces detectable TTVs. However, as expected, the orbit of the planet
in this case is primarily limited to the regions near first- and second-order mean-motion 
resonances 
(1:2 and 1:3 MMRs). The perturber can also produce detectable TTVs when in the habitable 
zone of the central star. 
For instance, as shown by the top panel of figure 13, a 1$M_\oplus$ perturber can
produces detectable TTVs when in the HZ of a 0.252$M_\odot$ star and near a 1:3 MMR 
with a 1$M_J$ transiting planet with a 
period larger than 40 days. 
As the mass of the star increases, the detectability in the HZ is shifted to farther orbits.

As expected, an outer perturber produces detectable TTVs in a larger range of distances.
The region of detectability in this case contains many first- and high-order MMRs.
We have shown some of these resonances in figures 13 and 14. As shown here, the lower
boundary is limited to the exterior 3:2 MMR whereas the upper boundary is slightly above the 2:1 MMR
and in the case of larger perturbers it may even reach to
the 3:1 MMR (bottom panel of figure 14). 
As shown in these figures, a small planet in the HZ of a typical M star can produce 
detectable TTVs on a transiting Jupiter-mass body in an orbit with a period larger than 15 (10) days 
corresponding to an Earth-mass (10$M_\oplus$ super-Earth) perturber.

\section{Summary and concluding remarks}

We carried out an extensive study of the possibility of the detection of small planets
in the HZ of low-mass stars with {\it Kepler} space telescope and 
transit timing variation method. With its high precision photometry, {\it Kepler}
has the capability of detecting small TTVs such as those created by terrestrial-class 
perturbers. We assumed that TTVs with amplitudes as small as  20 s can be detected by
{\it Kepler}, and identified regions of the parameter space where the variations in 
the time of the transit of a giant planet in a short-period orbit around a low-mass
star would be larger than this value.

In general, the lowest detectable amplitude
of TTVs depends upon several quantities including the magnitude of the central star,
the depth of the transit, the durations of ingress and egress, and the number of transits
observed. There will also be systematics that will affect the sensitivity of the
telescopes at different times. 
The 20 s detectability limit considered in this study corresponds to the best transit timing
precision obtained by Ford et al (2011) in their statistical analysis of the transit timing 
of {\it Kepler}'s planetary candidates. Analyzing the data obtained from 1235 potential transiting 
planets (Borucki et al 2011) during the first four months of the operation of this telescope,
Ford et al obtained a transit timing accuracy ranging from approximately 20 s to 100 minutes
(figure 1 of Ford et al 2011). 
The 20 s precision is also consistent with the analysis of the light curves
of Kepler-4 through Kepler-8 as given by Kipping and Bakos (2011). As shown by these authors,
the timing precision for the transiting planets in these systems ranges between 20 and 250 seconds.
It is worth noting that systems with TTVs larger than 15-20 s may also be detected by CoRoT.
This telescope has been monitoring more than 12000 dwarf stars with apparent
magnitudes ranging from 11 to 16. 

Although many non-resonant systems produced detectable TTVs, the results of our
study indicate that as expected, perturbers near mean-motion resonances with the
transiting planet show a higher prospect of detection. This can also be seen in
the system of Kepler-9 where the two TTV-detected planets of this system are near a 2:1 MMR
(Holman et al 2010).
Our study shows that interior perturbers with masses ranging from
1 to 10 $M_\oplus$ can produce detectable TTVs when near 1:2 and 1:3 MMRs.
However, in the majority of our systems, these interior perturbers were not in
the HZ. When in an outer orbit, these planets can be in the HZ and produce
detectable TTVs when in first- and second-order resonances. For 
instance, for a typical M star with a mass of approximately 1/3 of the Sun where the HZ 
extends between 0.1 and 0.2 AU, 
perturbers in 3:2, 2:1, 5:3, and 3:1 MMRs produced large-magnitude
TTVs when in the system's HZ. For the systems studied here, a perturber with a mass  
ranging from 1 to 10 Earth-masses can be detected in the HZ when the transiting planet
is in 15-80 day orbits.

In this study we considered the planetary system to be planar and consists of only
three bodies. These assumptions may limit the applicability of our results. For instance,
an inclined
perturber will cause the inclination of the orbit of the transiting planet to vary 
which in turn affects the variations in the times of its transits (Payne et al 2010). Also, as suggested
by the simulations of planet formation around low-mass stars, these stars may be hosts
to more than two planets (e.g., GJ 876 hosts four planets, and GL 581 has 4--6 super-Earths).
Depending on their masses and orbital elements, additional bodies may play a role in 
the magnitude of TTVs. 

Another assumption considered in our study is the existence of a giant planet in
a short-period orbit around a low-mass star. In general, the low masses of circumstellar 
disks around these stars, combined with the short lifetime of the gas in such disks,
casts doubt in the possibility that giant planets may exist in these systems and
in close-in orbits (Laughlin et al 2004). The fact that giant planets have been
observed around M stars [e.g., GJ 876 with two Jupiter-like planets and a Uranus-mass 
body in approximately 30, 60, and 120 days orbits (Rivera et al. 2005, 2010), or HIP
57050 with a Saturn-mass planet in a 42 days orbit (Haghighipour et al 2010)] 
suggests that these planets might have formed at the outer
regions of the disk where more material was available, and migrated to
their current orbits. As shown by Zhou et al (2005), Fog and Nelson 
(2005, 2006, 2007a \& b, 2009), Raymond et al (2006), Mandell et al (2007), 
and Kennedy and Kenyon (2008), terrestrial planets may form during the migration
of a giant planet around a Sun-like star. These planets may be captured
in resonance with the giant planet and migrate to smaller orbits.
Recently Haghighipour and Rastegar (2011) carried out similar simulations 
around M stars. By allowing the giant planet to migrate to orbits with periods
of 3-5 days, these authors showed that around low-mass stars,
resonance capture during the migration of the giant
planet seems to be the most viable mechanism for the formation of a system that consists of
a transiting giant planet and a small interior perturber. In case of an outer perturber,
although it is possible for small planets to migrate and capture larger bodies
in resonance (e.g., GJ 876 where the outermost planet is a Uranus-mass object in a
4:2:1 resonance with the two inner Jupiter-size planets), it is more probable that
the smaller, outer perturber is accompanied by another larger planet in an
exterior mean-motion resonance. How such a second exterior perturber affects the results 
presented in this paper is the subject of our currently on-going research.

\begin{acknowledgements}

Support is acknowledged for NH from NASA Astrobiology Institute (NAI) under Cooperative Agreement
NNA04CC08A at the Institute for Astronomy (IfA), University of Hawaii (UH), and NASA EXOB grant
NNX09AN05G. SK is thankful to the IfA and UH/NAI for their great hospitality during 
the course of this project, and to NAI central for travel support. We are especially thankful to 
the referee, Eric Agol, for his critically reading of our manuscript and for his insightful comments 
that greatly enhanced this article.

\end{acknowledgements}

\clearpage

\begin{figure}
\centering{
\includegraphics[width=7.9cm]{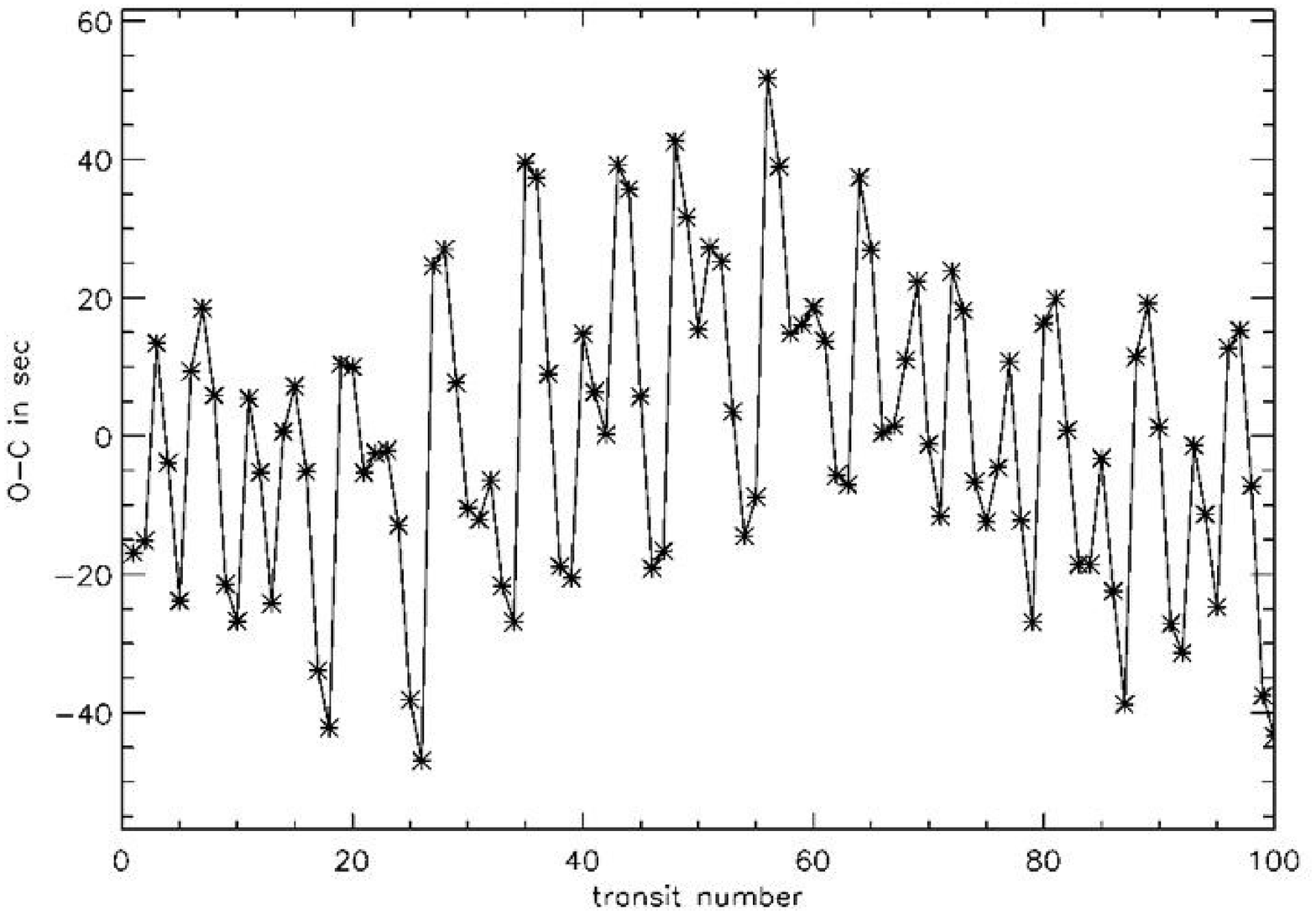}
\vskip 2pt
\includegraphics[width=7.9cm]{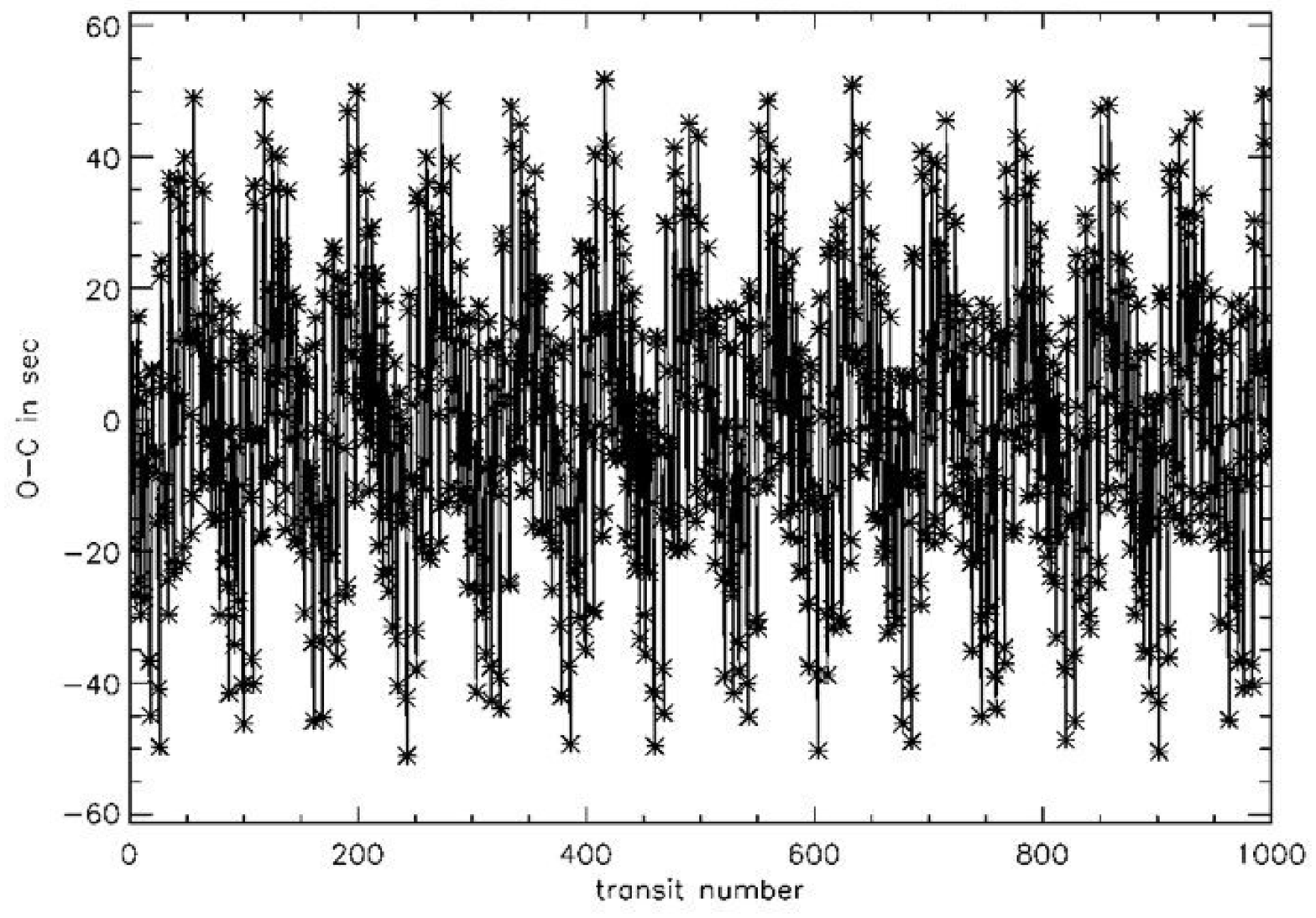}
\vskip 2pt
\includegraphics[width=7.9cm]{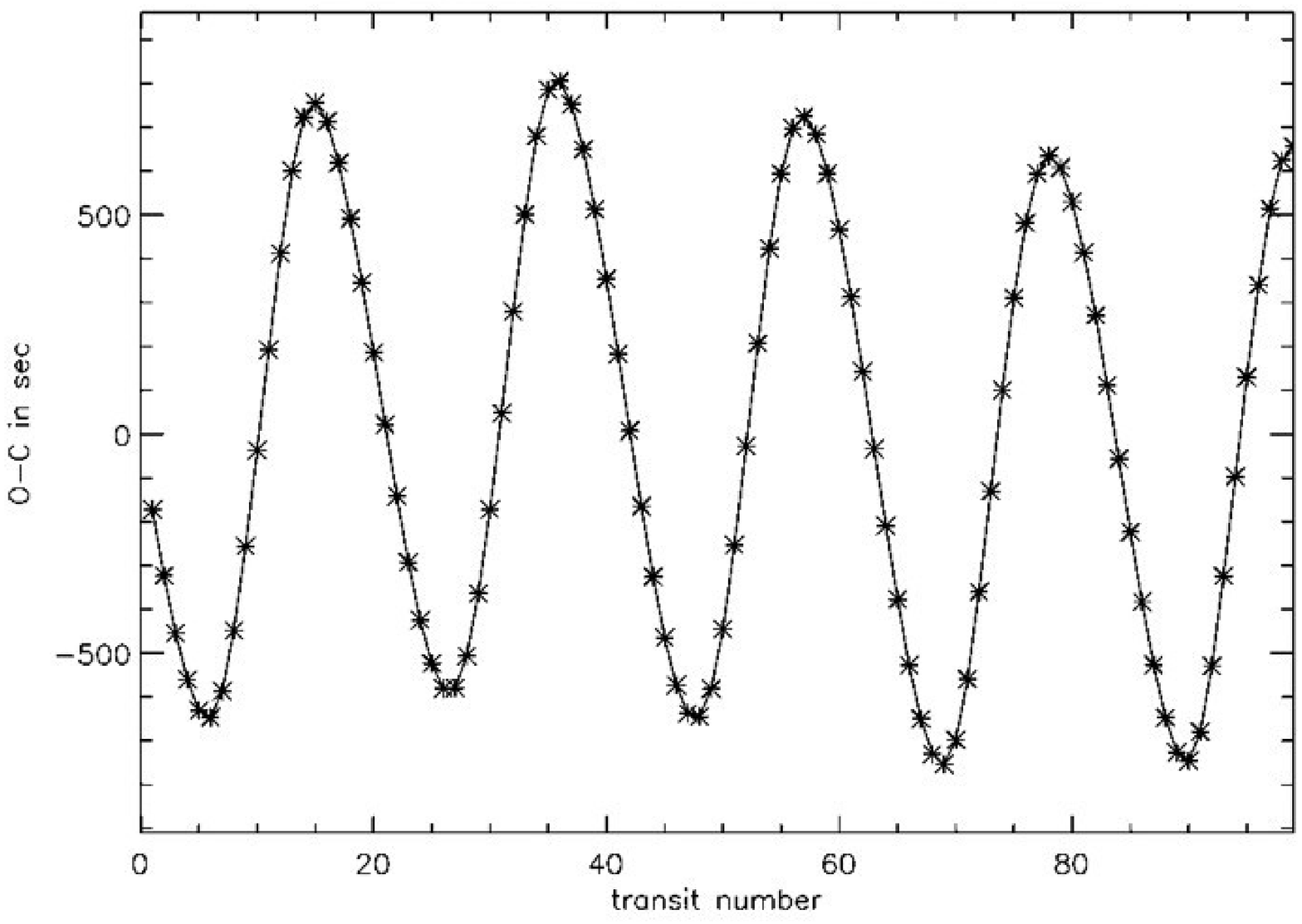}}
\caption{Graphs of TTVs for a three-body system with a 0.367$M_\odot$ star,
a Jupiter-mass transiting planet in a 10-day orbit, and an Earth-mass
perturber. The top and middle panels show the TTVs when the perturber is in 
a 16.11-day orbit. The bottom panel corresponds to near an interior 1:2 resonance.
As expected, the amplitude of TTV is enhanced by an order of magnitude 
in this case.} 
\end{figure}

\clearpage

\begin{figure}
\centering{
\includegraphics[width=8cm]{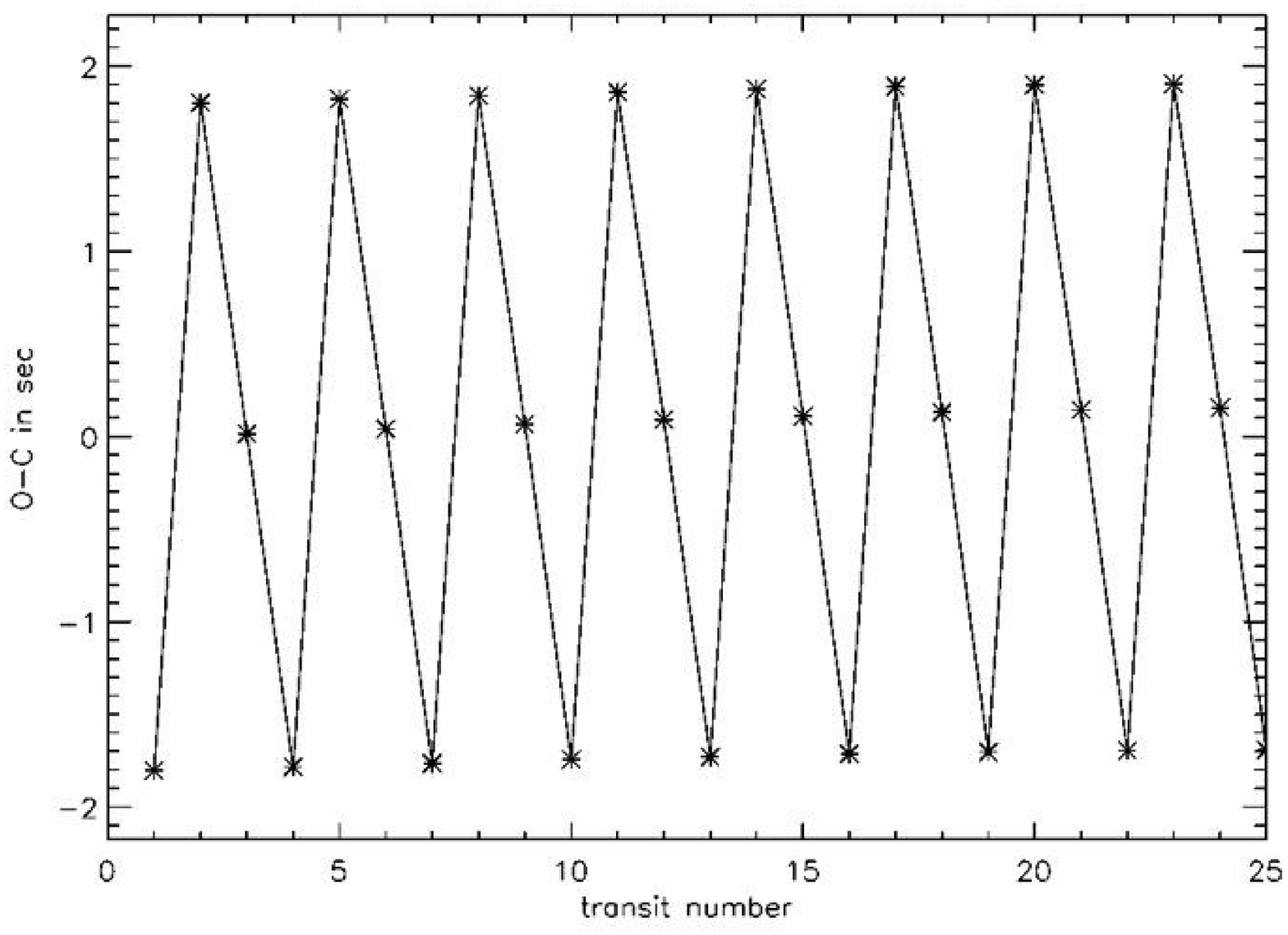}
\vskip 2pt
\includegraphics[width=8cm]{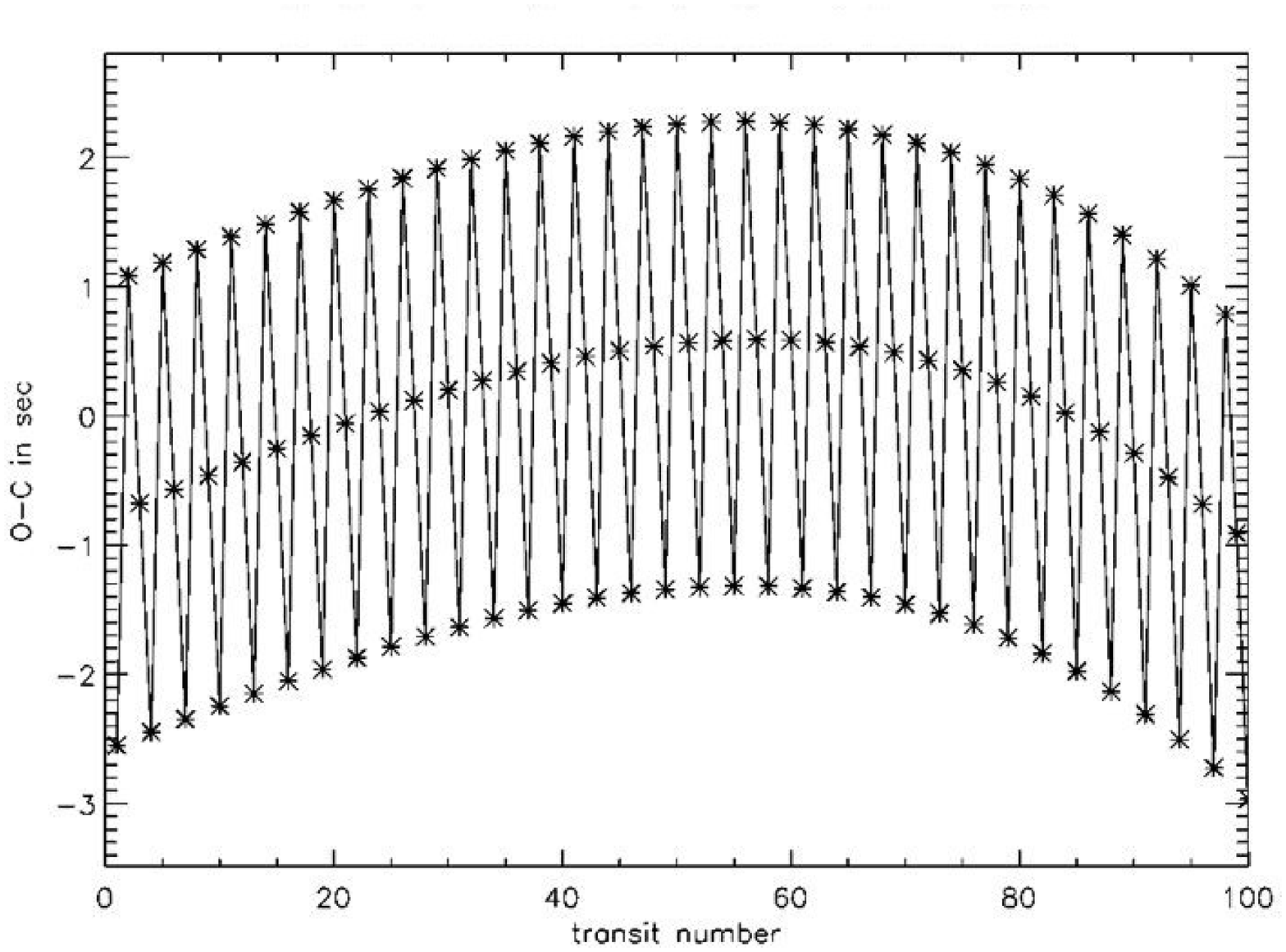}
\vskip 2pt
\includegraphics[width=8cm]{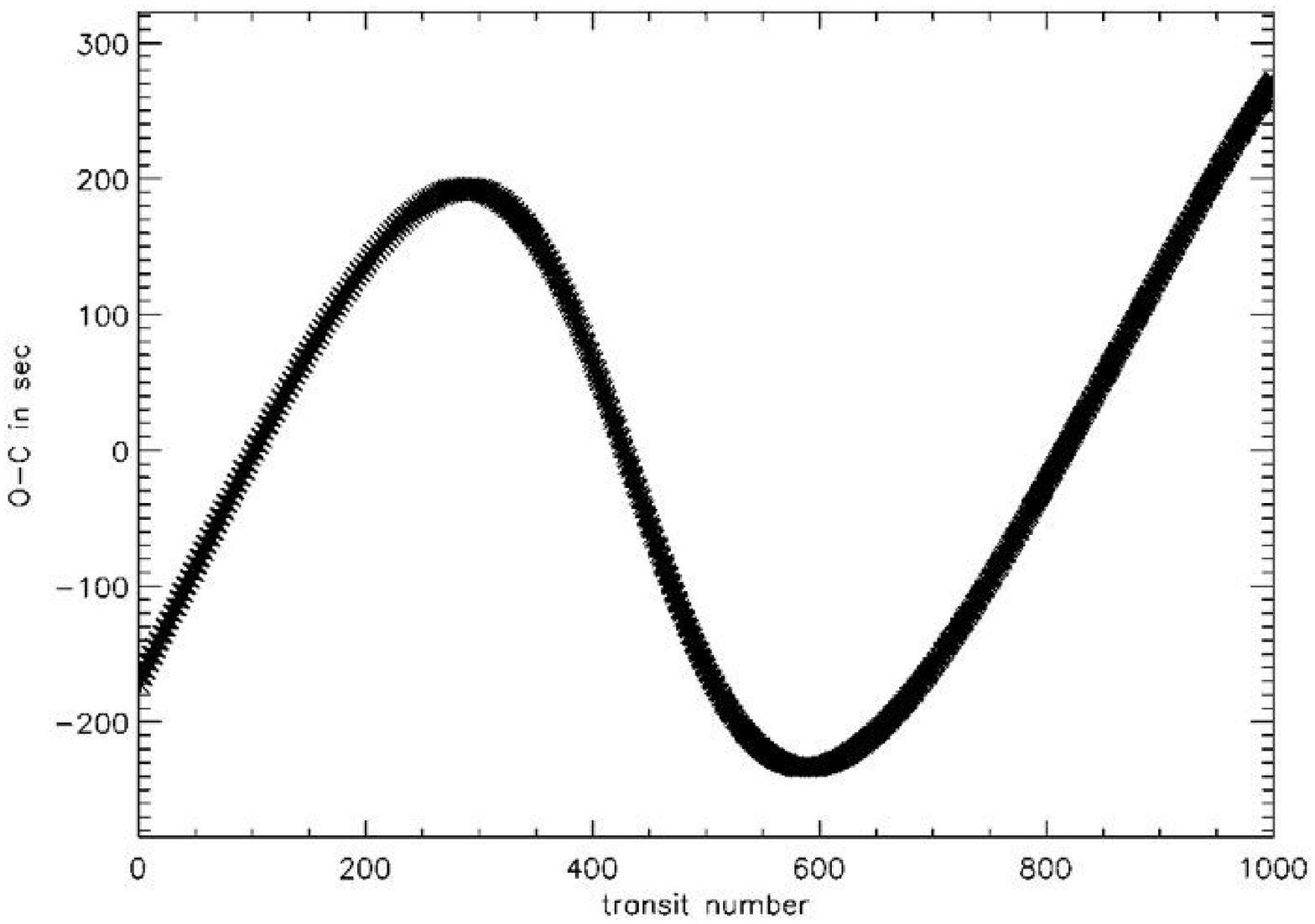}}
\caption{Graphs of TTVs for the system of figure 1 when the perturbing planet is
near an exterior 3:1 MMR. As shown here, increasing the time of integrations results
in the development of long-term secular variations in the amplitudes of TTV. The
top panel shows the results for 250 days, the middle panel is for 1000 days,
and the bottom panel is for $10^4$.}  
\end{figure}

\clearpage

\begin{figure}
\centering{
\vskip 30pt
\includegraphics[width=10cm]{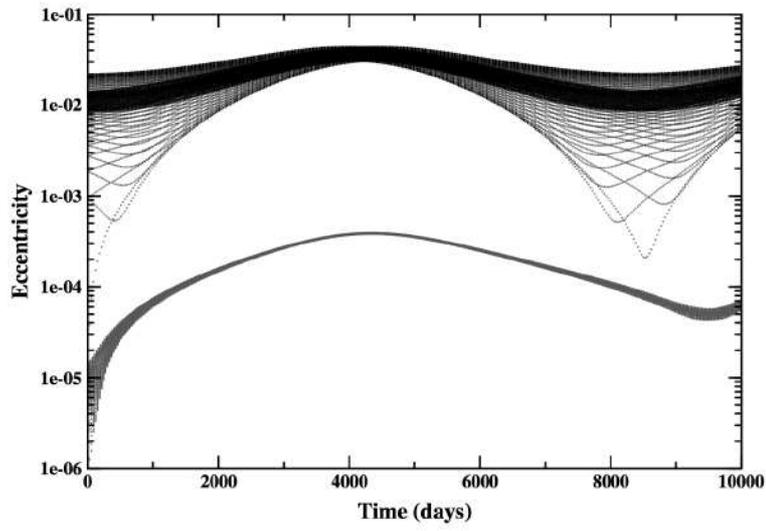}}
\vskip 10pt
\caption{The graph of the eccentricities of the transiting (gray) and
perturbing (black) planets for the system of figure 2.}  
\end{figure}

\clearpage

\begin{figure}
\centering{
\vskip 30pt
\includegraphics[width=10cm]{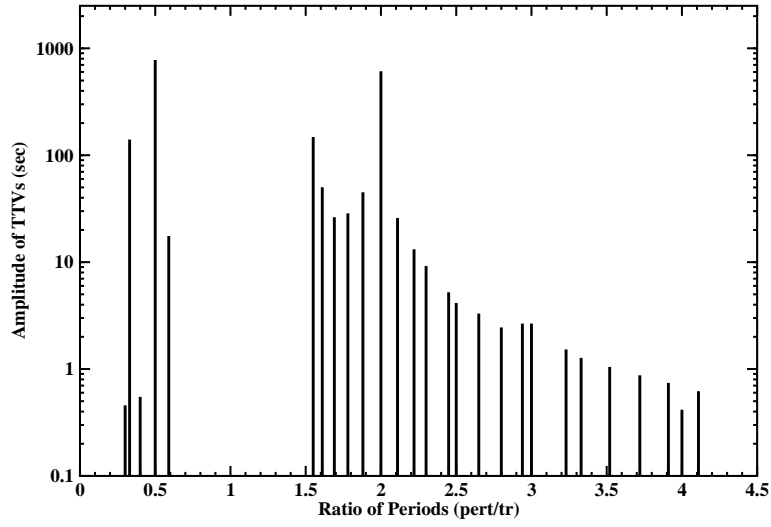}}
\vskip 30pt
\caption{The maximum amplitude of TTVs for the system of figure 1 when the perturber
is near different low-order interior and exterior mean-motion resonances.}  
\end{figure}

\clearpage

\begin{figure}
\centering{
\includegraphics[width=10cm]{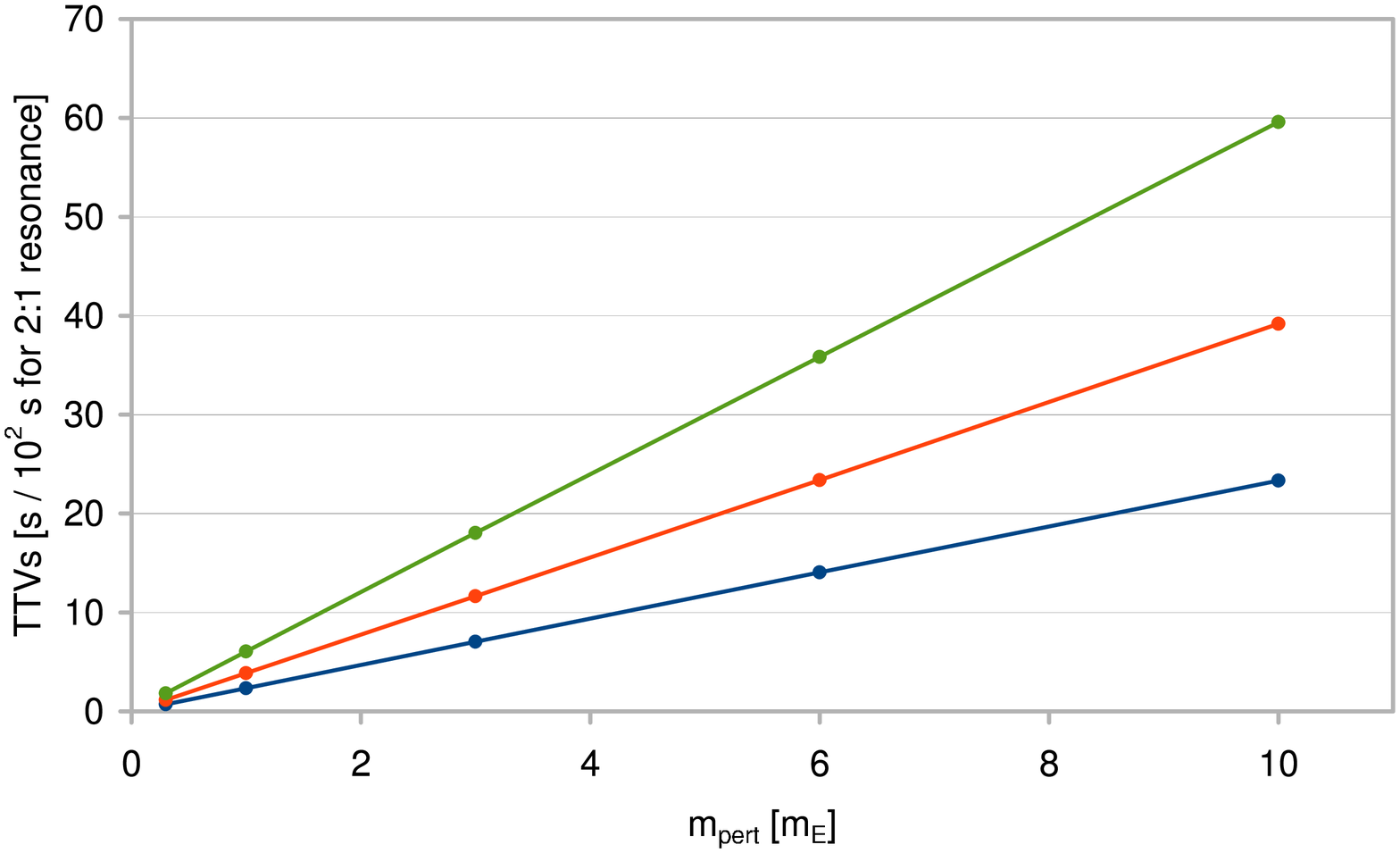}}
\caption{Graph of TTVs for a Jupiter-mass transiting planet in a 10-day orbit
around a 0.367$M_\odot$ star in terms of the mass of the perturber. 
From top to bottom, ${P_{\rm pr}}/{P_{\rm tr}}\sim$ 2, 2.5, 2.8, respectively.
Note that for the top graph where the orbits are near a 2:1 MMR, the values of TTVs
on the vertical axis are multiplied by 100 s.}  
\end{figure}

\clearpage

\begin{figure}
\centering{
\includegraphics[width=10cm]{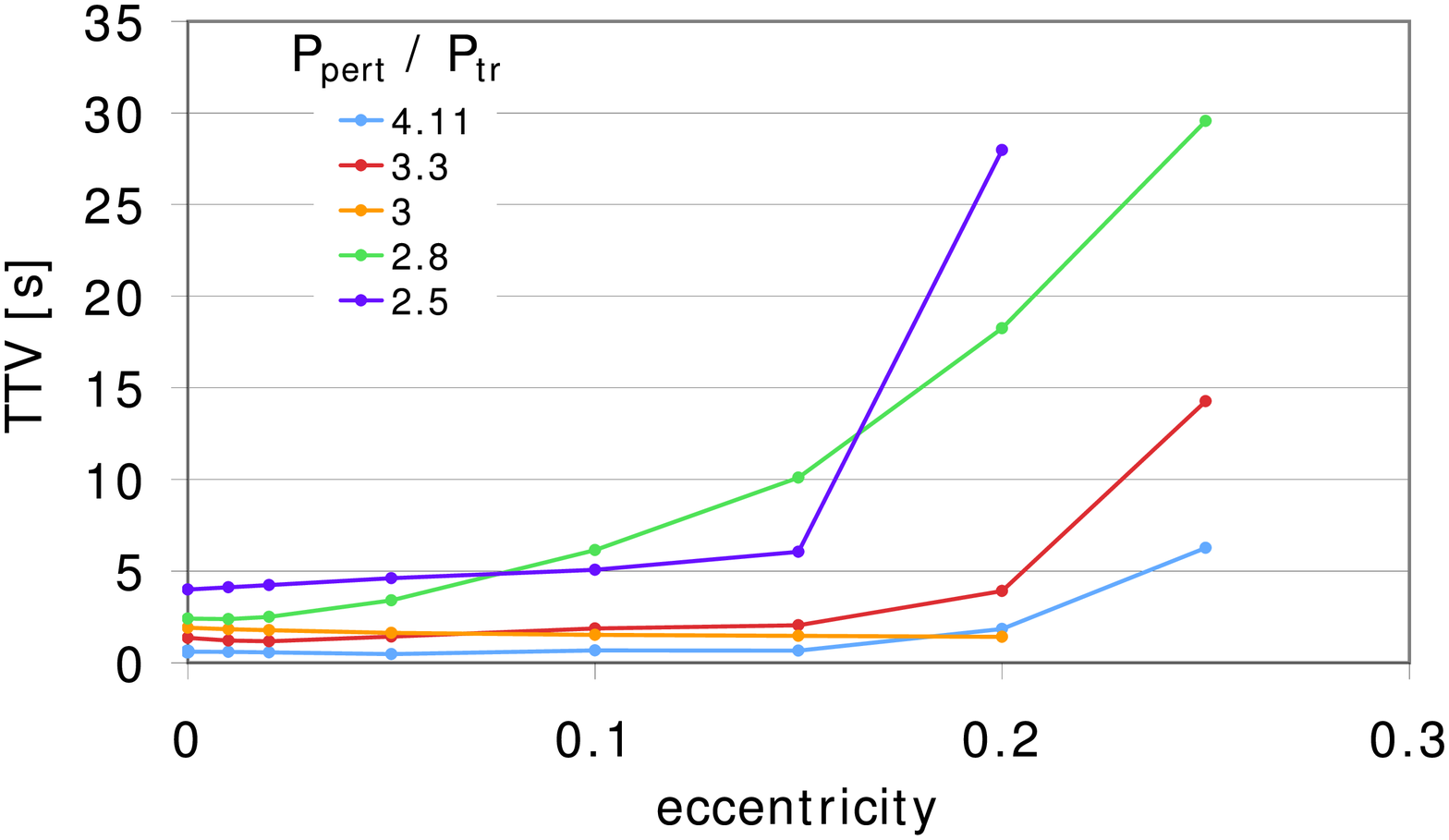}}
\caption{Graph of TTVs  for a Jupiter-mass transiting planet in a 10-day orbit
around a 0.367$M_\odot$ star in terms of the eccentricity of the perturber. The mass of
the perturber is 1$M_\oplus$. Note that 
the ratio of periods as given in the figure are approximate values.}  
\end{figure}

\clearpage

\begin{figure}
\centering{
\includegraphics[width=10cm]{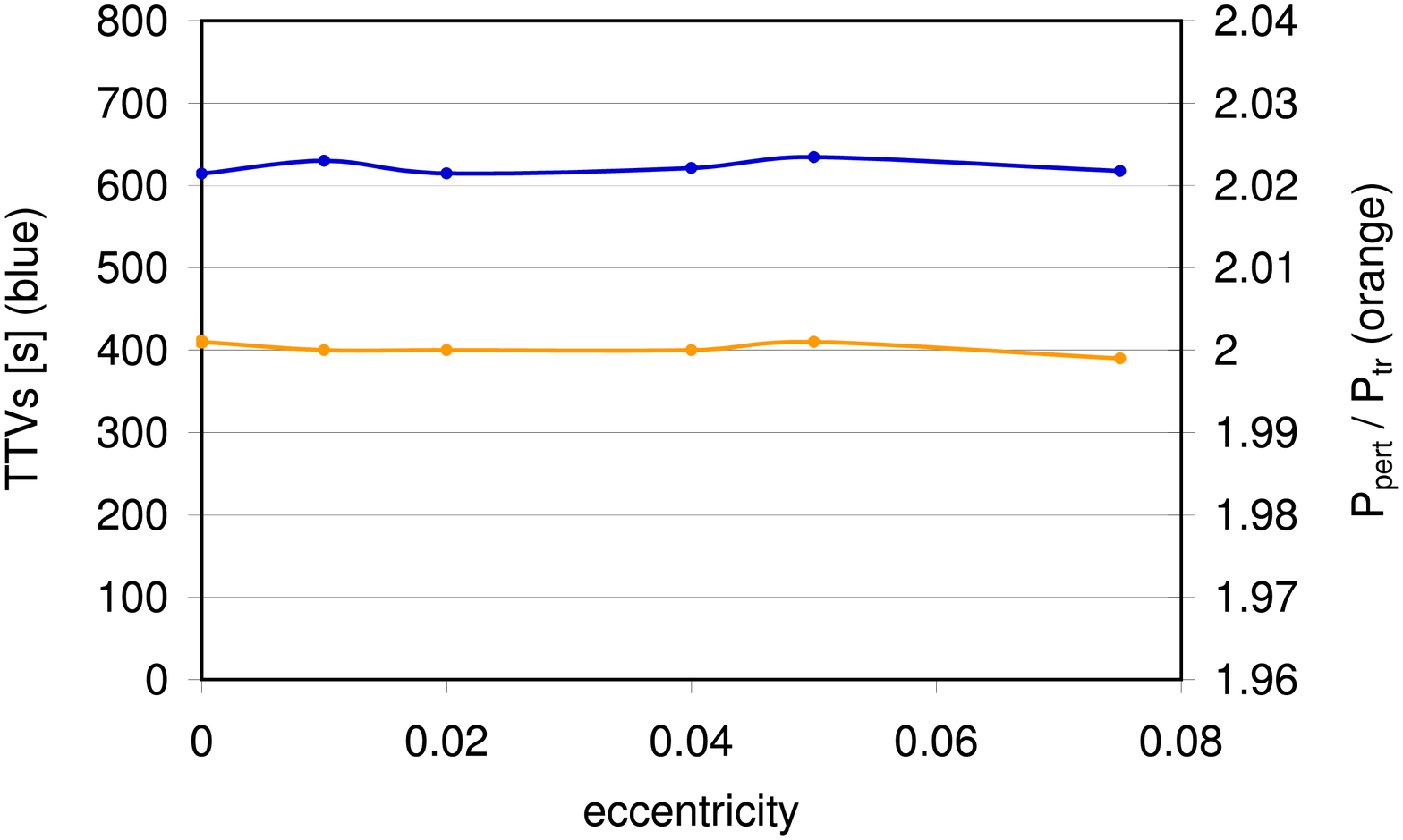}}
\caption{Graph of the amplitude of TTV (top curve) for the system of figure 6 when the
system is near a 2:1 MMR.}  
\end{figure}

\clearpage

\begin{figure}
\centering{
\includegraphics[width=10cm]{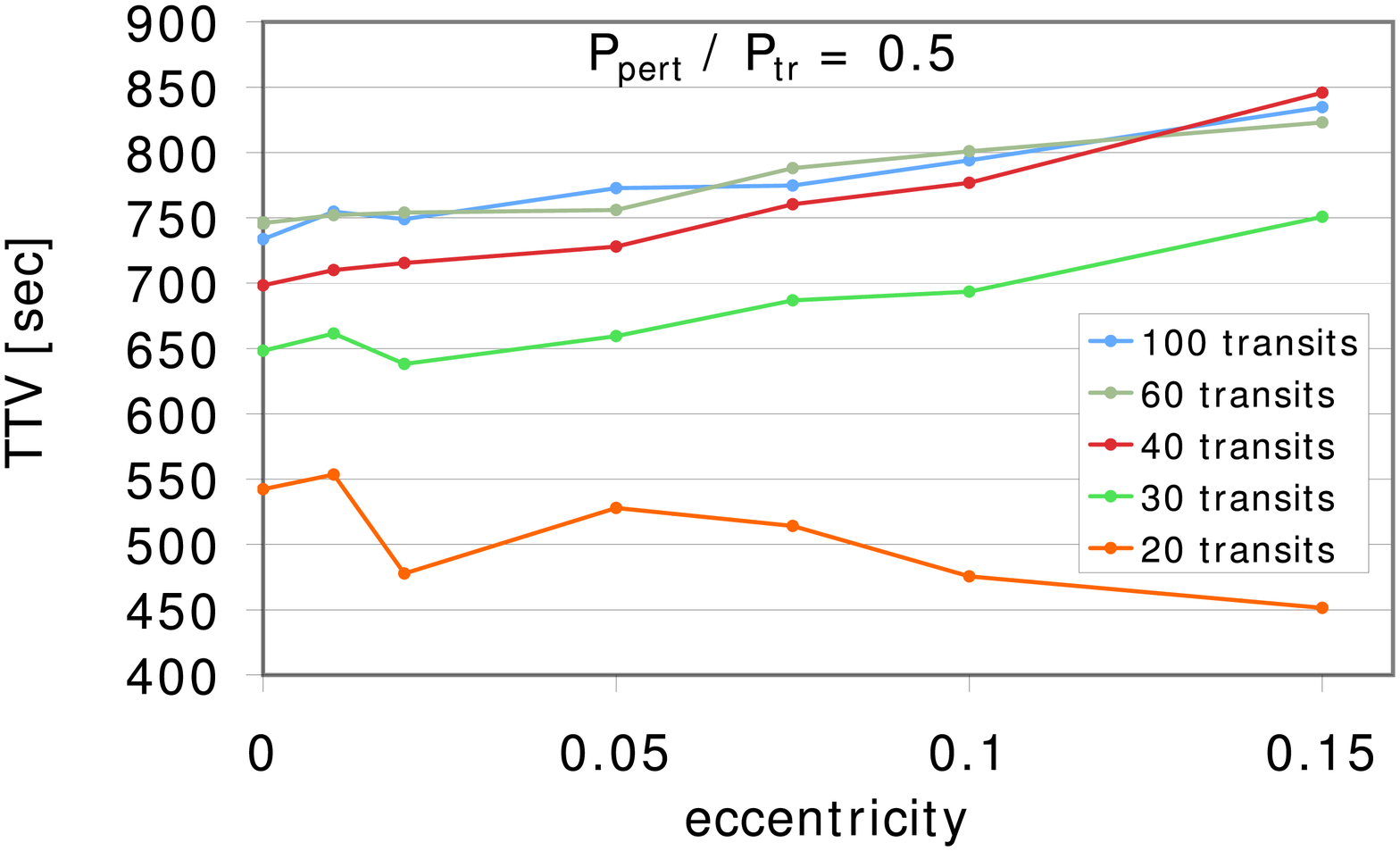}}
\caption{Amplitude of TTV in terms of the eccentricity of the perturber
for a system consisting of a 0.367$M_\odot$ star and
a Jupiter-mass transiting planet. The perturber is Earth-mass and 
near an interior 1:2 MMR. Each line represents different time of integration.}  
\end{figure}

\clearpage

\begin{figure}
\centering{
\includegraphics[width=8cm]{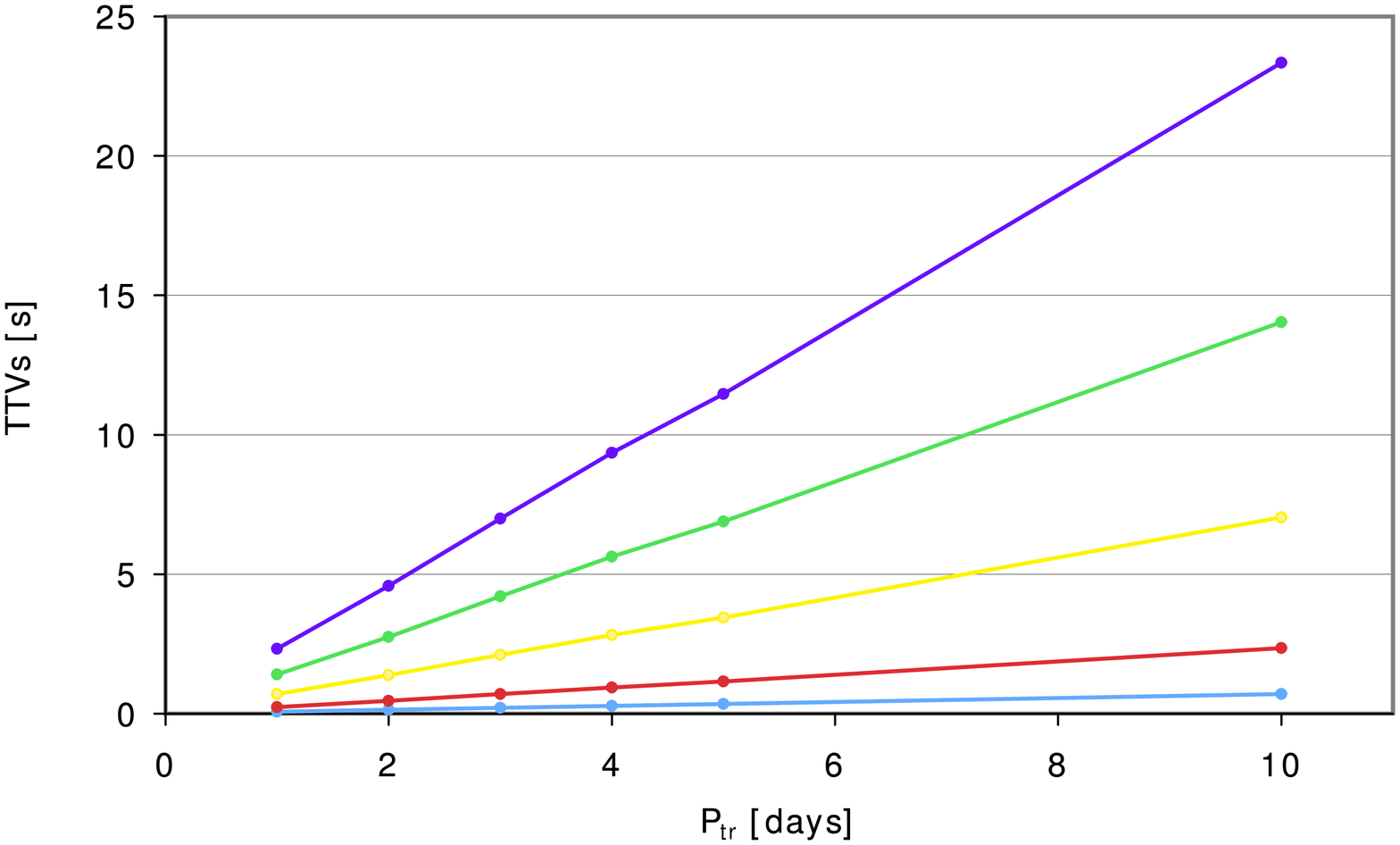}
\vskip 10pt
\includegraphics[width=8cm]{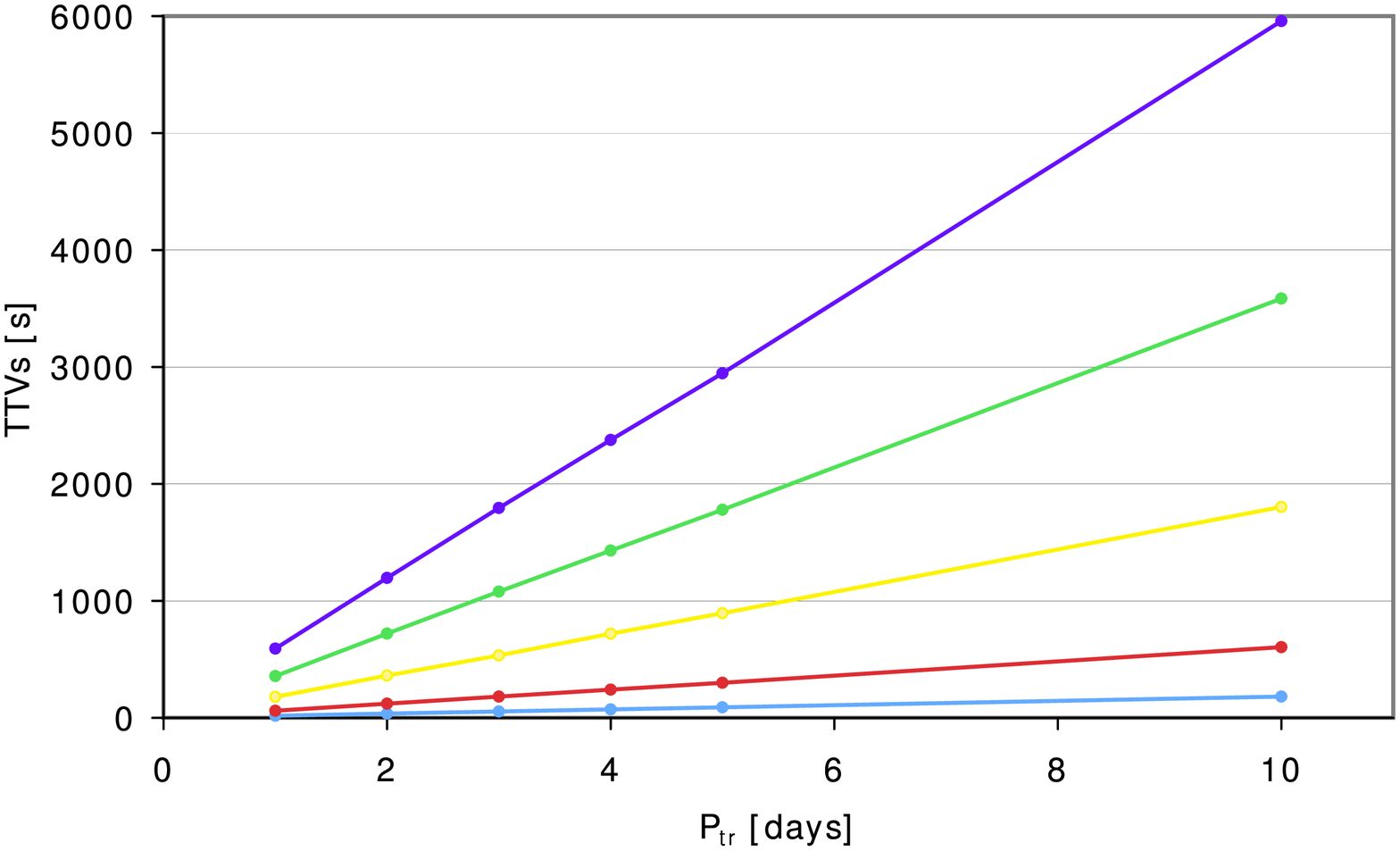}
}
\caption{Amplitude of TTV for different values of the period of a
Jupiter-mass transiting planet.The star is 0.367$M_\odot$. The graph on the
top corresponds to ${P_{\rm pr}}/{P_{\rm tr}}\sim$ 2.8, and the one on the bottom
is for ${P_{\rm pr}}/{P_{\rm tr}}\sim$ 2. The lines in each graph correspond to
different values of the mass of the perturber (from top to bottom: 10, 6, 3, 1, 0.3
$M_\oplus$).}  
\end{figure}

\clearpage

\begin{figure}
\centering{
\includegraphics[width=10cm]{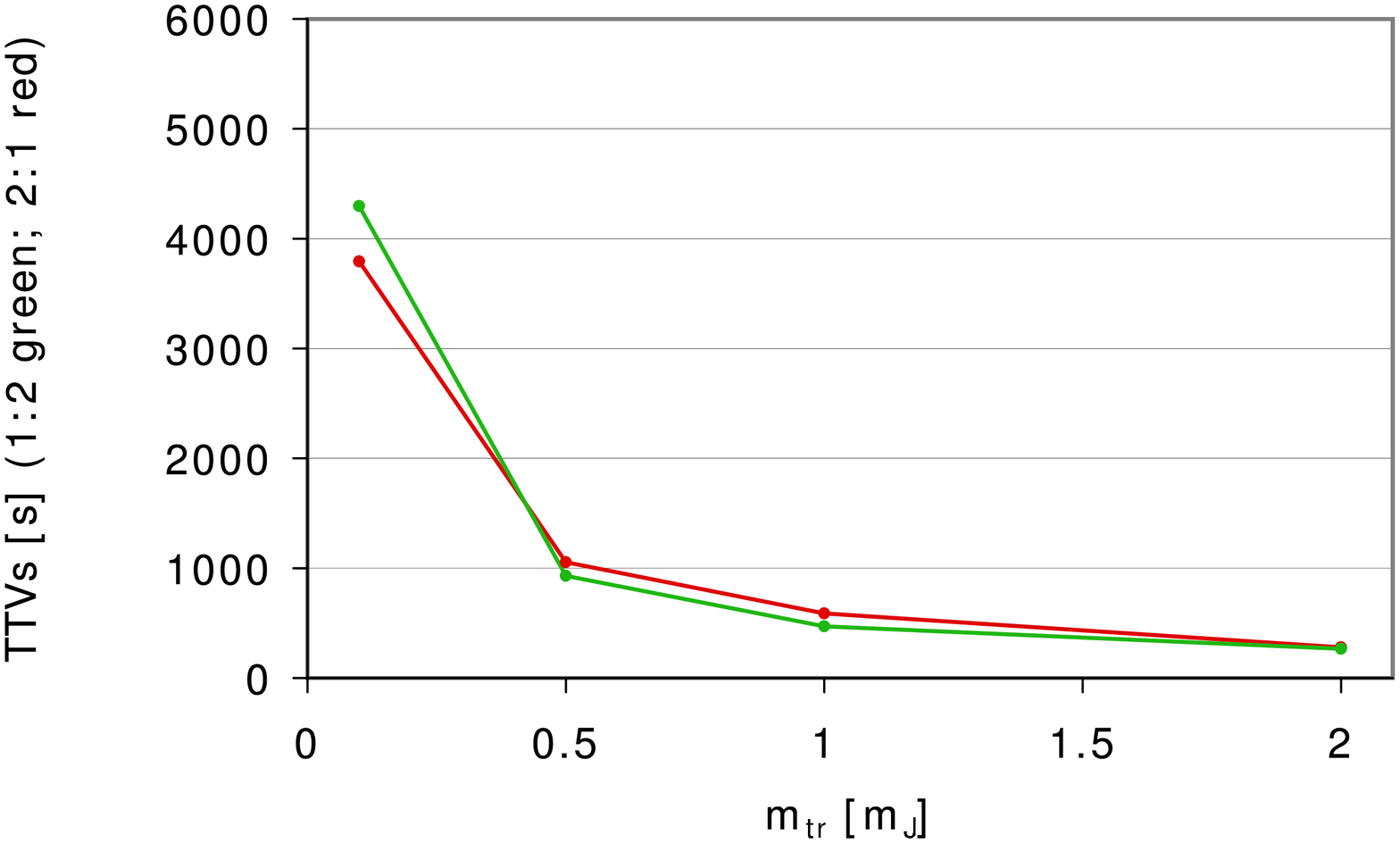}}
\caption{The graph of the amplitude of TTV in terms of the mass of the transiting planet
for two cases of near 1:2 and 2:1 MMR. The star is 0.367$M_\odot$ and the 
perturber is Earth-mass. The transiting planet is in a 10-day orbit.}  
\end{figure}

\clearpage

\begin{figure}
\centering{
\includegraphics[width=10cm]{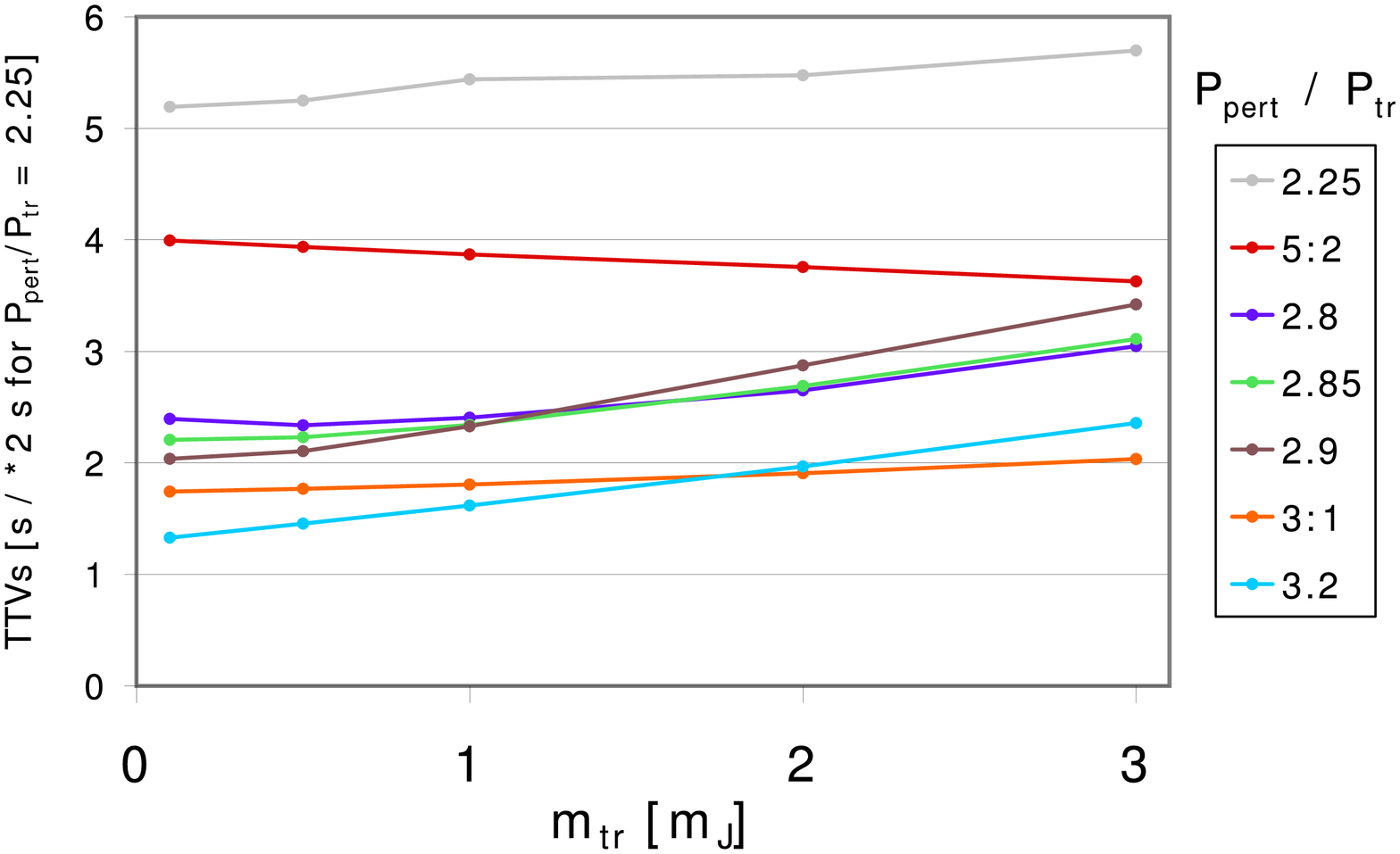}
\vskip 10pt
\includegraphics[width=10.5cm]{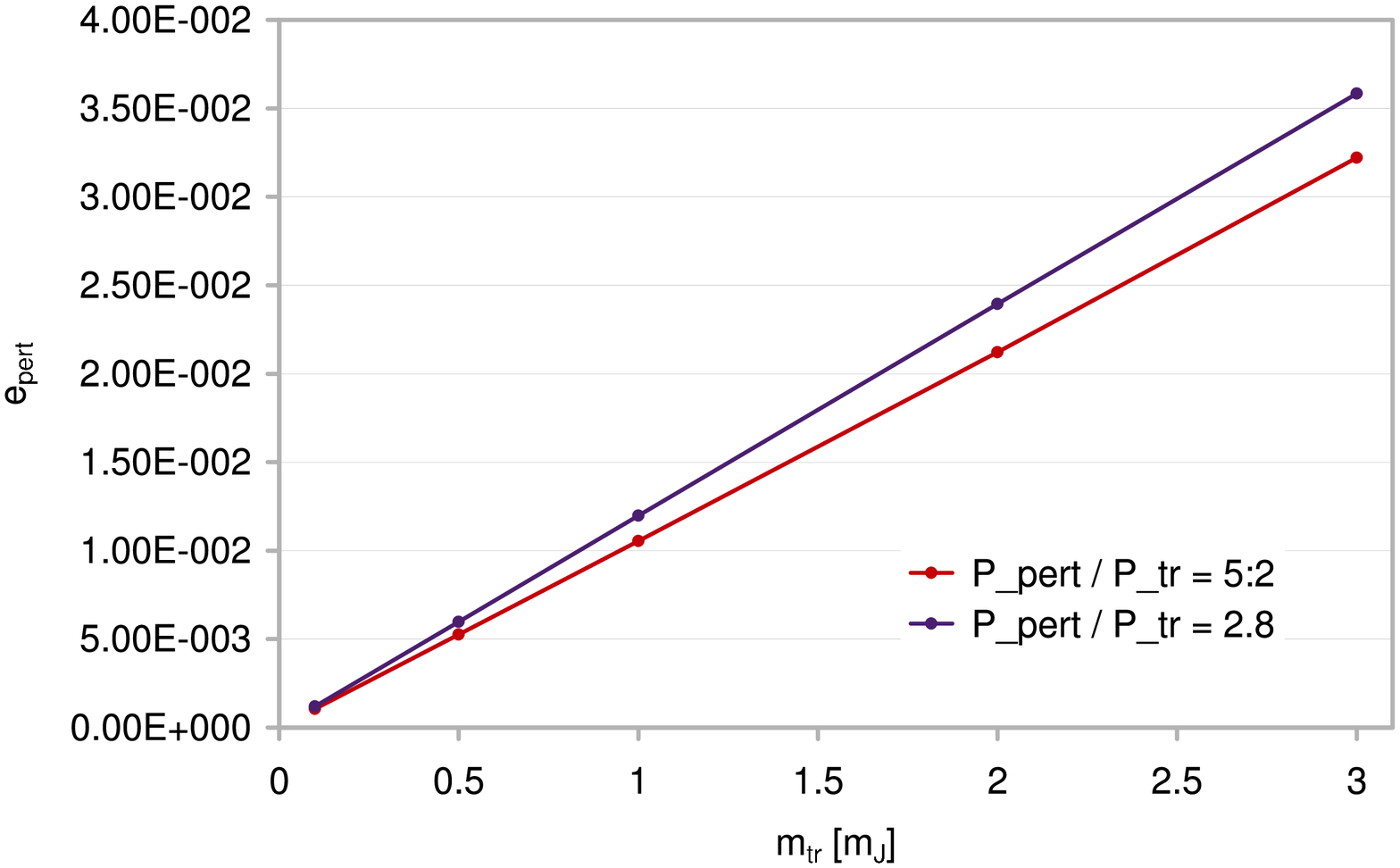}
}
\caption{Top: The graph of the amplitude of TTV for the system of figure 10 and for
different non-resonant cases, and high order resonant orbits. Note that  in the case 
of ${P_{\rm pr}}/{P_{\rm tr}}\sim$ 2.25, the values on the vertical axis are multiplied 
by 2. Bottom: Graph of the median eccentricity of the perturbing planet for a non-resonant 
system with a period-ratio of 2.8 and for the system in 5:2 MMR.}  
\end{figure}

\clearpage

\begin{figure}
\centering{
\includegraphics[width=9cm]{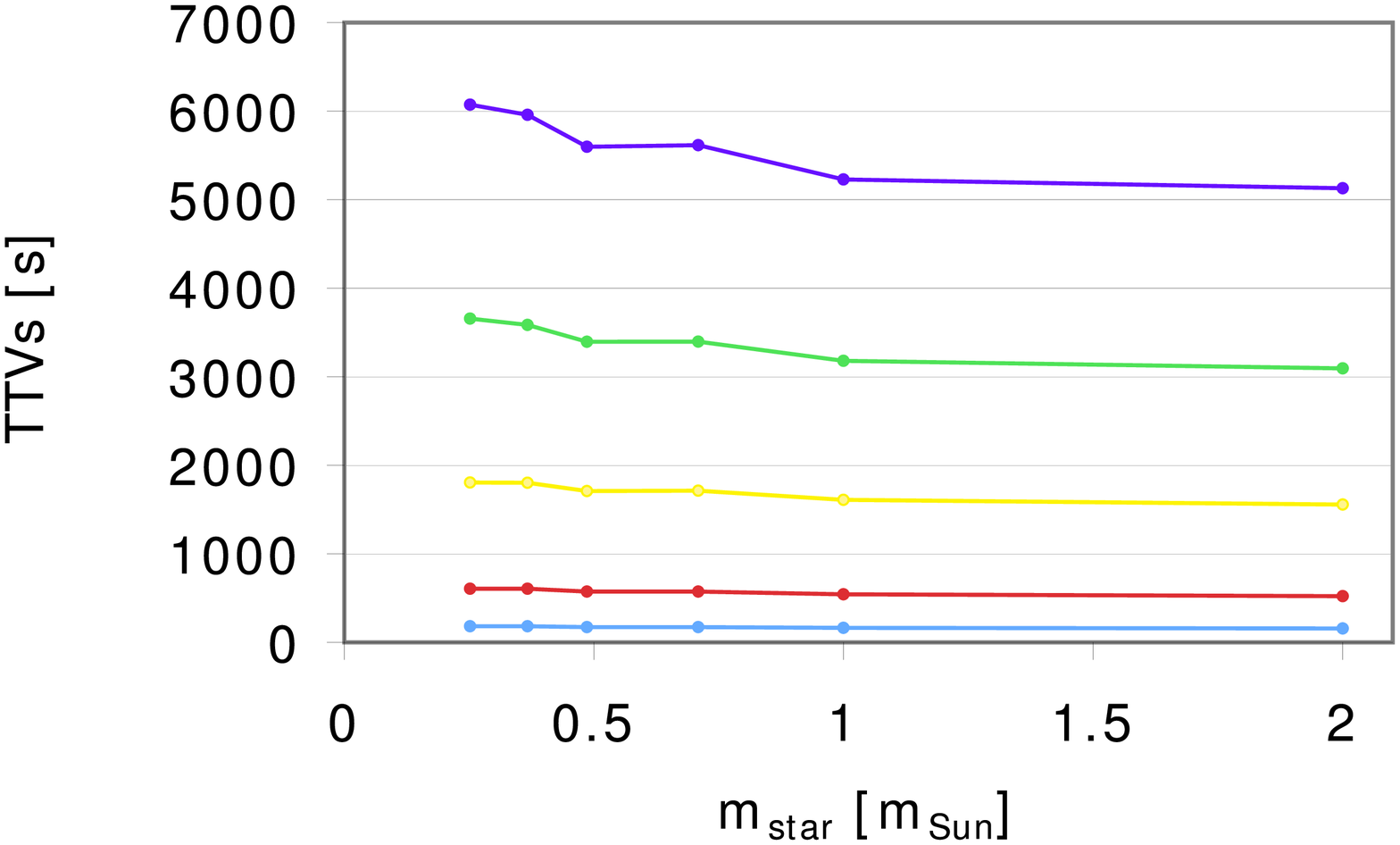}
\vskip 10pt
\includegraphics[width=8cm]{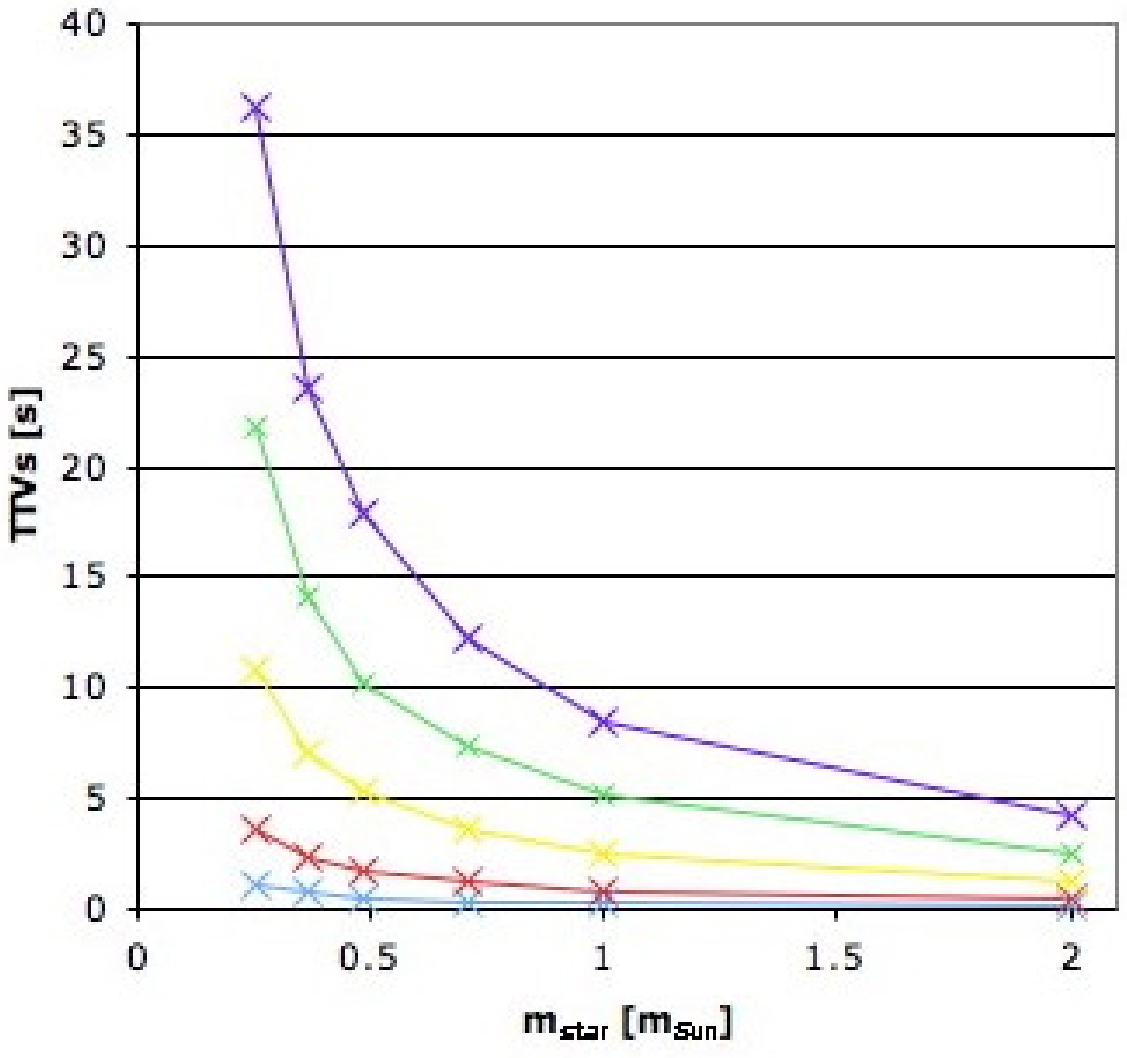}
}
\caption{Graphs of the amplitude of TTV in terms of the mass of the central star.
The perturber is a Jupiter-mass planet in a 10-day orbit. The top panel corresponds
to a near exterior 2:1 MMR and the bottom panel shows a non-resonant system where
${P_{\rm pr}}/{P_{\rm tr}}\sim$ 2.83. The graphs in each panel correspond to
a perturber with a mass of 10, 6, 3, 1, and 0.3 $M_\oplus$, from top to bottom.}  
\end{figure}

\clearpage

\begin{figure}
\centering{
\includegraphics[width=12cm]{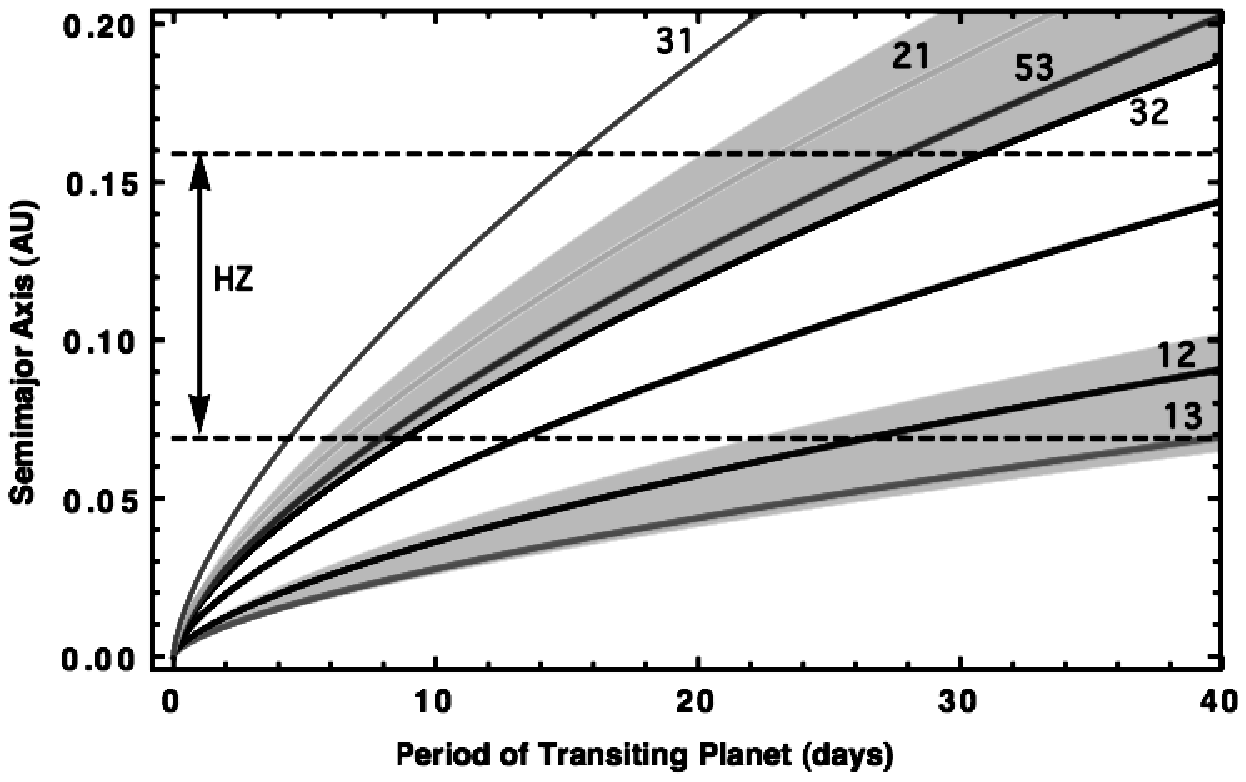}
\vskip 10pt
\includegraphics[width=12cm]{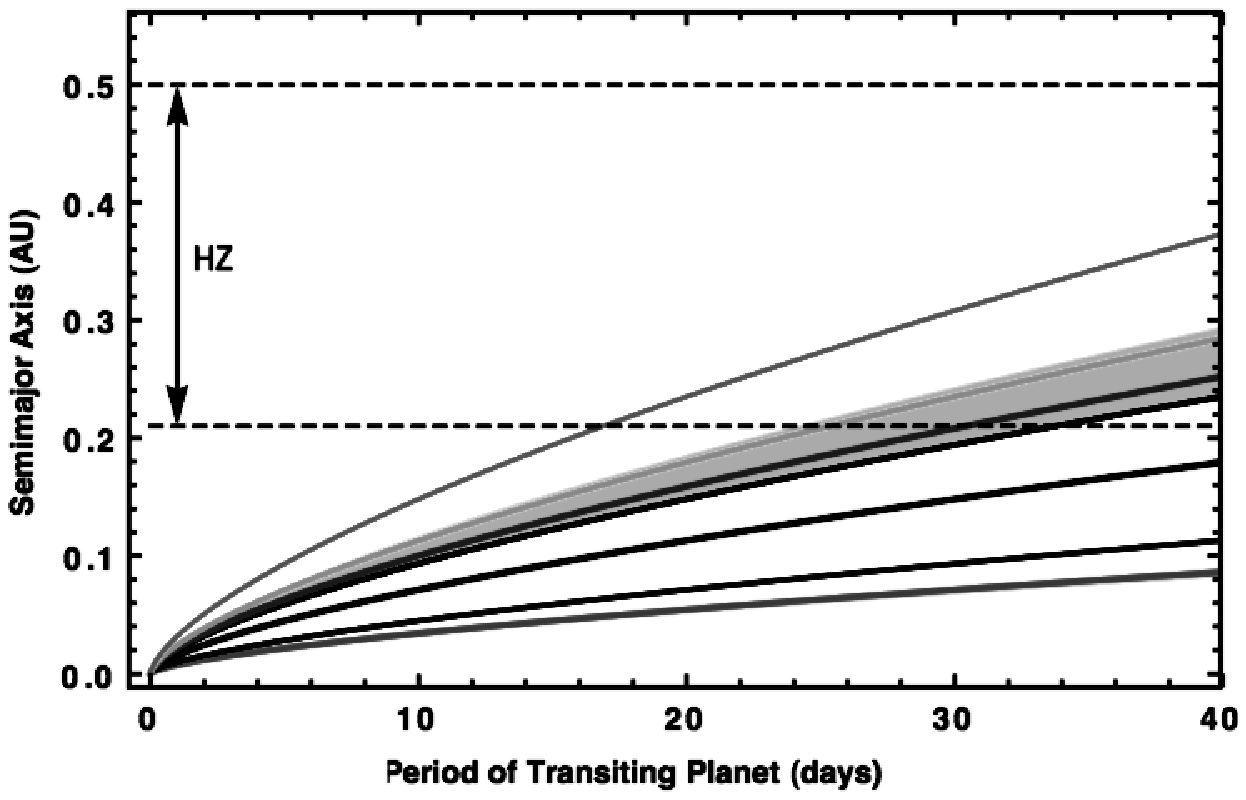}
}
\caption{Graphs of the regions where a 1$M_\oplus$ perturber creates
TTVs with amplitudes of 20 s or larger on a Jupiter-mass planet. The mass of the
central star is 0.252$M_\odot$ in the top panel and 0.486$M_\odot$ in the bottom.
The numbers on each graph are the mean-motion resonances. The un-numbered curve is the
transiting planet.}  
\end{figure}

\clearpage

\begin{figure}
\centering{
\includegraphics[width=12cm]{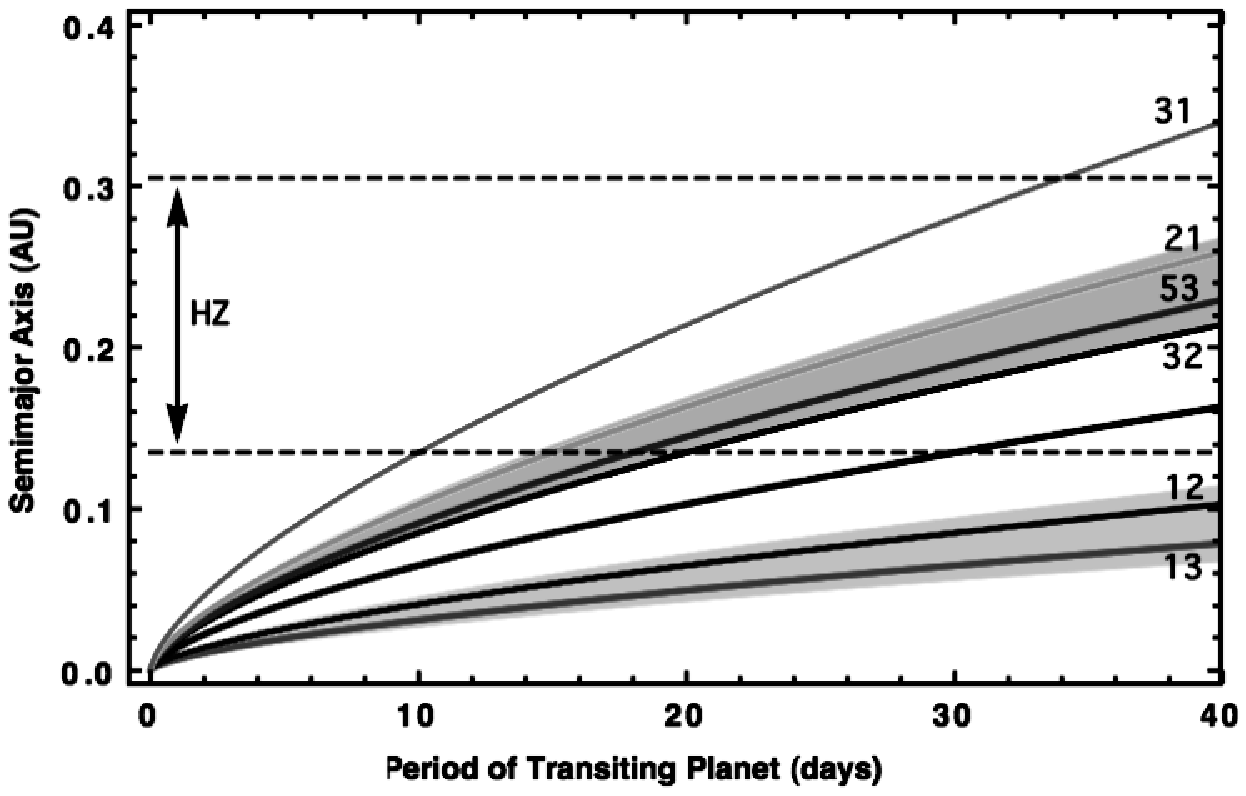}
\vskip 10pt
\includegraphics[width=12cm]{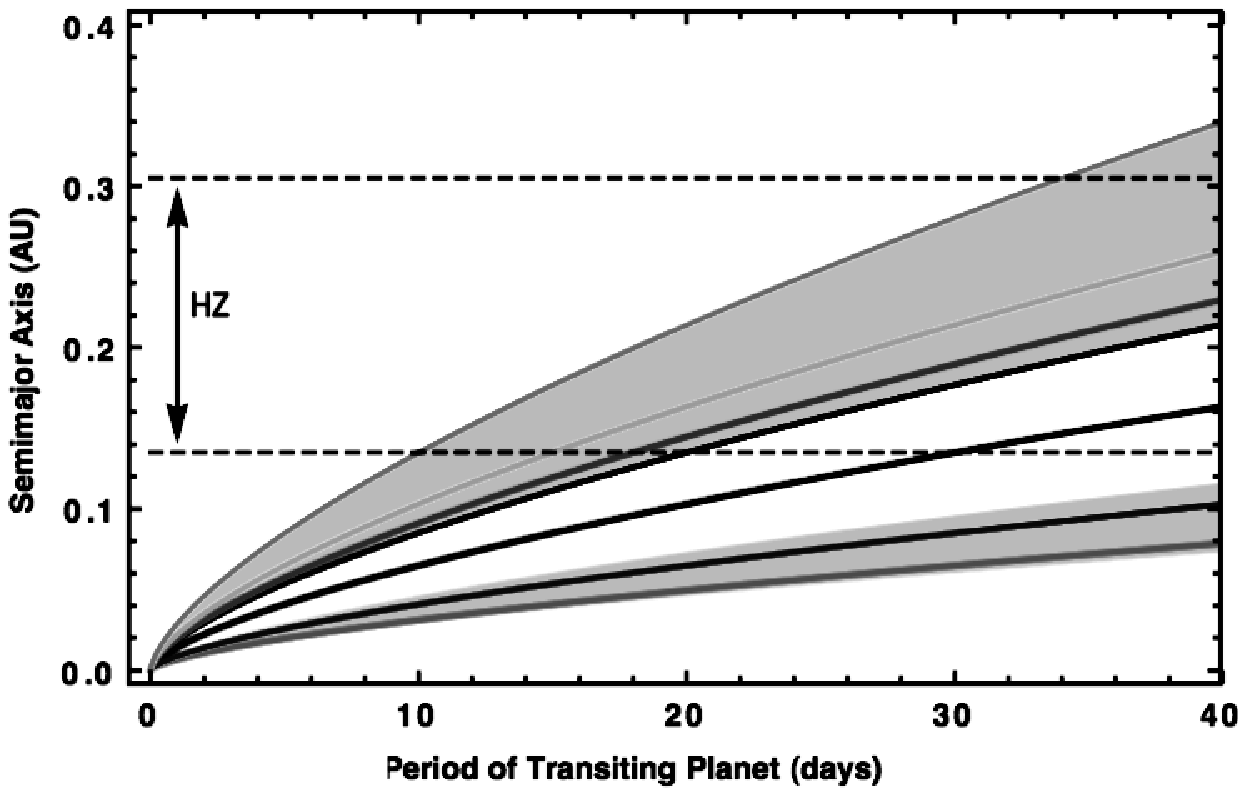}
}
\caption{Similar graph as in figure 13 for a 0.367$M_\odot$ central star.
The top panel corresponds to an Earth-mass perturber, and the bottom panel
is for a 10$M_\oplus$ super-Earth.The numbers on each graph are the mean-motion resonances. 
The un-numbered curve is the transiting planet.}  
\end{figure}

\clearpage

\begin{figure*}
\begin{center}
\hbox{
\includegraphics[width=6cm]{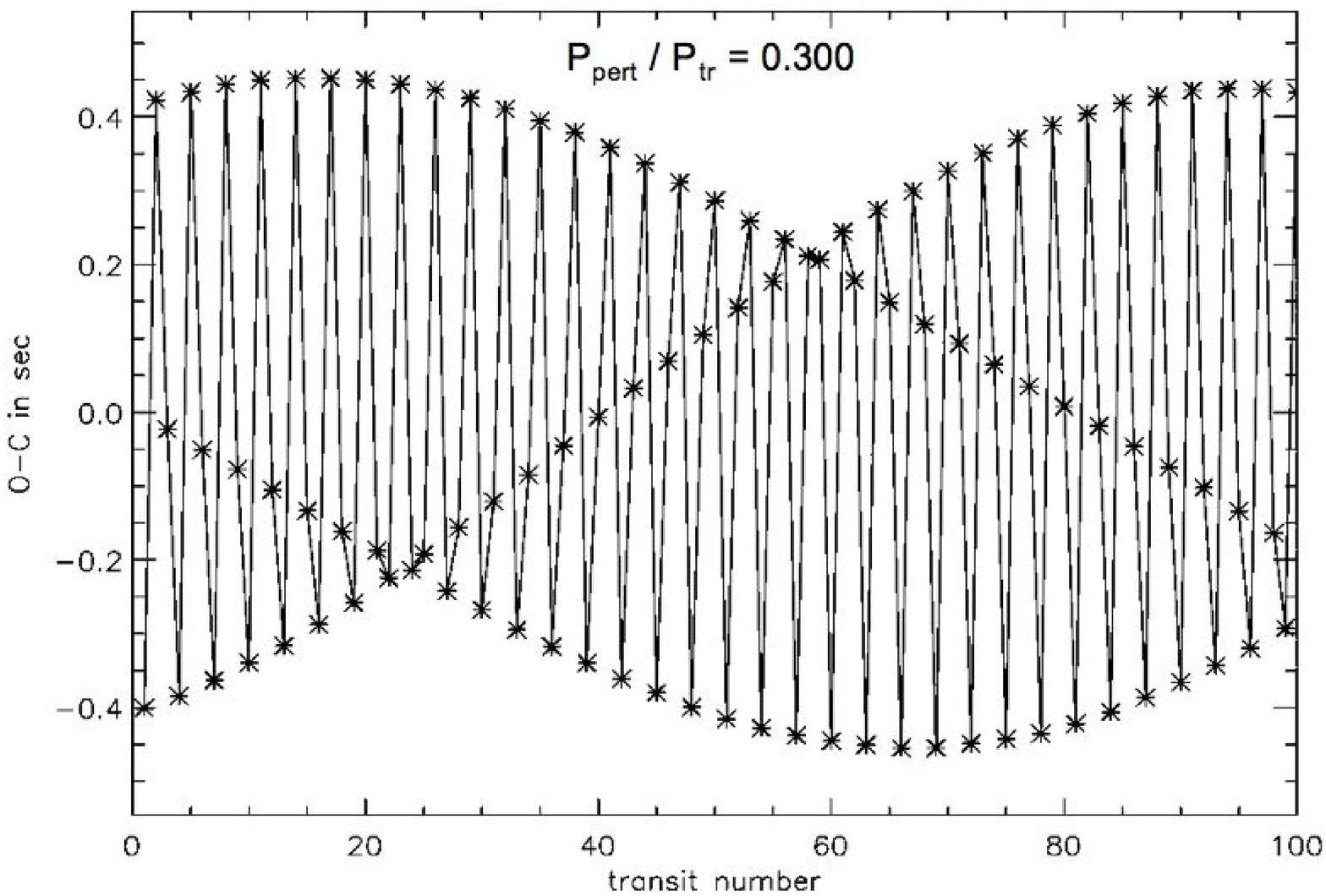}
\includegraphics[width=6cm]{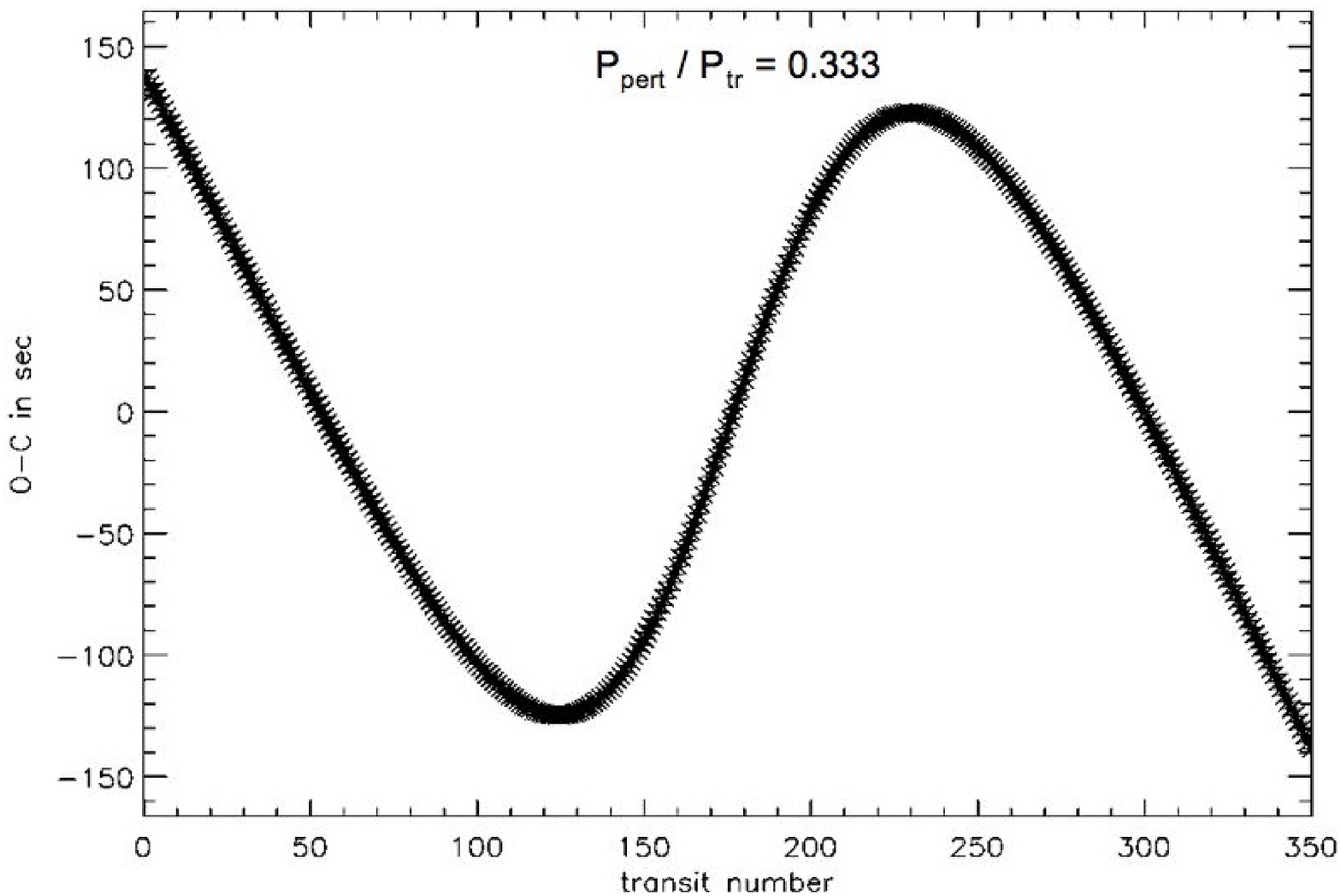}
}
\vskip 2pt
\hbox{
\includegraphics[width=6cm]{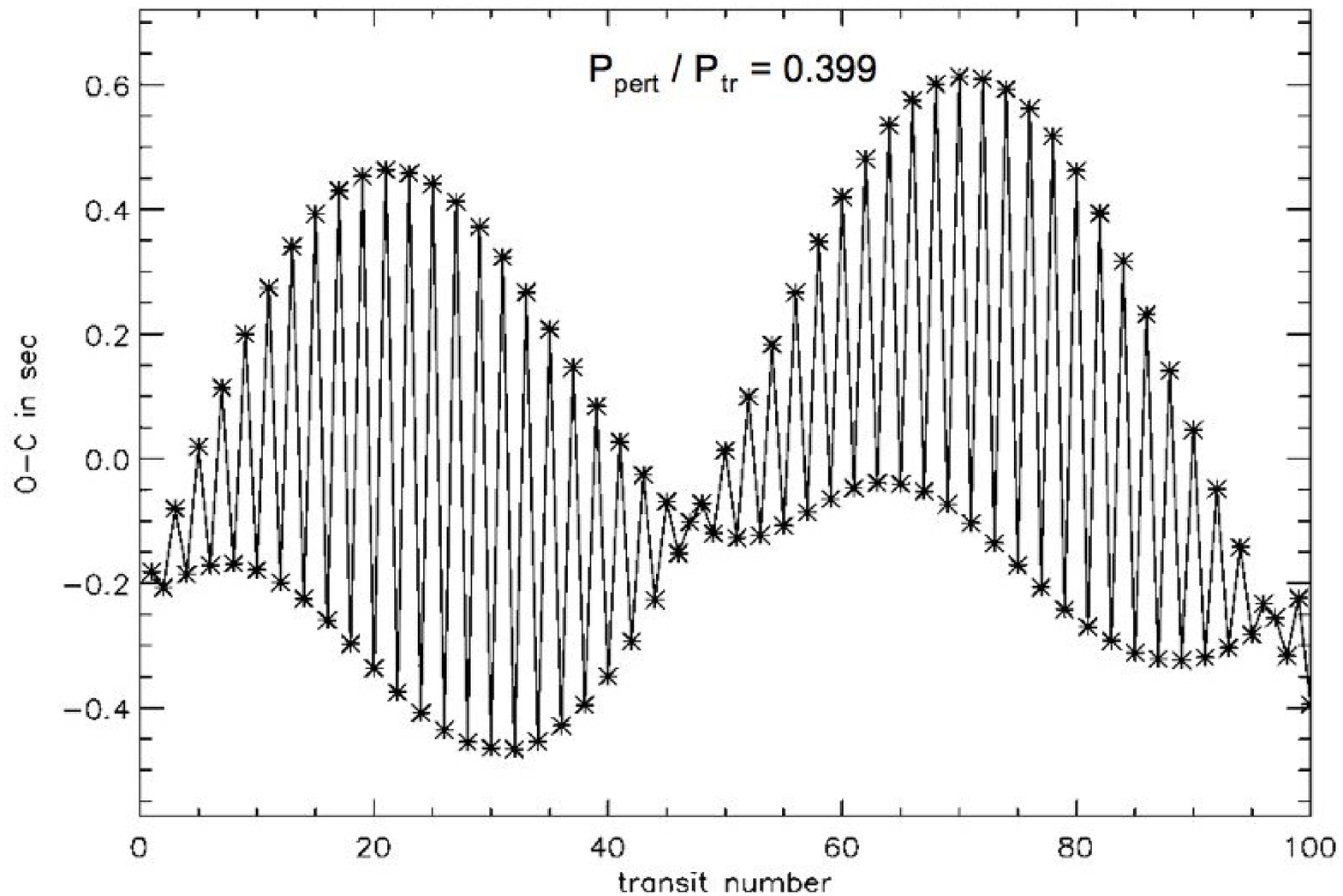}
\includegraphics[width=6cm]{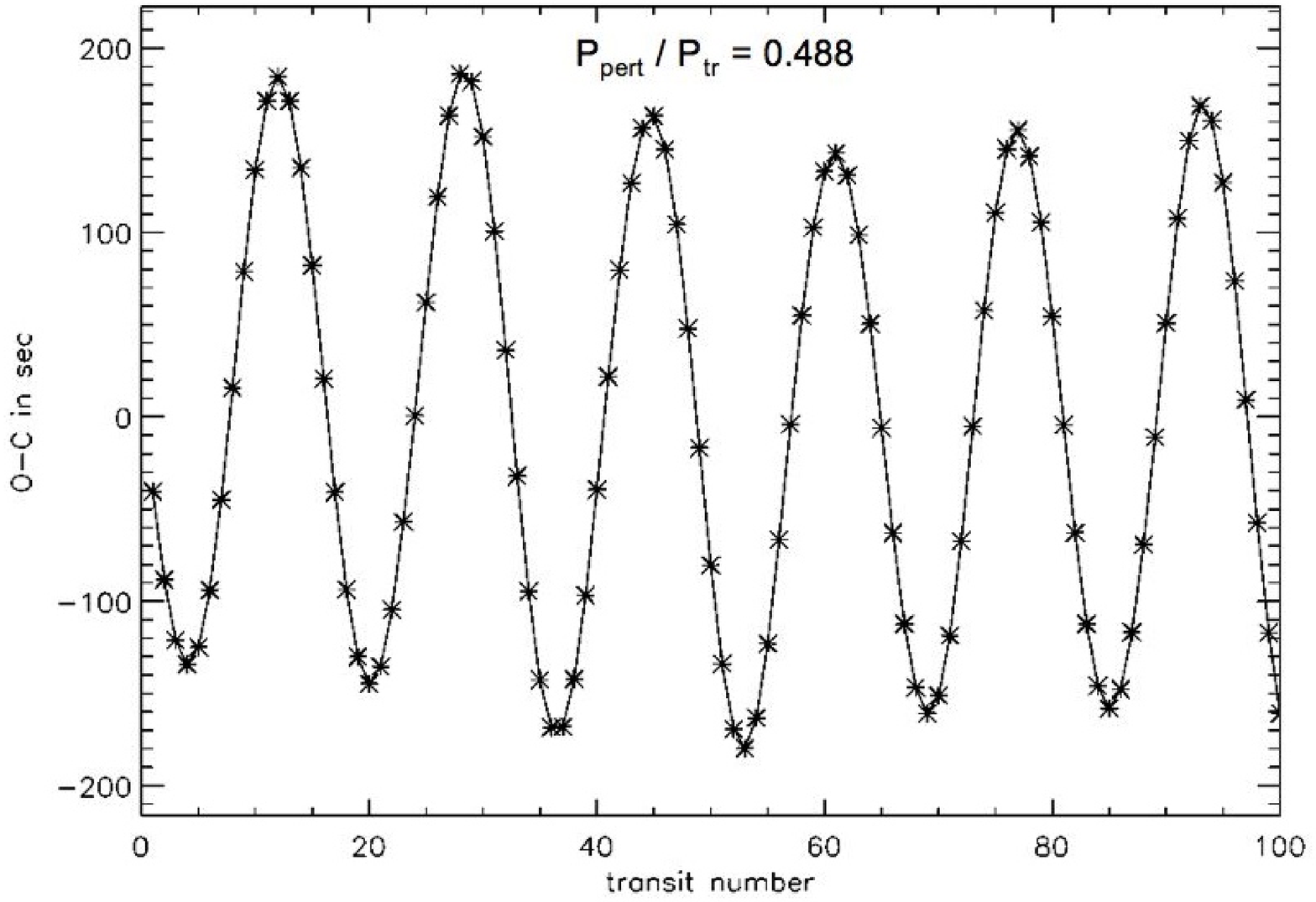}
}
\vskip 2pt
\hbox{
\includegraphics[width=6cm]{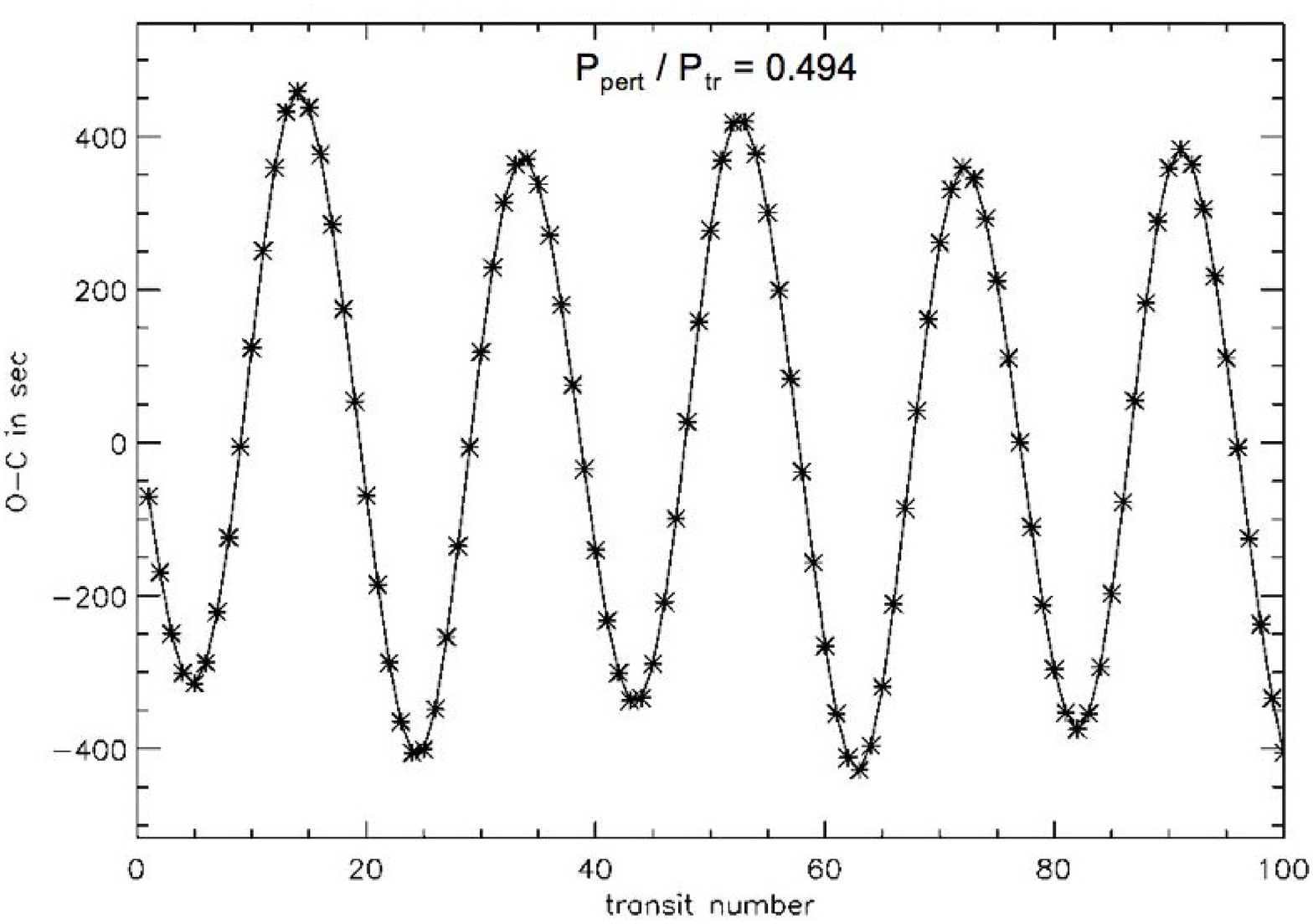}
\includegraphics[width=6cm]{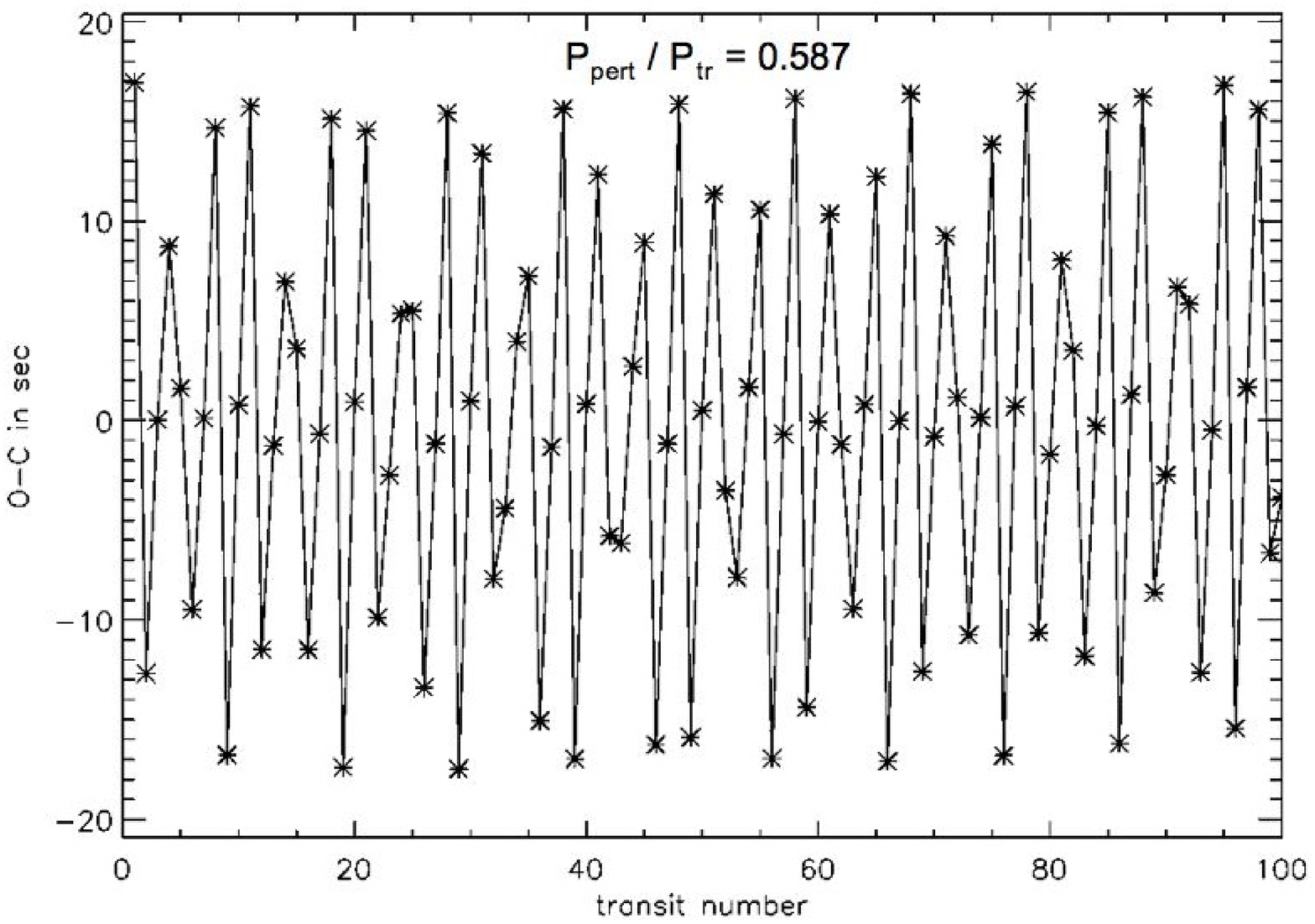}
}
\vskip 2pt
\hbox{
\includegraphics[width=6cm]{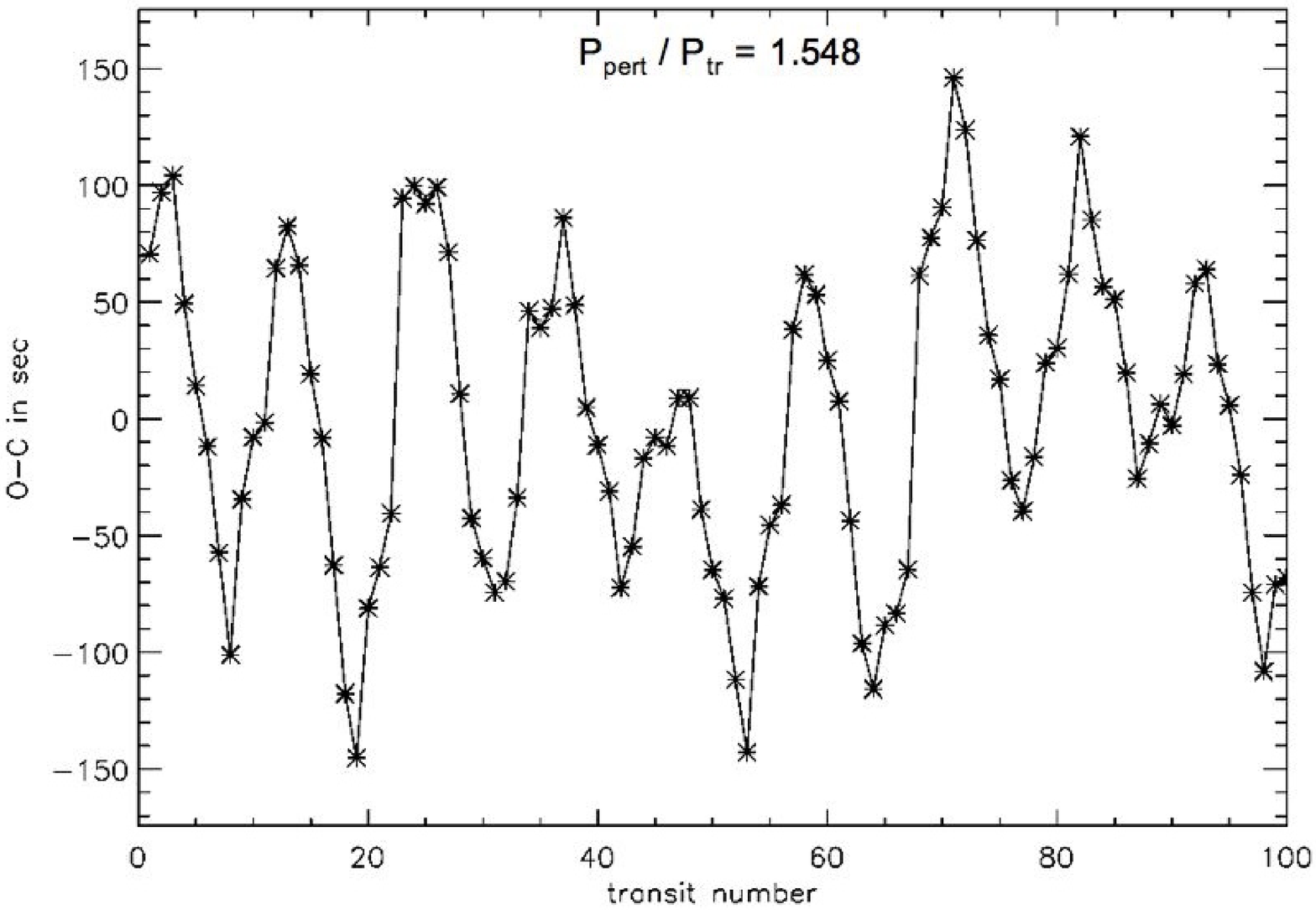}
\includegraphics[width=6cm]{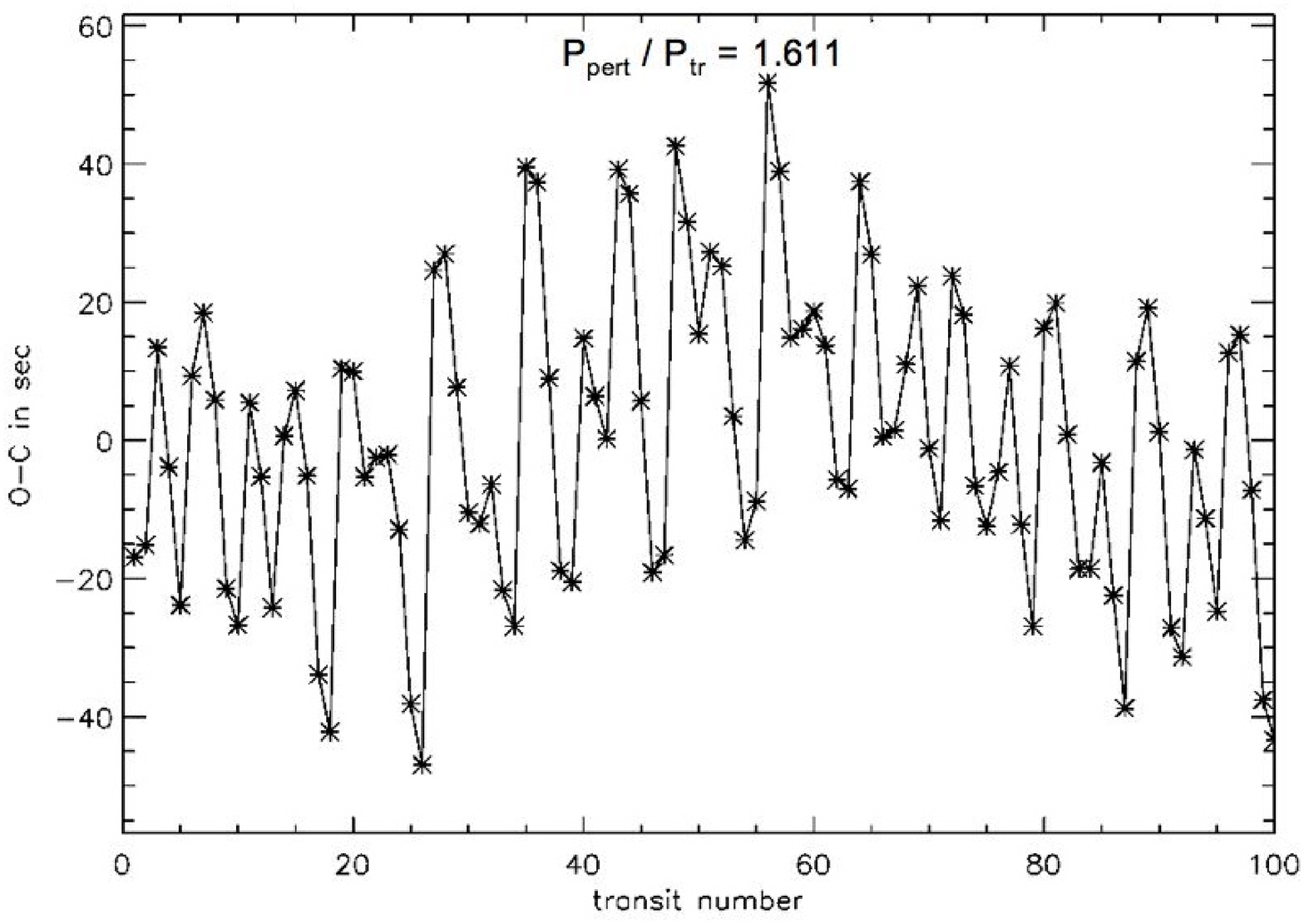}
}
\hbox{
{\bf Appendix}: TTVs for a system with ${m_{\rm star}}={M_\odot},
{m_{\rm tr}}=1 {M_J},\, {m_{\rm pr}}= 1 {M_\oplus}, {P_{\rm tr}}=10$days.}
\end{center}
\end{figure*}

\clearpage

\begin{figure*}
\begin{center}
\hbox{
\includegraphics[width=6cm]{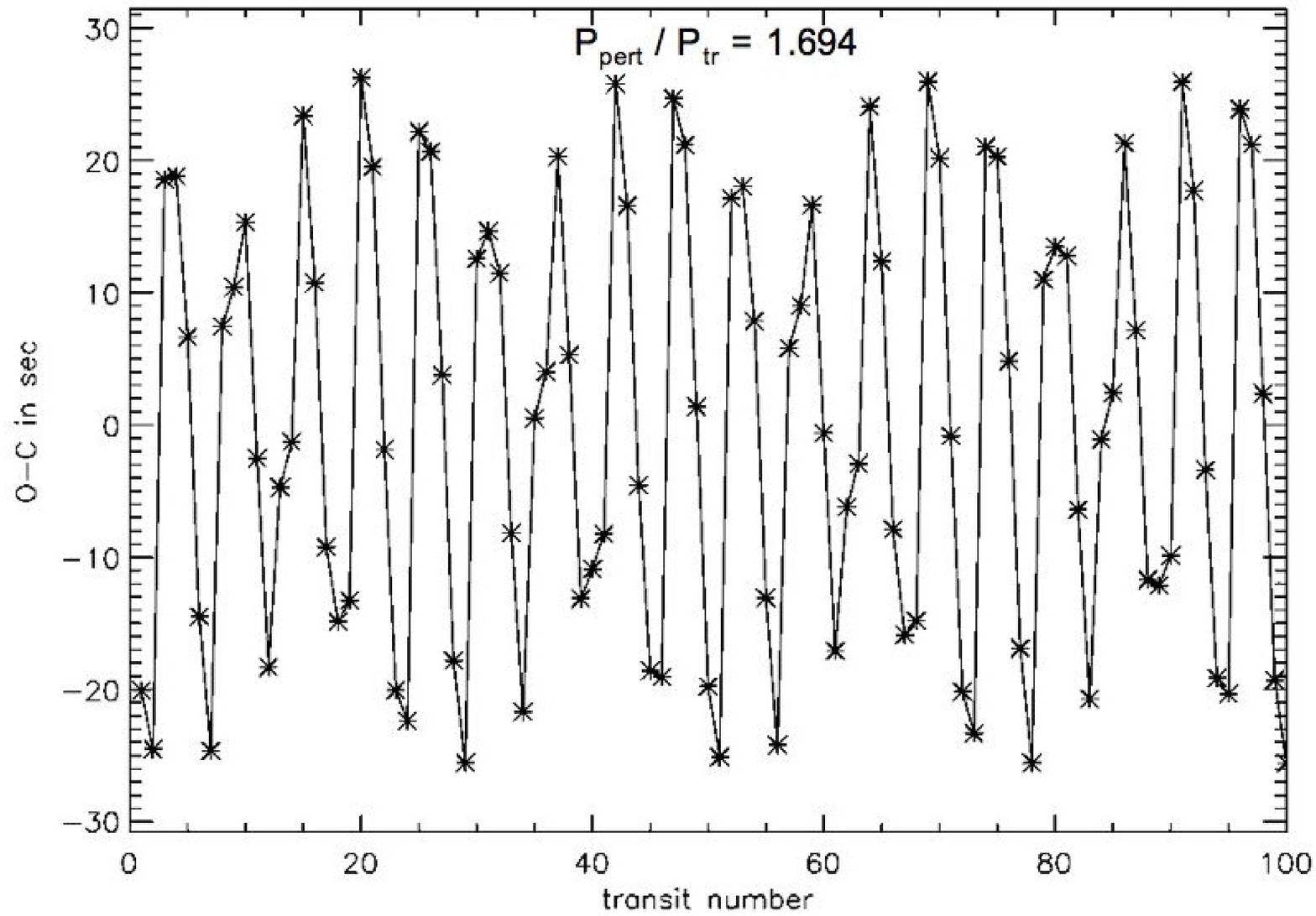}
\includegraphics[width=6cm]{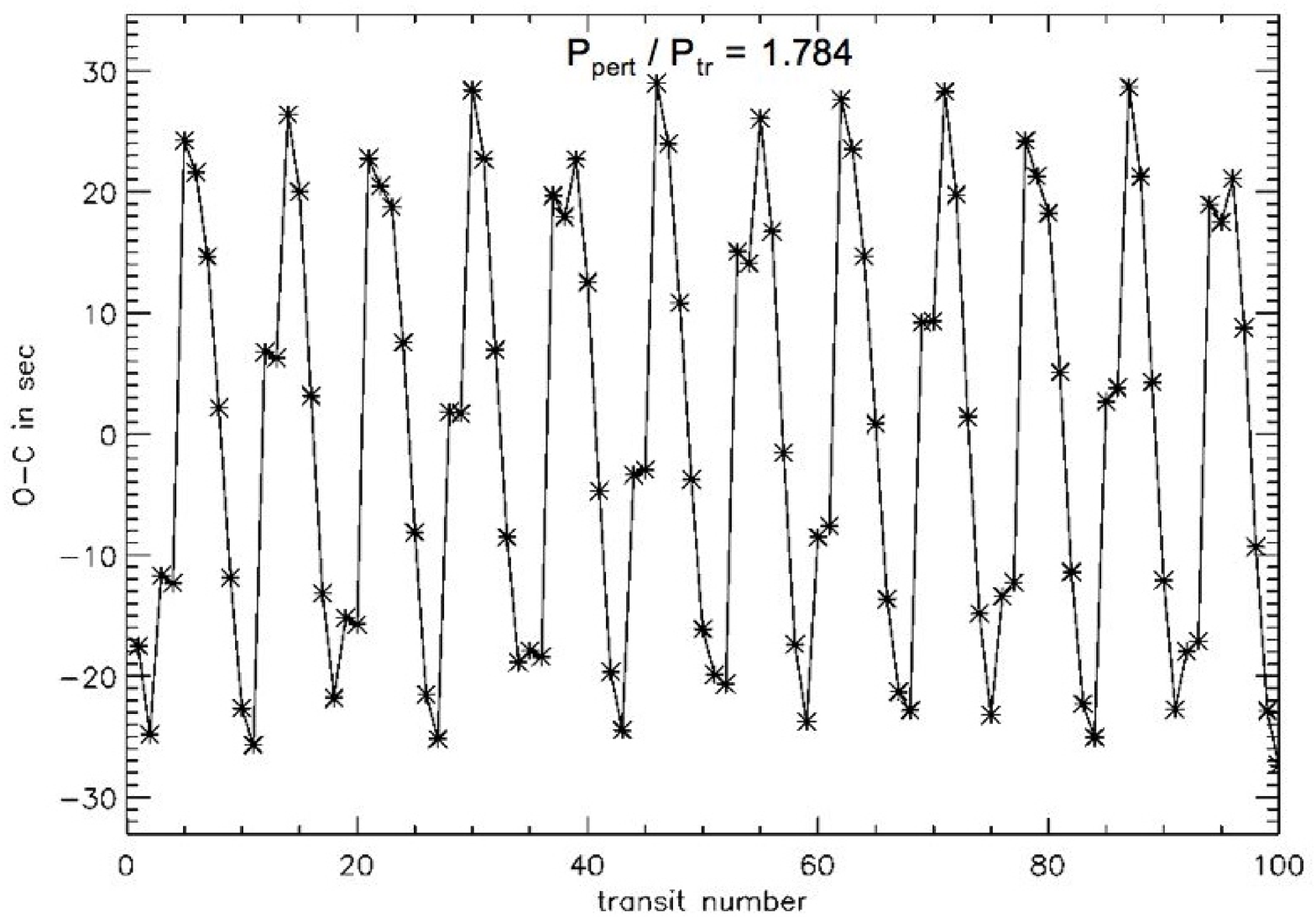}
}
\vskip 2pt
\hbox{
\includegraphics[width=6cm]{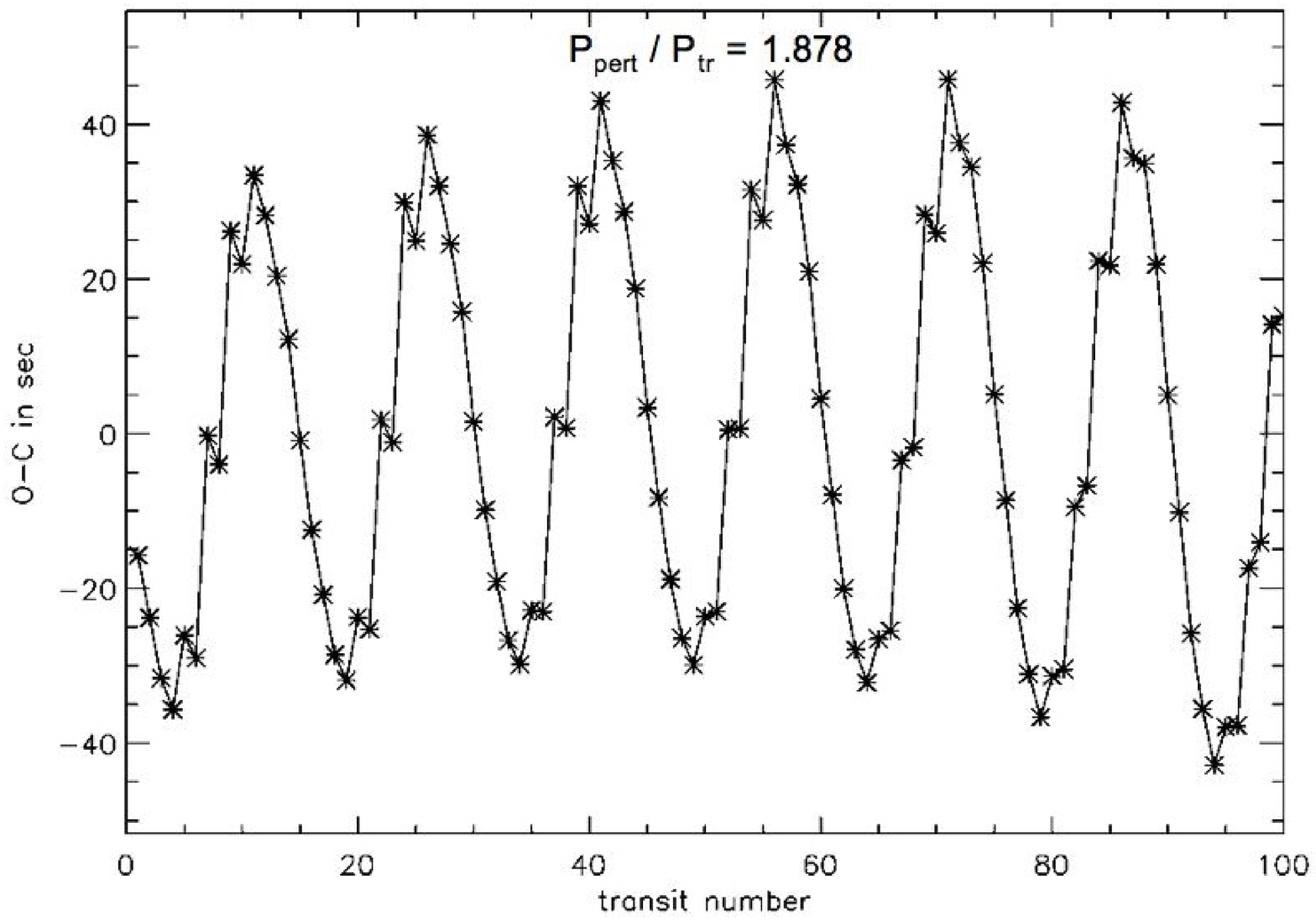}
\includegraphics[width=6cm]{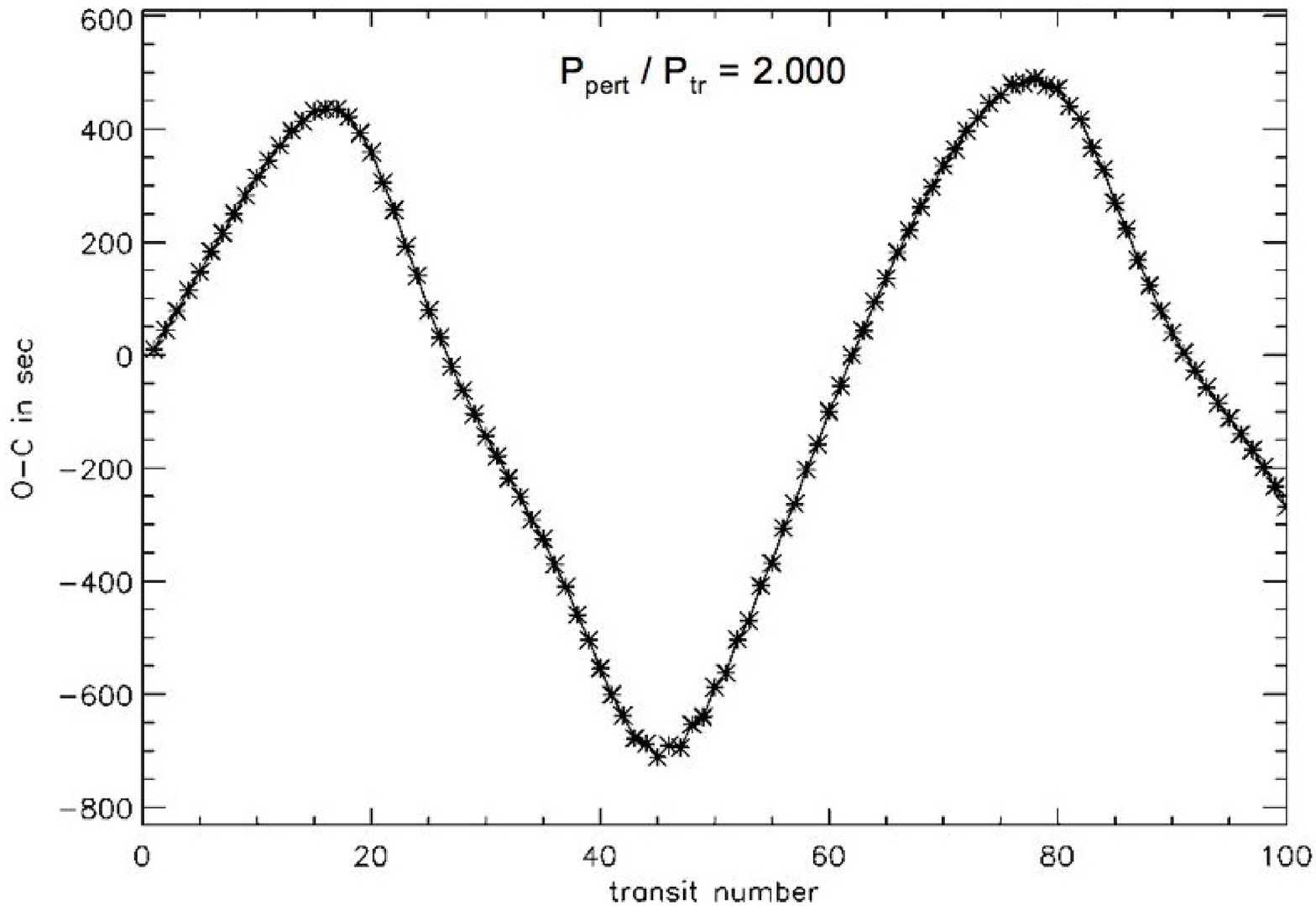}
}
\vskip 2pt
\hbox{
\includegraphics[width=6cm]{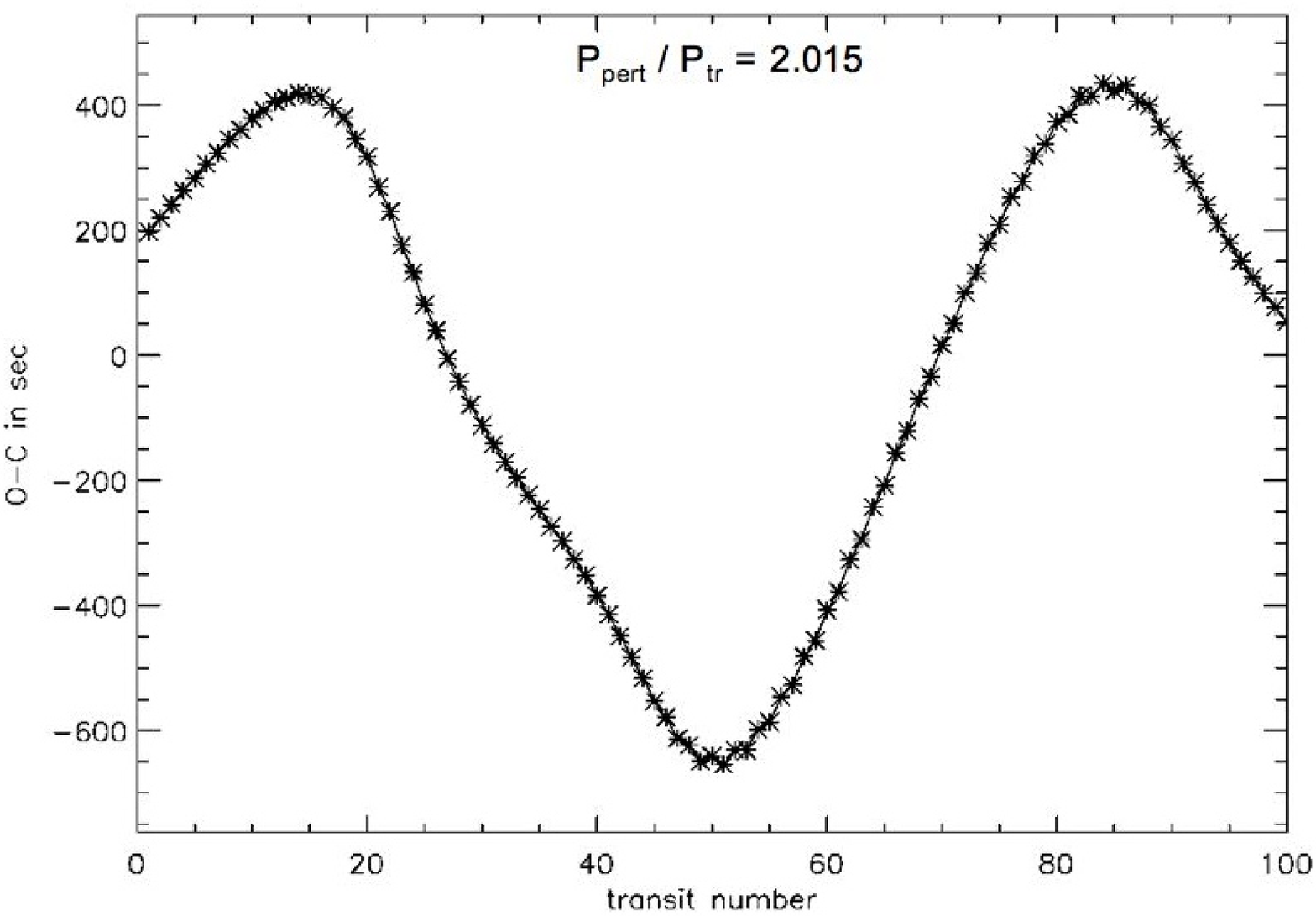}
\includegraphics[width=6cm]{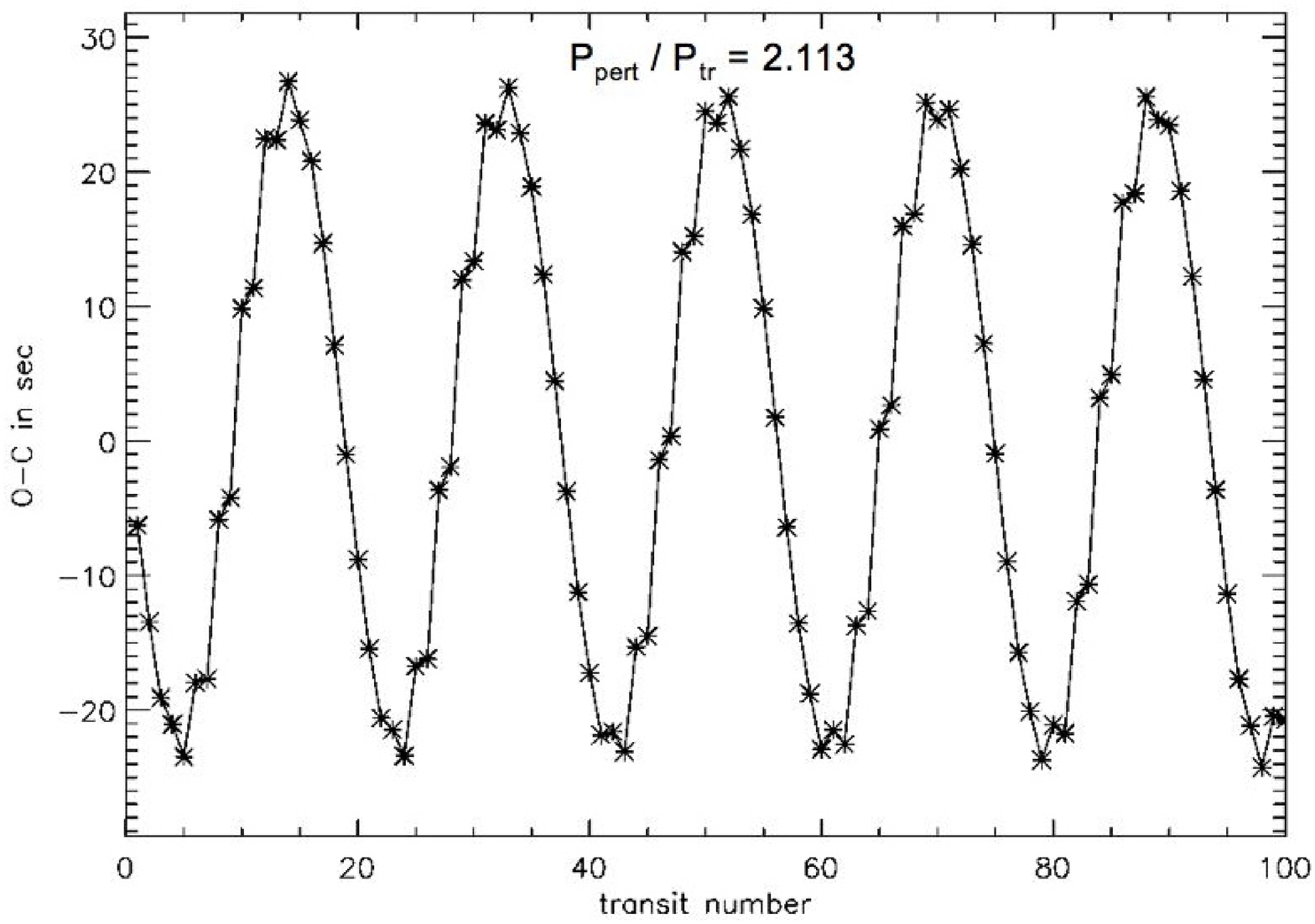}
}
\vskip 2pt
\hbox{
\includegraphics[width=6cm]{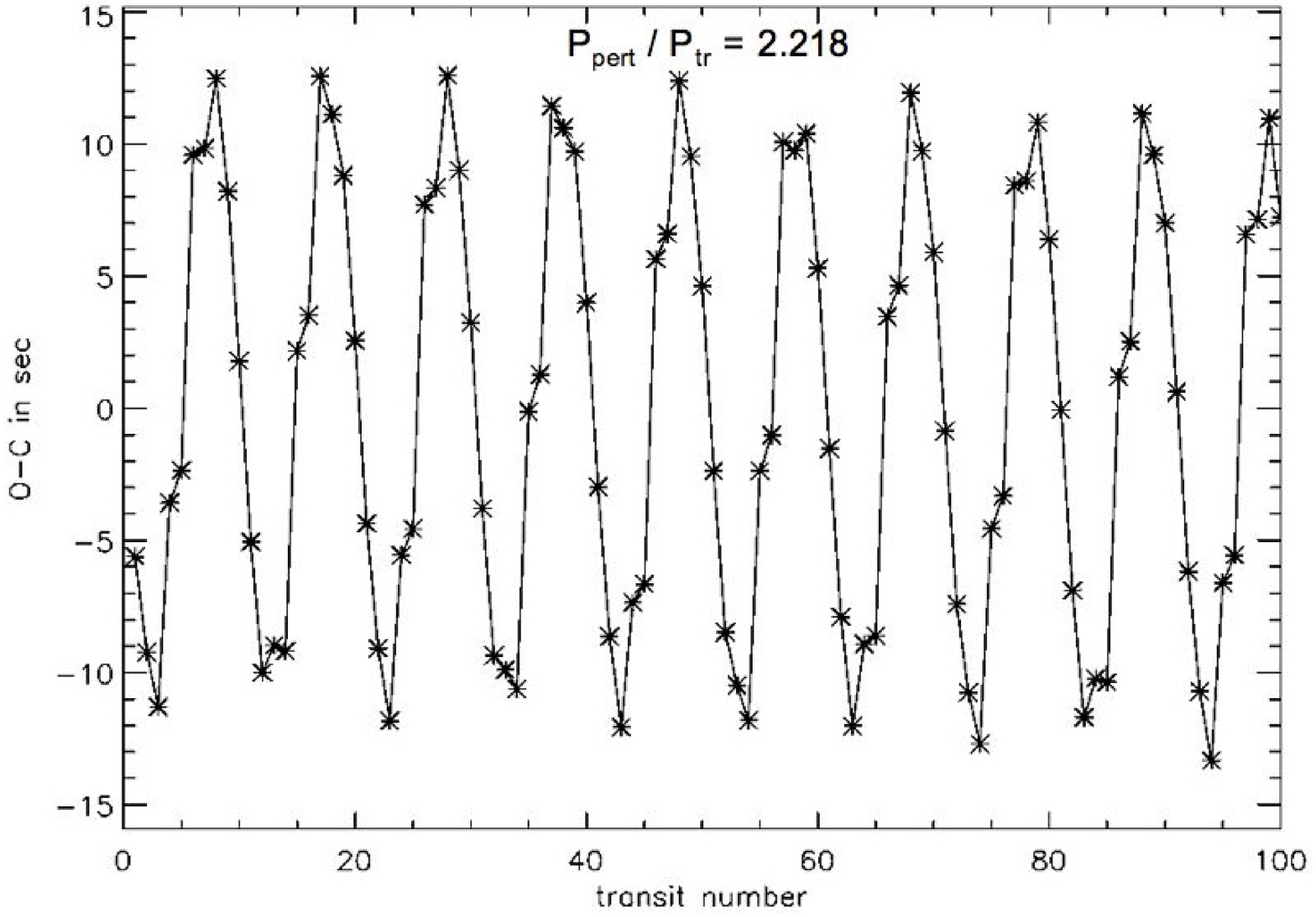}
\includegraphics[width=6cm]{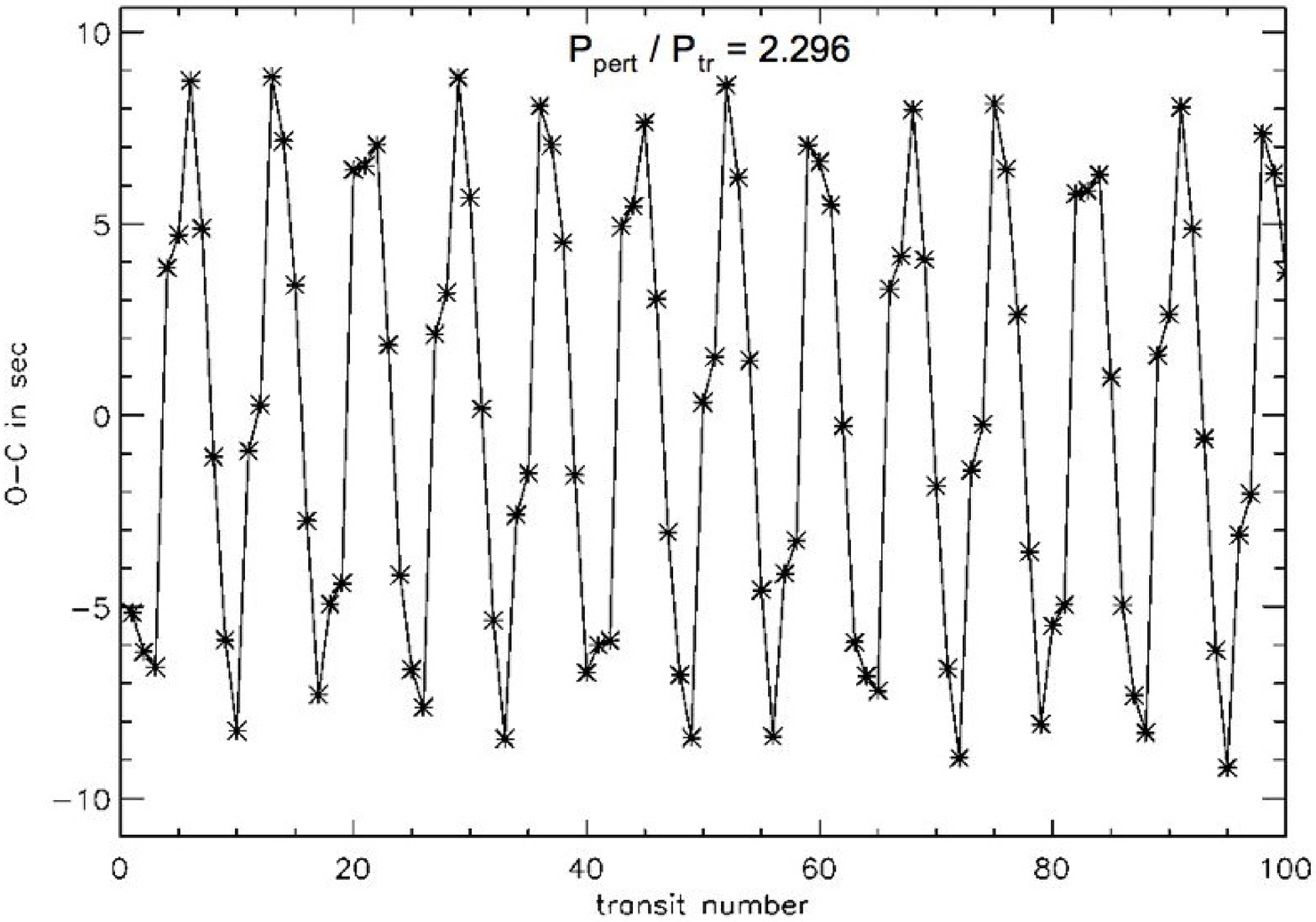}
}
\hbox{
{\bf Appendix}: TTVs for a system with ${m_{\rm star}}={M_\odot},
{m_{\rm tr}}=1 {M_J},\, {m_{\rm pr}}= 1 {M_\oplus}, {P_{\rm tr}}=10$days.}
\end{center}
\end{figure*}

\clearpage

\begin{figure*}
\begin{center}
\hbox{
\includegraphics[width=6cm]{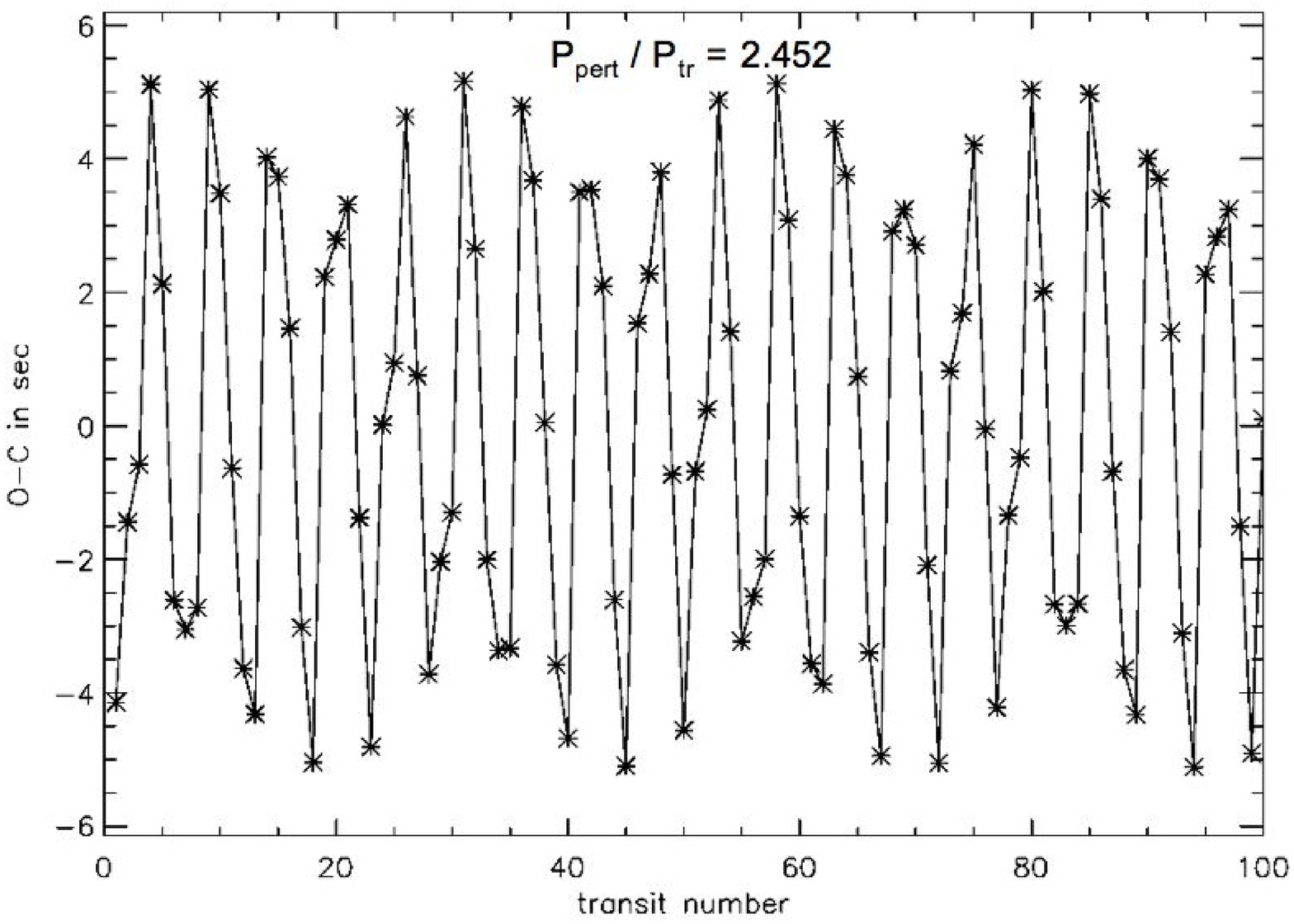}
\includegraphics[width=6cm]{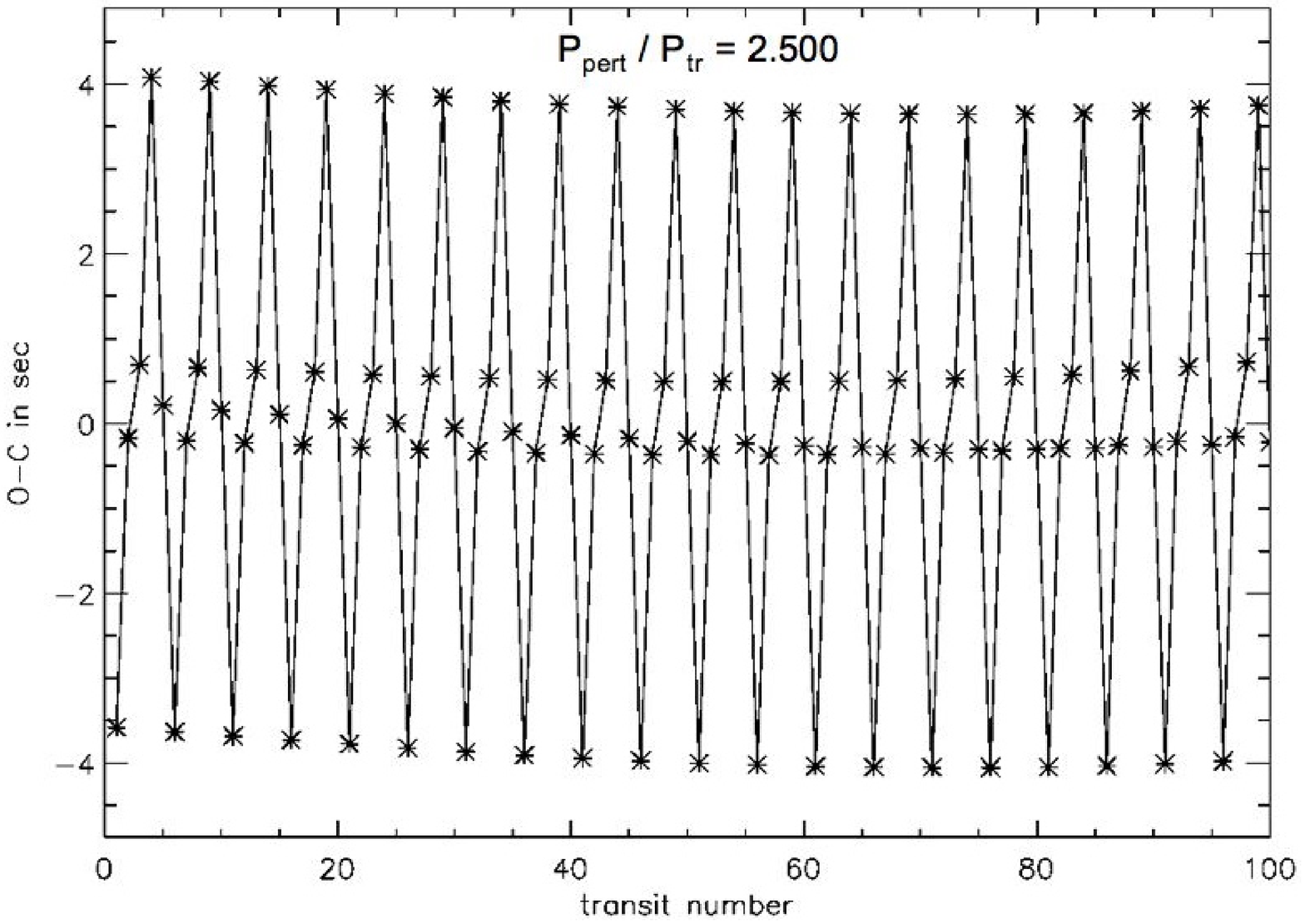}
}
\vskip 2pt
\hbox{
\includegraphics[width=6cm]{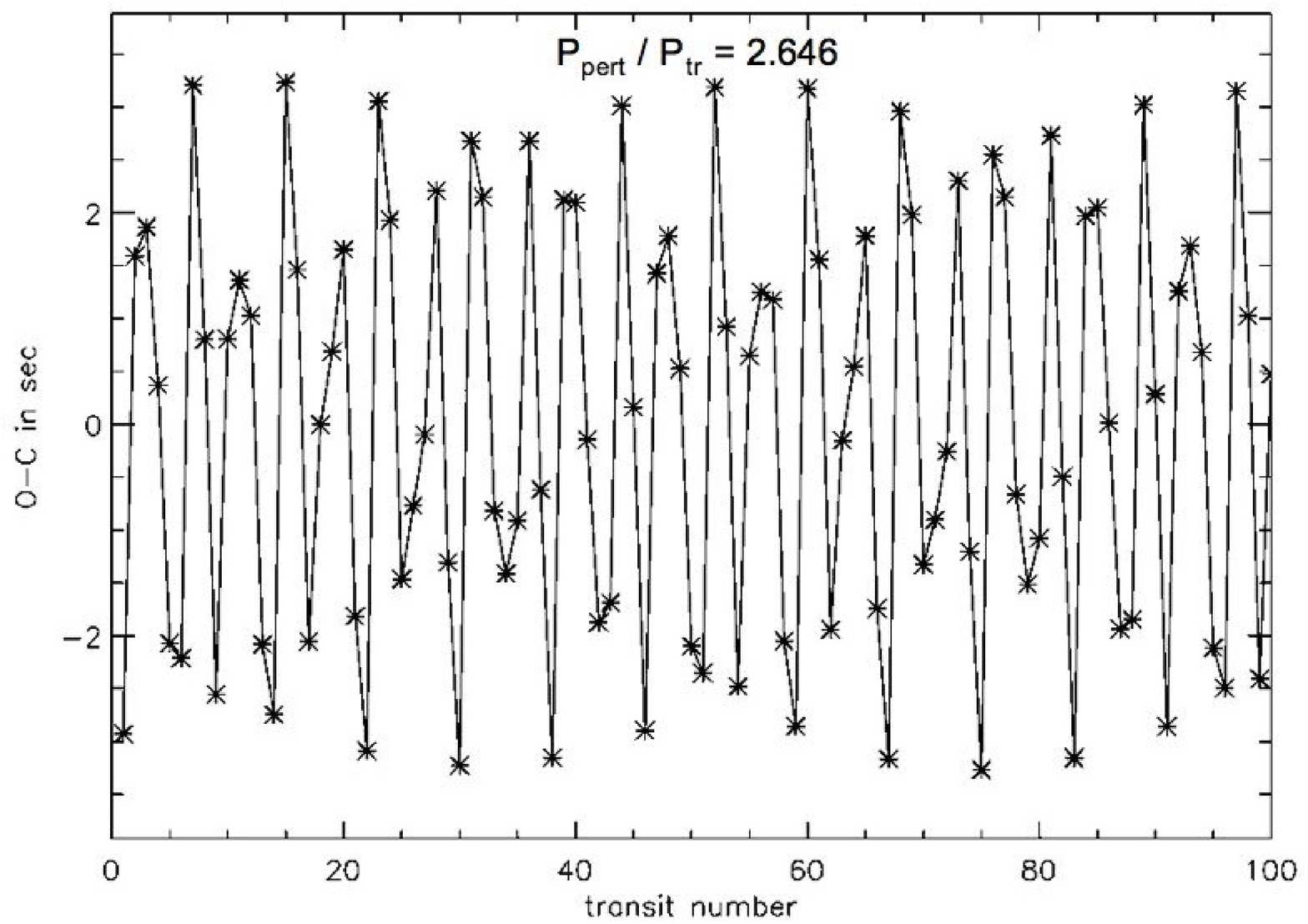}
\includegraphics[width=6cm]{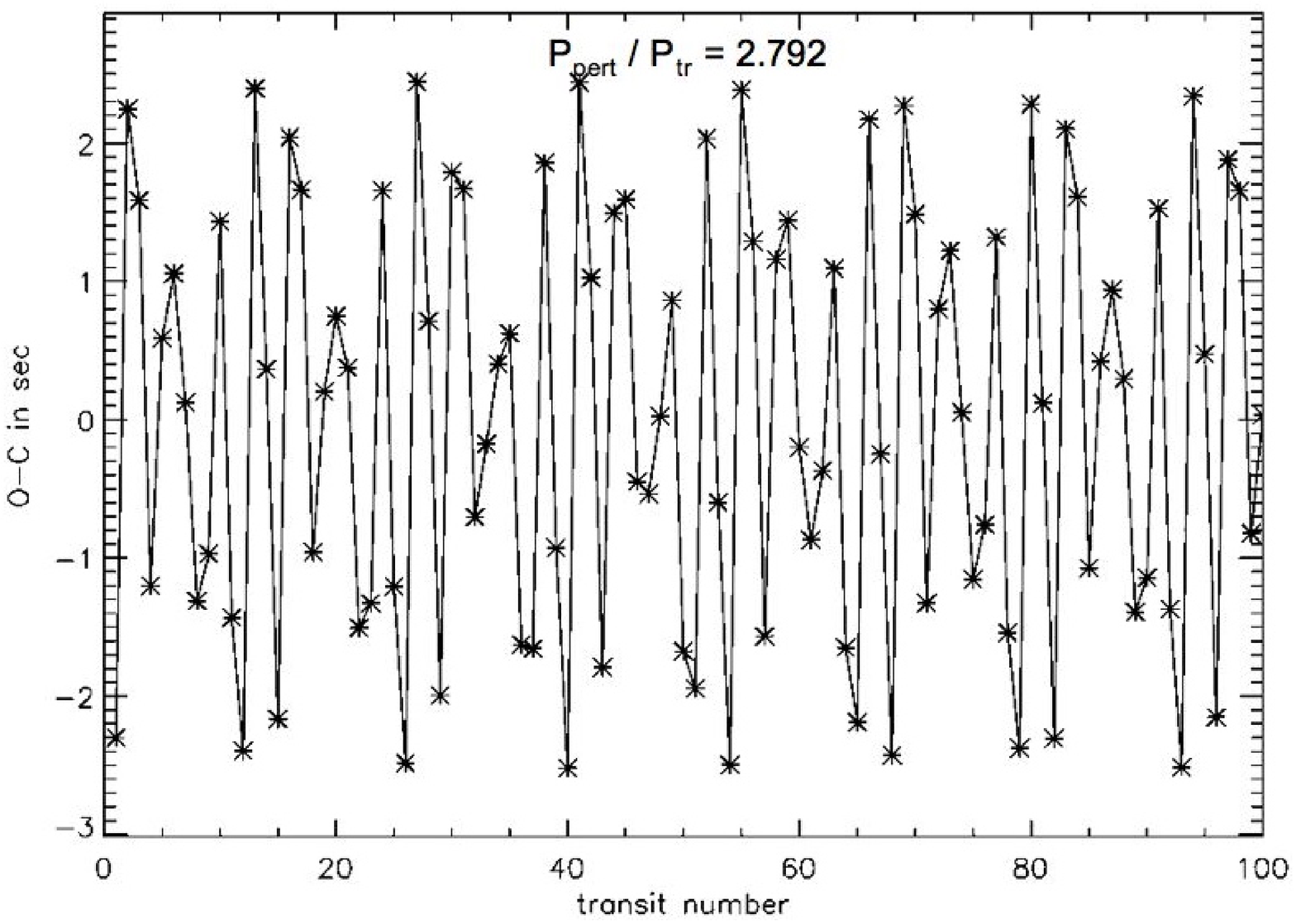}
}
\vskip 2pt
\hbox{
\includegraphics[width=6cm]{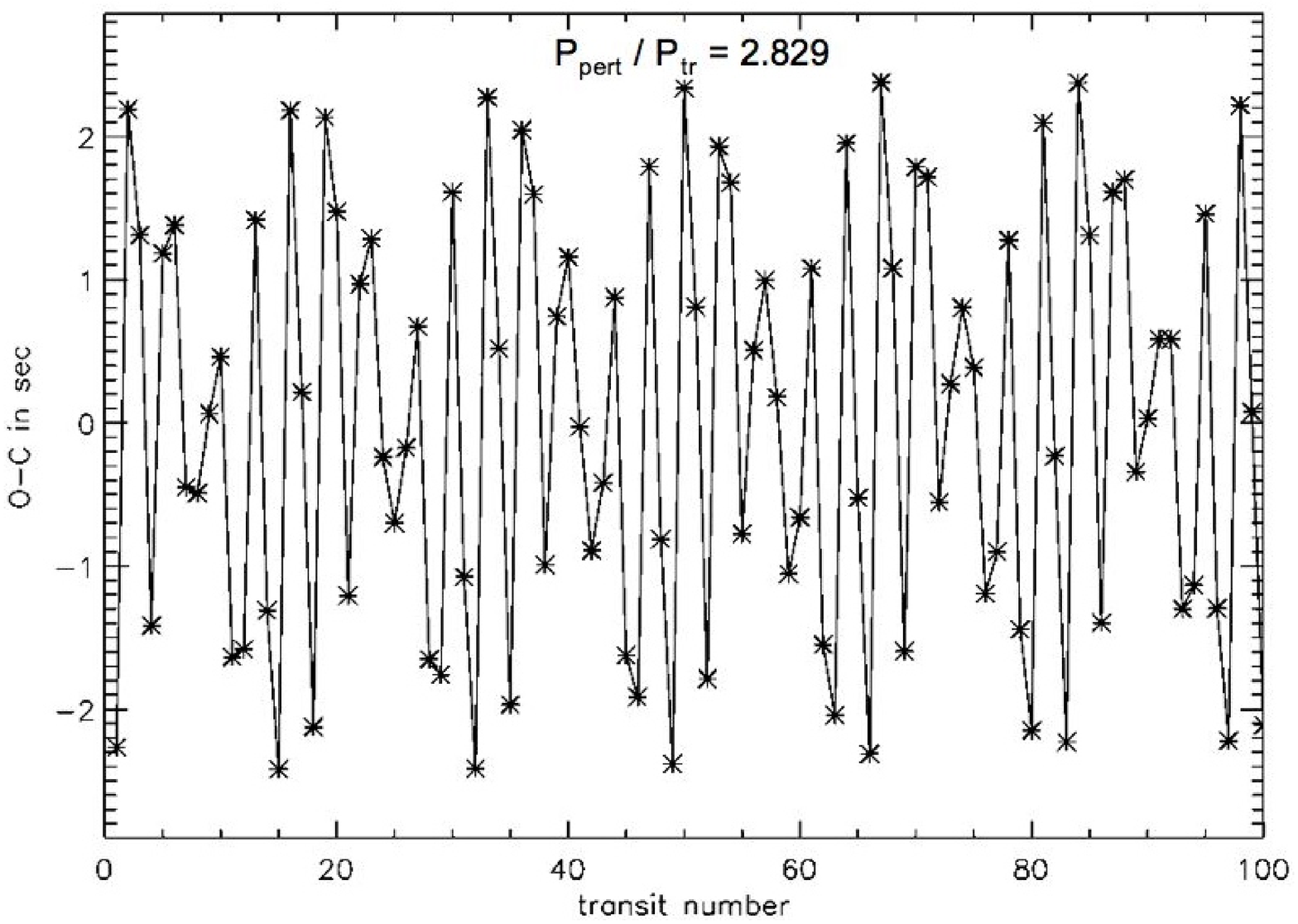}
\includegraphics[width=6cm]{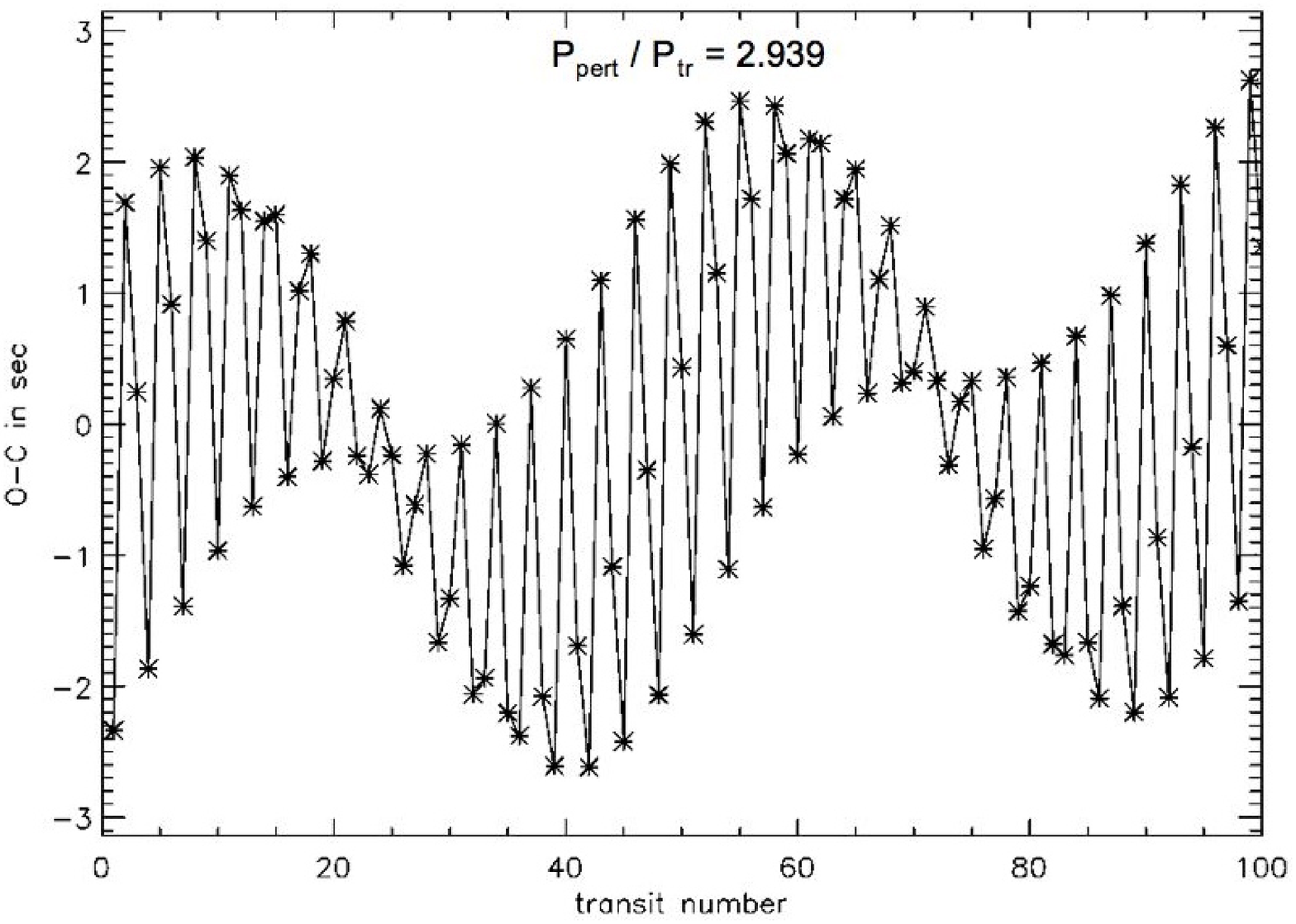}
}
\vskip 2pt
\hbox{
\includegraphics[width=6cm]{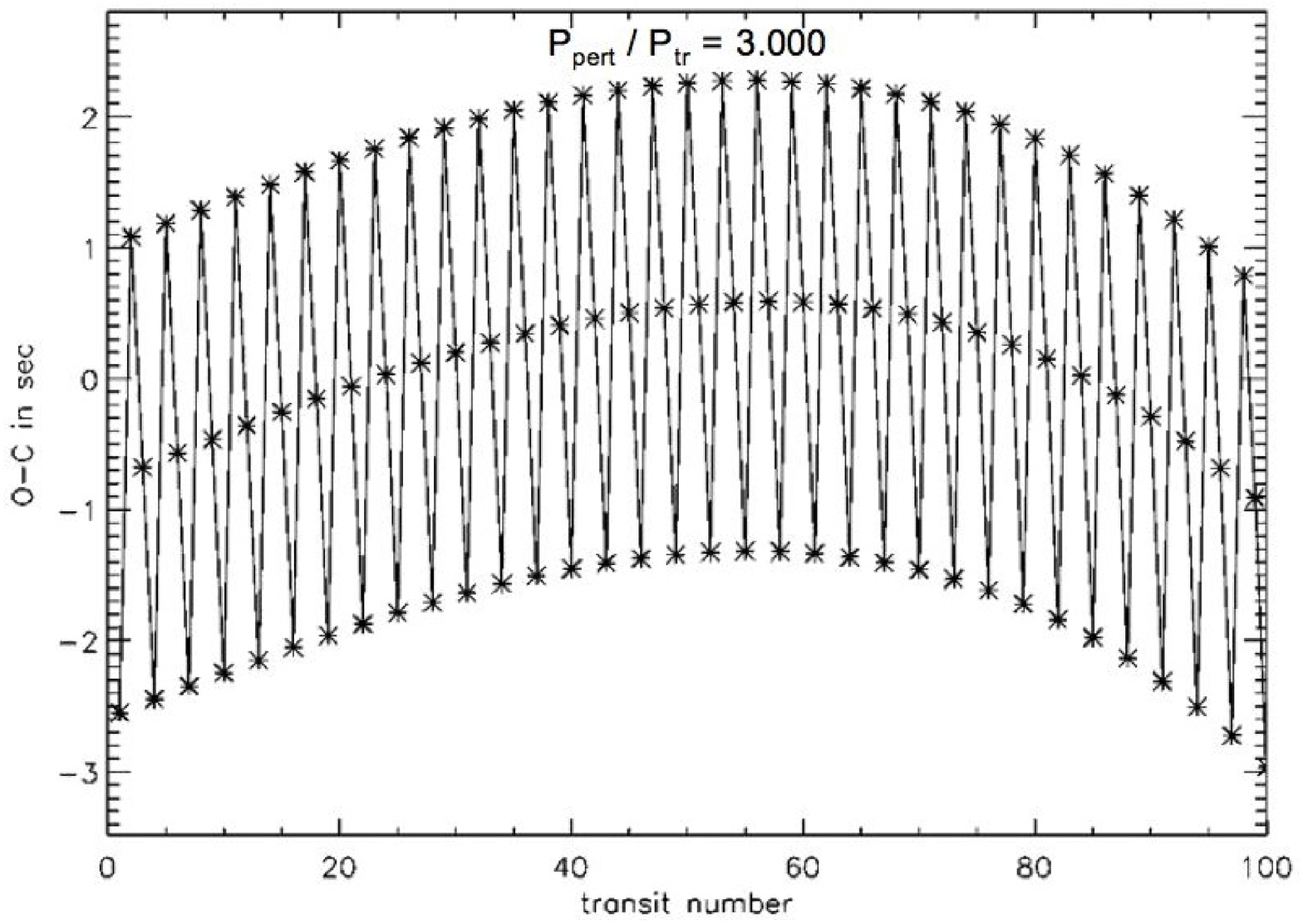}
\includegraphics[width=6cm]{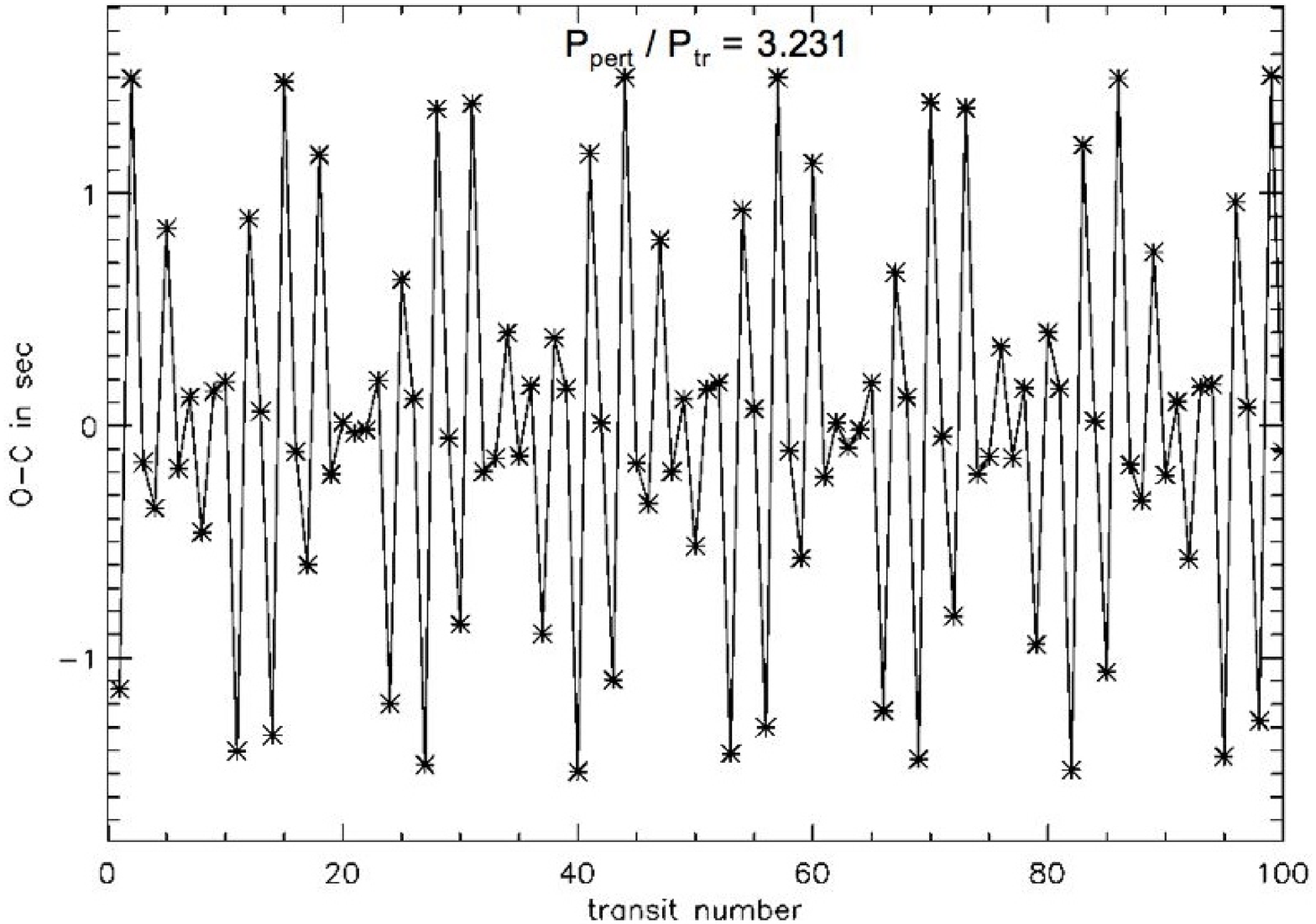}
}
\hbox{
{\bf Appendix}: TTVs for a system with ${m_{\rm star}}={M_\odot},
{m_{\rm tr}}=1 {M_J},\, {m_{\rm pr}}= 1 {M_\oplus}, {P_{\rm tr}}=10$days.}
\end{center}
\end{figure*}

\clearpage

\begin{figure*}
\begin{center}
\hbox{
\includegraphics[width=6cm]{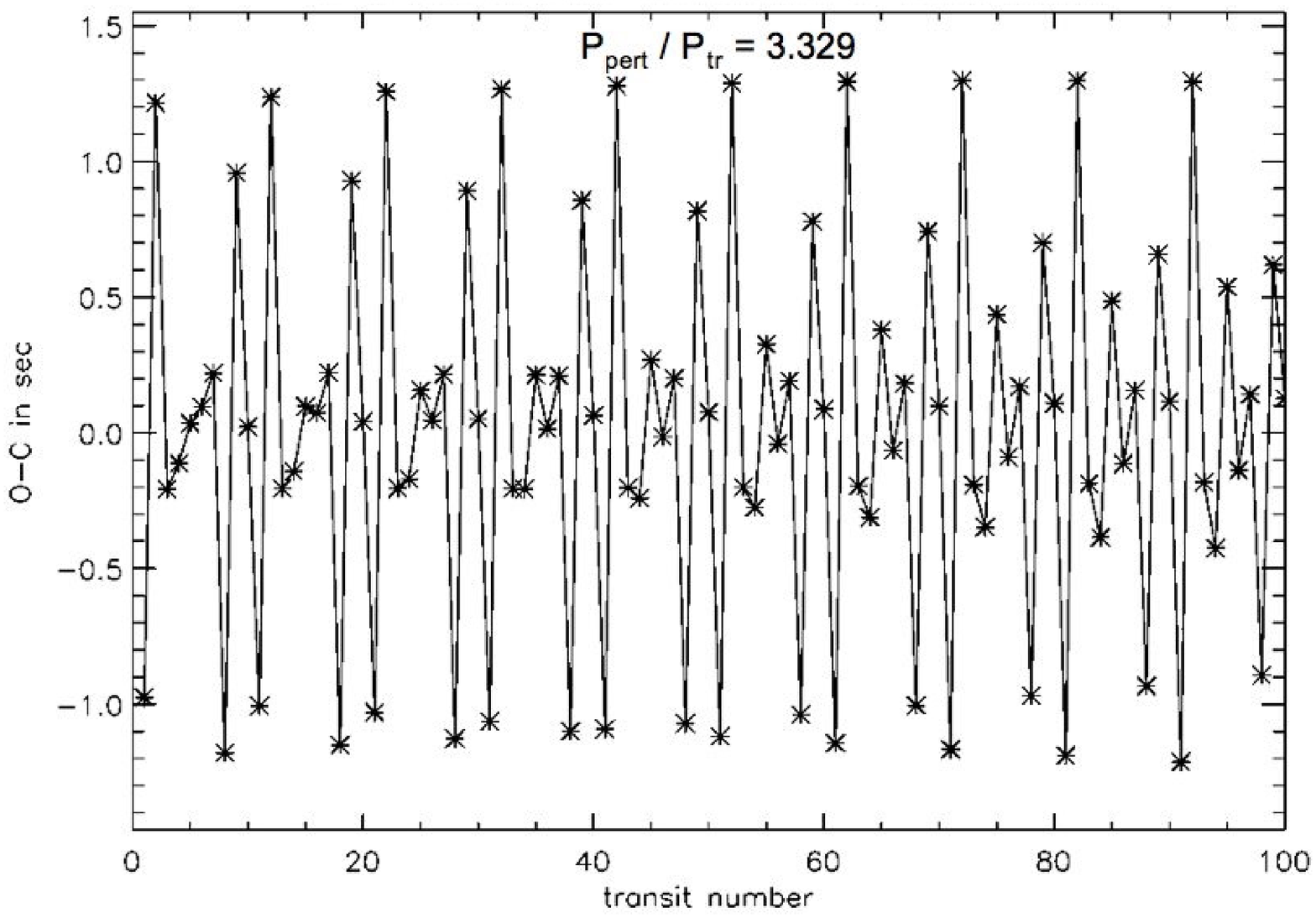}
\includegraphics[width=6cm]{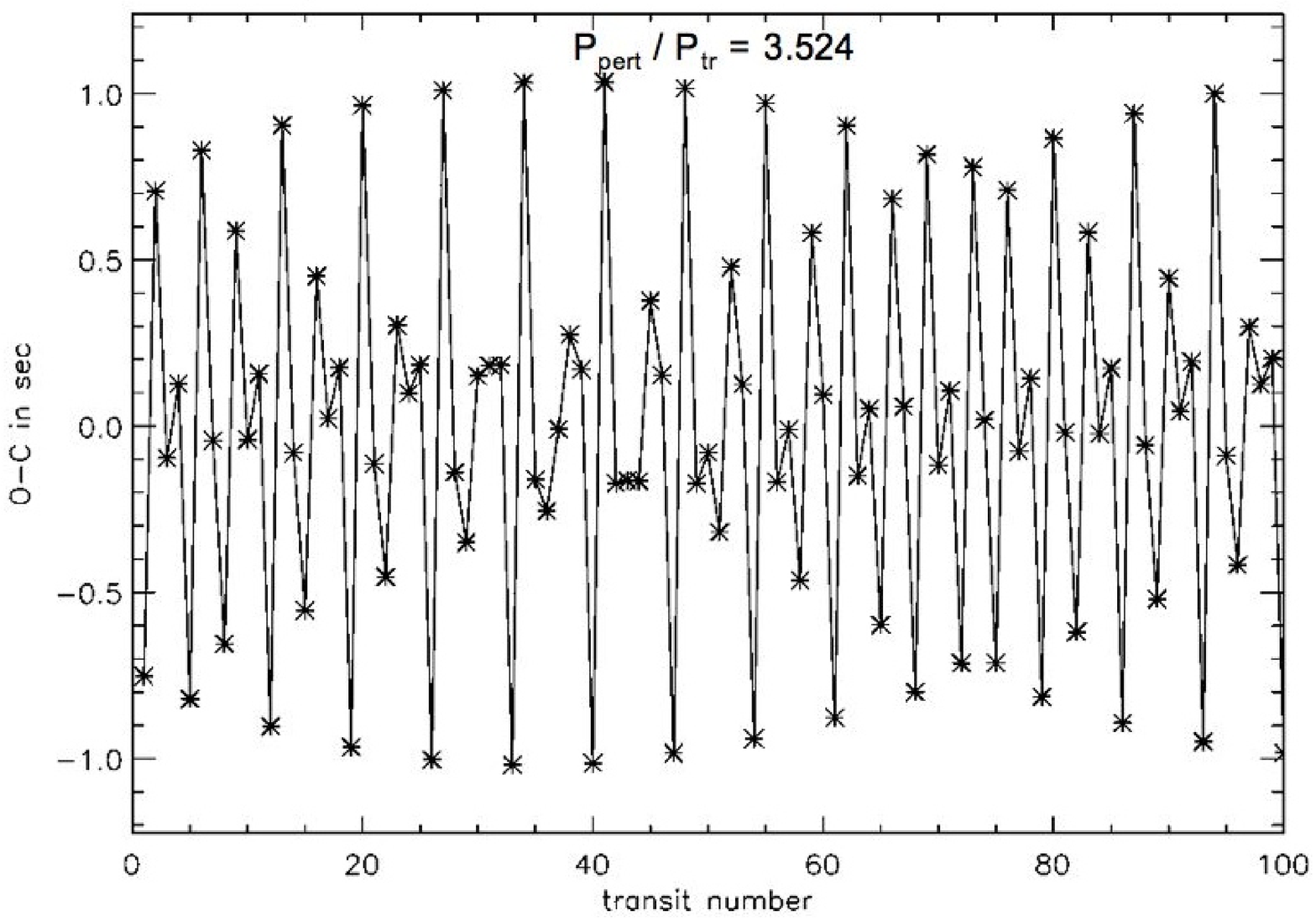}
}
\vskip 2pt
\hbox{
\includegraphics[width=6cm]{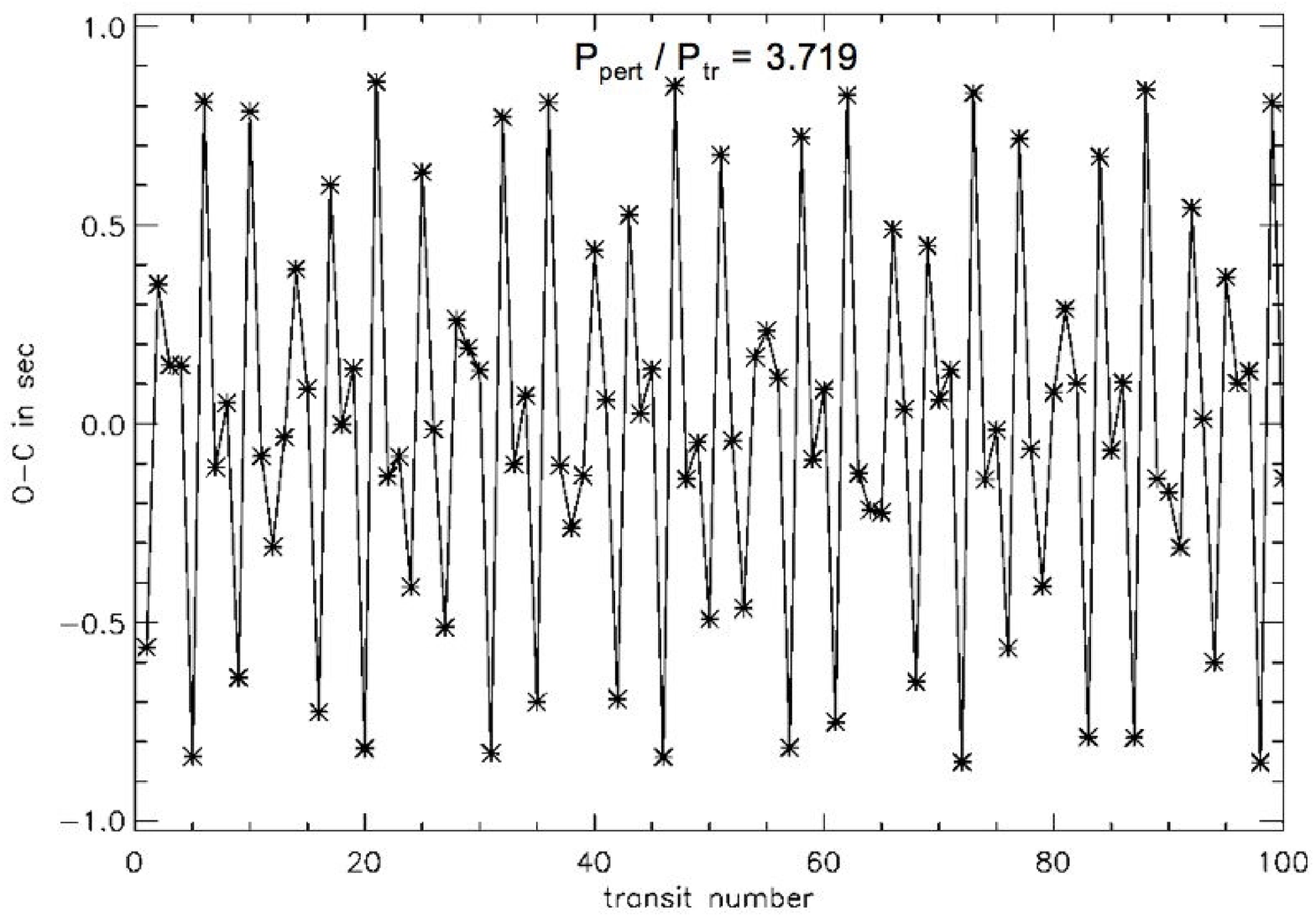}
\includegraphics[width=6cm]{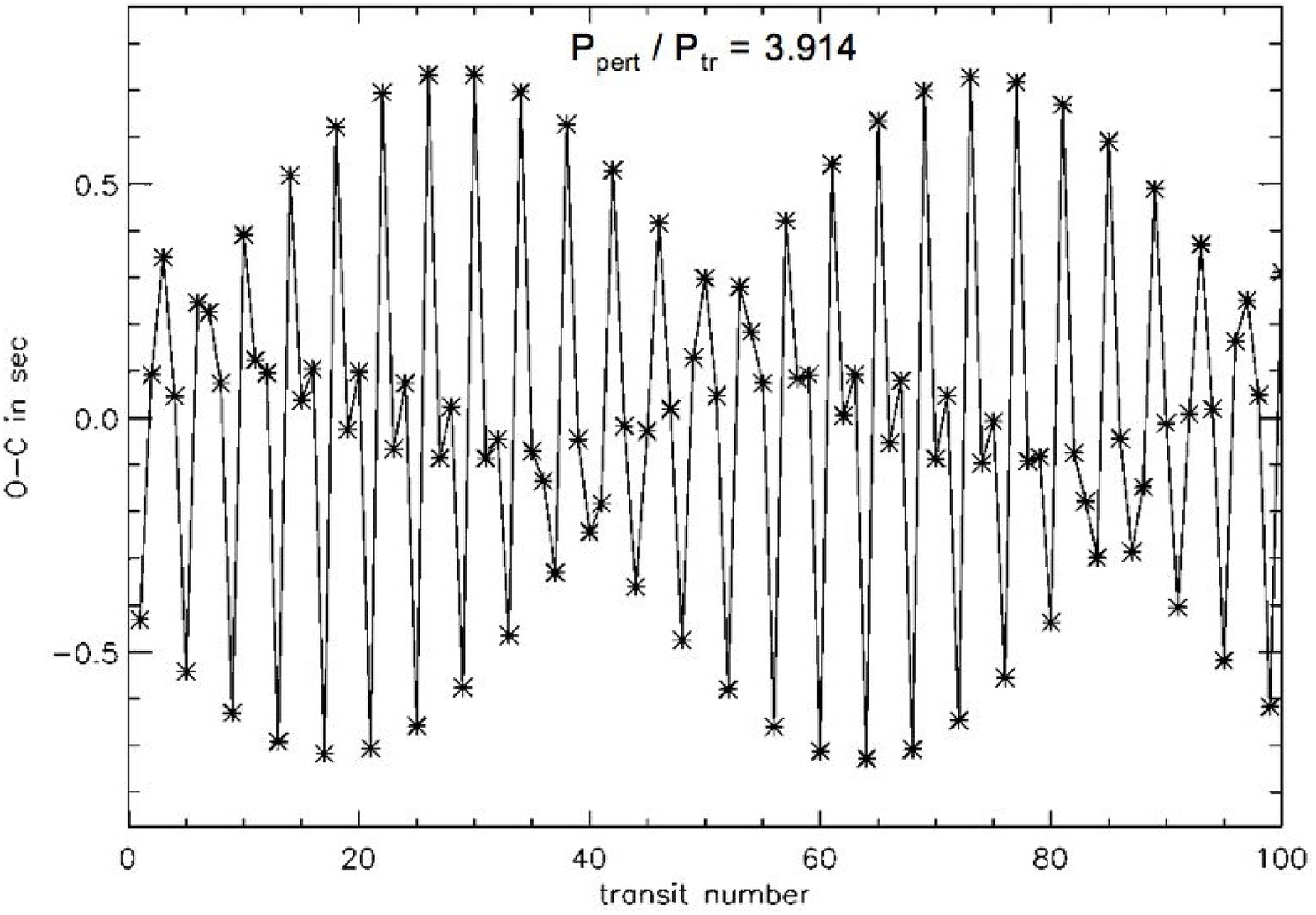}
}
\vskip 2pt
\hbox{
\includegraphics[width=6cm]{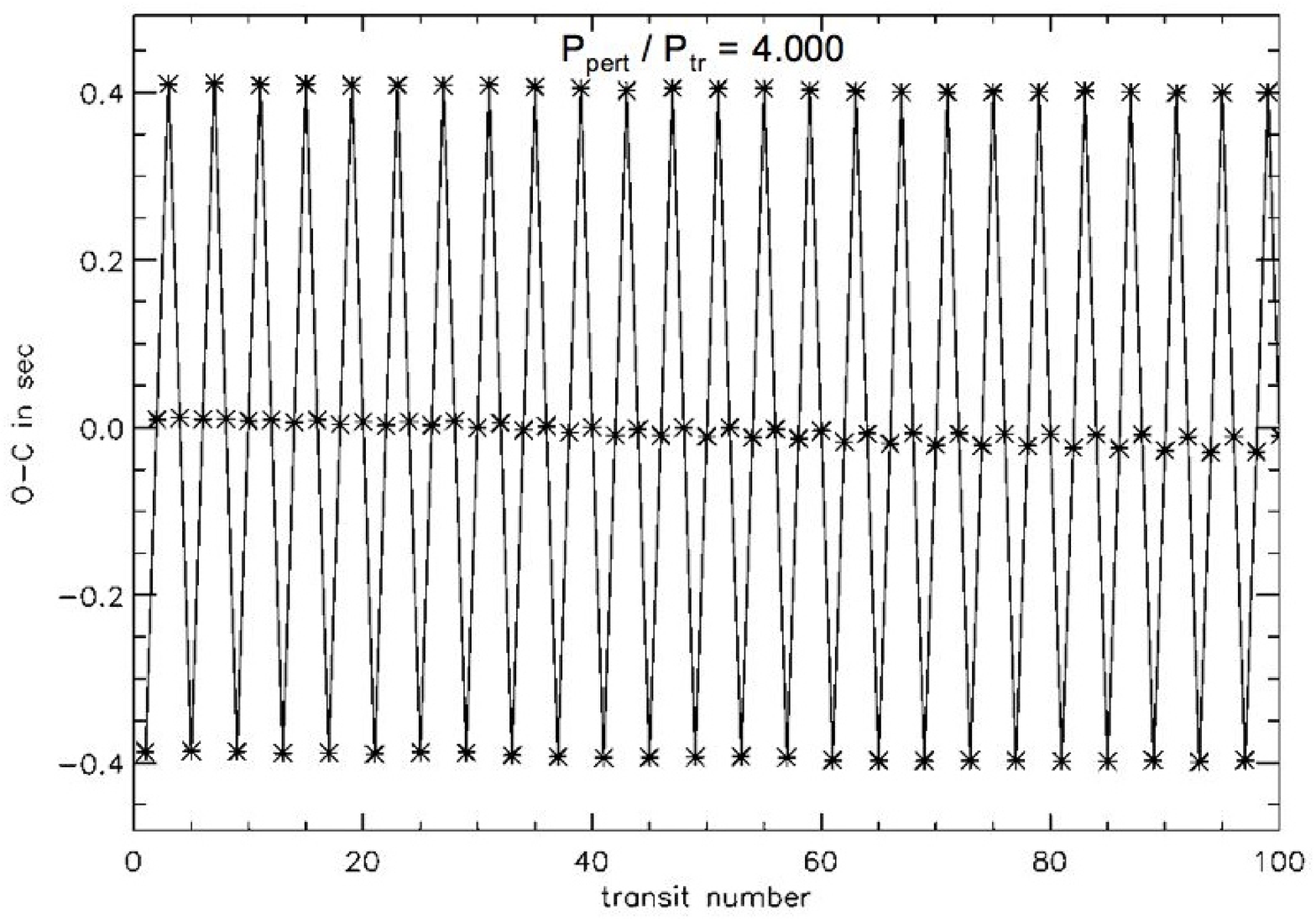}
\includegraphics[width=6cm]{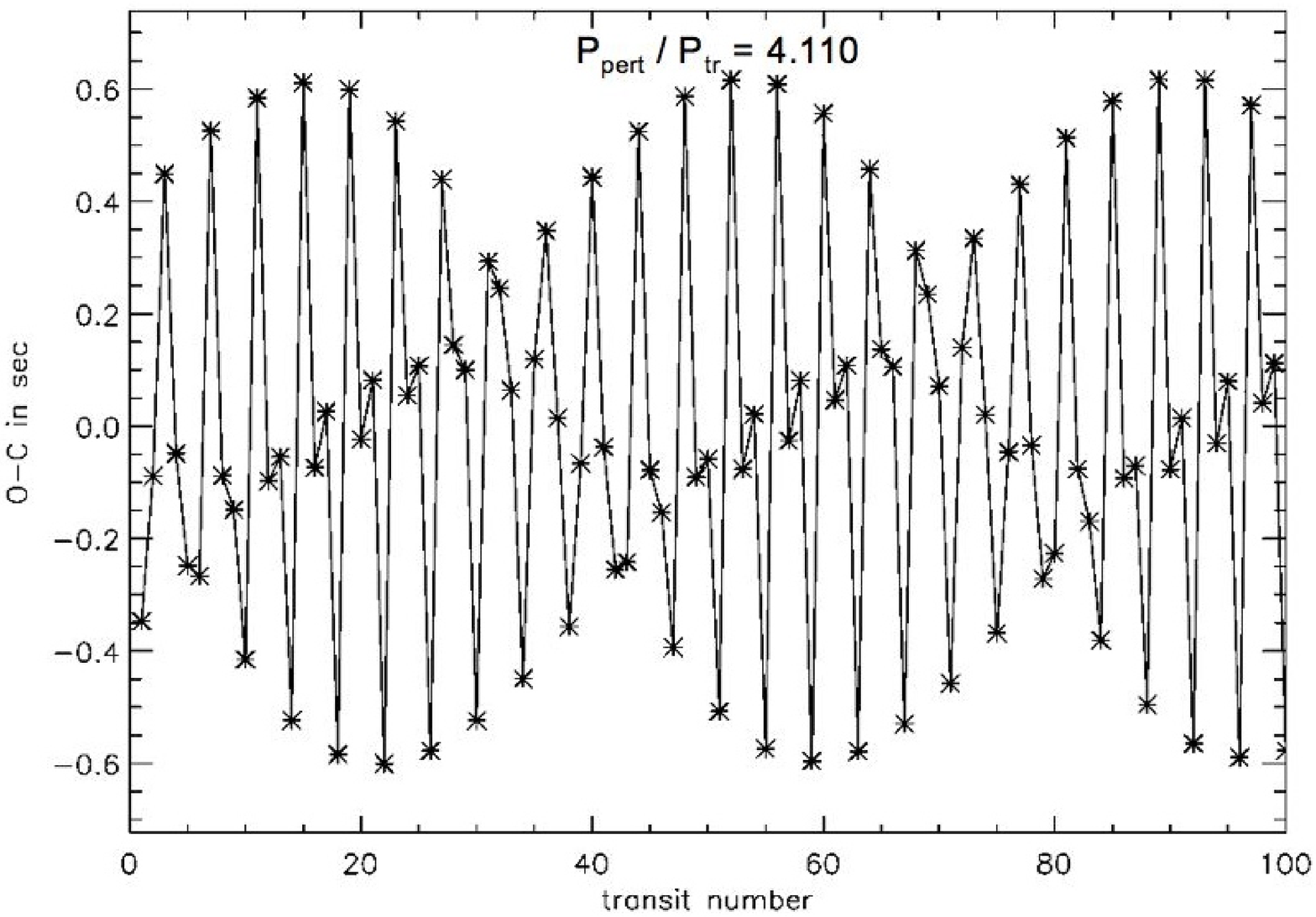}
}
\hbox{
{\bf Appendix}: TTVs for a system with ${m_{\rm star}}={M_\odot},
{m_{\rm tr}}=1 {M_J},\, {m_{\rm pr}}= 1 {M_\oplus}, {P_{\rm tr}}=10$days.}
\end{center}
\end{figure*}


\begin{thebibliography}{}

\bibitem{Agol05}
Agol, E., Steffen, J., Saari, R. and Clarkson, W.:On detecting terrestrial planets with timing of giant 
planet transits, MNRAS, {\bf 359}, 567-579 (2005)

\bibitem{Agol07}
Agol, E. and Steffen, J. H.: A limit on the presence of Earth-mass planets around a Sun-like star,
MNRAS, {\bf 374}, 941-948 (2007)

\bibitem{Batalha10}
Batalha, N. M., et al: Selection, prioritization, and characteristics of Kepler target stars,
ApJ, {\bf 713}, L109-L114 (2010)

\bibitem{Bonfils05}
Bonfils, X., et al: The HARPS search for southern extra-solar planets. VI. A Neptune-mass planet 
around the nearby M dwarf Gl 581, A\& A, {\bf 443}, L15-L18 (2005)

\bibitem{Borucki11}
Borucki, W. J., et al.: Characteristics of planetary candidates observed by Kepler, II: Analysis 
of the first four months of data, ApJ {\bf 736}, 19-40 (2011)

\bibitem{Brown01}
Brown, T. M., Charbonneau, D., Gilliland, R. L., Noyes, R. W. and Burrows, A.: 
Hubble space telescope time-series photometry of the transiting planet of HD 209458,
ApJ, {\bf 552}, 699-709 (2001)

\bibitem{Butler04}
Butler, R. P., et al: A Neptune-Mass Planet Orbiting the Nearby M Dwarf GJ 436,
ApJ, {\bf 617}, 580-588 (2004)

\bibitem{Chambers99}
Chambers, J. E.: A hybrid symplectic integrator that permits close encounters between massive bodies,
MNRAS, {\bf 304}, 793-799 (1999)

\bibitem{Charbonneau09}
Charbonneau, D., et al: A super-Earth transiting a nearby low-mass star, Nature, {\bf 462},
891-894 (2009)

\bibitem{Correia10}
Correia, A. C. M., et al: The HARPS search for southern extrasolar planets. XIX.
Characterization and dynamics of the GJ 876 planetary system, A\& A, {\bf 511}, A21 (2010)

\bibitem{Coughlin11}
Coughlin, J. L., L\'opez-Morales, M., Harrison, T. E., Ule, N. and Hoffman, D. I.:
Low-mass eclipsing binaries in the initial Kepler data release, AJ, {\bf 141}, id.78 (2011)

\bibitem{Deeg08}
Deeg, H. J., Ocana, B., Kozhevnikov, V. P., Charbonneau, D., O'Donovan, F. T. and Doyle, L. R.:
Extrasolar planet detection by binary stellar eclipse timing: evidence for a third body around CM Draconis,
A\& A, {\bf 480}, 563-571 (2008)

\bibitem{Doyle98}
Doyle, L. R., et al: Detectability of Jupiter-to-brown-dwarf-mass companions around small eclipsing 
binary systems. In: R. Rebolo, E. L. Martin and M. R. Z. Osorio (eds) 
Brown dwarfs and extrasolar planets, ASP Conference Series {\bf 134}, 224 (1998) 

\bibitem{Doyle03}
Doyle, L. R. and Deeg, H-J.: Timing detection of eclipsing binary planets and transiting extrasolar moons.
In R. Norris, and F. Stootman (eds) Bioastronomy 2002: Life Among the Stars, 
Proceedings of IAU Symposium {\bf 213}, 80 (2003)

\bibitem{Duric04}
Duric, N.: Advanced Astrophysics. Cambridge University Press, Cambridge, pp. 19 (2004)

\bibitem{Fogg05}
Fogg, M. J. and Nelson, R. P.:Oligarchic and giant impact growth of terrestrial planets in the 
presence of gas giant planet migration, A\& A, {\bf 441}, 791-806 (2005)

\bibitem{Fogg06}
Fogg, M. J. and Nelson, R. P.: On the possibility of terrestrial planet formation in hot-Jupiter systems,
Inter. J. Astrobio., {\bf 5}, 199-209 (2006)

\bibitem{Fogg07a}
Fogg, M. J. and Nelson, R. P.: On the formation of terrestrial planets in hot-Jupiter systems,
A\& A, {\bf 461}, 1195-1208 (2007a)

\bibitem{Fogg07b}
Fogg, M. J. and Nelson, R. P.: The effect of type I migration on the formation of terrestrial planets 
in hot-Jupiter systems, A\& A, {\bf 472}, 1003-1015 (2007b)

\bibitem{Fogg09}
Fogg, M. J. and Nelson, R. P.: Terrestrial planet formation in low-eccentricity warm-Jupiter systems,
A\& A, {\bf 498}, 575-589 (2009)

\bibitem{Ford06}
Ford, E. B. and Gaudi, B. S.: Observational constraints on Trojans of transiting extrasolar planets,
ApJ, {\bf 652}, L137-L140 (2006)

\bibitem{Ford07}
Ford, E. B. and Holman, M. J.: Using transit timing observations to search for Trojans of 
transiting extrasolar planets, ApJ, {\bf 664}, L51-L54 (2007)

\bibitem{Ford11}
Ford, E. B., et al: Transit timing observations from Kepler: I. Statistical analysis of the first four months,
arXiv:1102.0544 (2011)

\bibitem{Gillon07}
Gillon, M., et al: Detection of transits of the nearby hot Neptune GJ 436 b,
A\& A, {\bf 472}, L13-L16 (2007)

\bibitem{Haghighipour10a}
Haghighipour, N., Vogt, S. S., Butler, R. P., Rivera, E., Laughlin, G., Meschiari, S. and Henry, G. W.:
The Lick-Carnegie Exoplanet Survey: A Saturn-mass planet in the habitable zone of the nearby M4V star HIP 57050,
ApJ, {\bf 715}, 271-276 (2010)

\bibitem{Haghighipour10b}
Haghighipour, N. and Rastegar S.: Implications of the TTV-detection of close-in terrestrial planets around
M stars for their origin and dynamical evolution, In  F. Bouchy, R. F. Diaz, and C. Moutou (eds) Detection and 
Dynamics of Transiting Exoplanets, EPJ Web of Conferences Series, {\bf 11}, Article id:04004

\bibitem{Hayl07}
Heyl, J. S. and Gladman, B. J.: Using long-term transit timing to detect terrestrial planets,
MNRAS, {\bf 377}, 1511-1519 (2007)

\bibitem{Holman2005}
Holman, M. J. and Murray, N. W.: The use of transit timing to detect terrestrial-mass extrasolar planets,
Science, {\bf 307}, 1288-1291 (2005)

\bibitem{Holman10}
Holman, M. J., et al: Kepler-9: A system of multiple planets transiting a sun-like star 
confirmed by timing variations, Science, {\bf 330}, 51-54 (2010)

\bibitem{Jones05}
Jones, B. W., Underwood, D. R., Sleep, P. N.: Prospects for habitable ``Earths'' in known exoplanetary systems,
ApJ, {\bf 622}, 1091-1101 (2005)

\bibitem{Jones06}
Jones, B. W.,  Sleep, P. N.,  Underwood, D. R.: Habitability of known exoplanetary systems based 
on measured stellar properties, ApJ, {\bf 649}, 1010-1019 (2006)

\bibitem{Kasting93}
Kasting, J. F., Whitmire, D. P. and Reynolds R. T.: Habitable zones around main sequence 
stars, Icarus, {\bf 101}, 108-128 (1993)

\bibitem{Kennedy08}
Kennedy, G. M. and Kenyon, S. J.: Planet formation around stars of various masses: Hot Super-Earths,
ApJ, {\bf 682}, 1264-1276 (2008)

\bibitem{Kipping09a}
Kipping, D. M.: Transit timing effects due to an exomoon, MNRAS, {\bf 396}, 1797-1804 (2009a)

\bibitem{Kipping09b}
Kipping, D. M.: Transit timing effects due to an exomoon - II, MNRAS, {\bf 392}, 181-189 (2009b)

\bibitem{Kipping11}
Kipping, D. and Bakos, G.: An independent analysis of Kepler-4b through Kepler-8b, ApJ, 
{\bf 730}, id.50 (2011) 

\bibitem{Laughlin04}
Laughlin, G., Bodenheimer, P. and Adams, F. C. : The core accretion model predicts few Jovian-mass 
planets orbiting red dwarfs, ApJ, {\bf 612}, L73-L76 (2004)

\bibitem{Lissauer11}
Lissauer, J. J., et al: A closely packed system of low-mass, low-density planets transiting Kepler-11,
Nature, {\bf 470}, 53-58 (2011)

\bibitem{Mandell07}
Mandell, A. M., Raymond, S. N. and Sigurdsson, S.: Formation of Earth-like planets during and after 
giant planet migration, ApJ, {\bf 660}, 823-844 (2007)

\bibitem{Mayor09}
Mayor, M., et al: The HARPS search for southern extra-solar planets. XVIII. An Earth-mass planet 
in the GJ 581 planetary system, A\& A, {\bf 507}, 487-494 (2009)

\bibitem{Menou03}
Menou, K. and Tabachnik, S.: Dynamical habitability of known extrasolar planetary systems,
ApJ, {\bf 583}, 473-488 (2003)

\bibitem{Meschiari10}
Meschiari, S. and Laughlin, G. P.: Systemic: A testbed for characterizing the detection of extrasolar 
planets. II. Numerical approaches to the transit timing inverse problem, ApJ, {\bf 718}, 543-550 (2010) 

\bibitem{Miralda-Escude02}
Miralda-Escud\'e, J.: Orbital perturbations of transiting planets: a possible method to measure 
stellar quadrupoles and to detect Earth-mass planets, ApJ, {\bf 564}, 1019-1023 (2002) 

\bibitem{Nesvorny08}
Nesvorn\'y, D. and Morbidelli, A.: Mass and orbit determination from transit timing variations of exoplanets,
ApJ, {\bf 688}, 636-646 (2008)

\bibitem{Nesvorny09}
Nesvorn\'y, D.: Transit Timing Variations for Eccentric and Inclined Exoplanets,
ApJ, {\bf 701}, 1116-1122 (2009)

\bibitem{Nesvorny10}
Nesvorn\'y, D. and Beaug\'e, C.: Fast inversion method for determination of planetary parameters 
from transit timing variations, ApJ, {\bf 709}, L44-L48 (2010)

\bibitem{oshagh10}
Oshagh, M., Haghighipour, N. and Santon, N.: A Survey of M Stars in the Field of View of Kepler Space 
Telescope, to appear in proceedings of the IAU Symposium No.276 (arXiv:1012.2234)  

\bibitem{Pan04}
Pan, M. and Sari, R.: A generalization of the Lagrangian points: studies of resonance for highly 
eccentric orbits, AJ, {\bf 128}, 1418-1429 (2004)

\bibitem{Payne10}
Payne, M. J., Ford, E. B. and Veras, D.: Transit timing variations for inclined and retrograde 
exoplanetary systems, ApJ, {\bf 712}, L86-L92 (2010)

\bibitem{Raymond06}
Raymond, S. N., Mandell, A. M. and Sigurdsson, S.: Exotic Earths: Forming habitable worlds with 
giant planet migration, Science, {\bf 313}, 1413-1416 (2006)

\bibitem{Rivera05}
Rivera, E. J., et al: A $\sim 7.5 {M_\oplus}$ planet orbiting
the nearby star, GJ 876, ApJ, {\bf 634}, 625-640 (2005)

\bibitem{Rivera10}
Rivera, E. J., Laughlin, G., Butler, R. P., Vogt, S. S., Haghighipour, N.
and Mechiari, S.: The Lick-Carnegie Exoplanet Survey: A Uranus-mass fourth planet for
GJ 876 in an extrasolar Laplace configuration, ApJ, {\bf 719}, 890-899 (2010)

\bibitem{Saha92}
Saha, P. and Tremaine, S.: Symplectic integrators for solar system dynamics, AJ, 
{\bf 104}, 1633-1640 (1992) 

\bibitem{Sartoretti99}
Sartoretti, P. and Schneider, J.: On the detection of satellites of extrasolar planets with the 
method of transits, A\& A, {\bf 134}, 553-560 (1999)

\bibitem{Schneider90}
Schneider, J. and Chevreton, M.: The photometric search for earth-sized extrasolar planets by 
occultation in binary systems, A\&A, {\bf 232}, 251-257 (1990)

\bibitem{Schneider95}
Schneider, J. and Doyle, L. R.: Ground-based detection of terrestrial extrasolar planets by photometry: 
The case for CM Draconis, EM\&P, {\bf 71}, 153-173 (1995)

\bibitem{Schneider03}
Schneider, J.: Multi-planet system detection by transits. In: F. Combes, D. Barret, T. Contini, 
and L. Pagani (eds) SF2A-2003, EdP-Sciences, Conference Series, p. 149 (2003) 

\bibitem{Schwarz11}
Schwarz, R., Haghighipour, N., Eggl, S., Pilat-Lohinger, E. and Funk, B.:
Prospects of the detection of circumbinary planets with {\it Kepler} and CoRoT using the variations of eclipse timing,
MNARS, {\bf 414}, 2763-2770 (2011)

\bibitem{Simon07}
Simon, A., Szatm\'ary, K. and Szab\'o, Gy. M.:
Determination of the size, mass, and density of ``exomoons'' from photometric transit timing variations,
A\& A, {\bf 470}, 727-731 (2007)

\bibitem{Steffen05}
Steffen, J. H. and Agol, E.: An analysis of the transit times of TrES-1b,
MNRAS, {\bf 364}, L96-L100 (2005)

\bibitem{Steffen10}
Steffen J. H., et al: Five Kepler target stars that show multiple transiting exoplanet candidates, 
ApJ, {\bf 725}, 1226-1241 (2010)

\bibitem{Sybilski10}
Sybilski, P., Konacki, M. and Kozlowski, S.: Detecting circumbinary planets using eclipse timing of 
binary stars - numerical simulations, MNRAS, {\bf 405}, 657-665 (2010)

\bibitem{Udry07}
Udry, S., et al: The HARPS search for southern extrasolar planets. XI. Super-Earths 
(5 and 8 $M_\oplus$) in a 3-planet system, A\& A, {\bf 469}, L43-L47 (2007)

\bibitem{Veras11}
Veras, D., Ford, E. B. and Payne, M. J.: Quantifying the challenges of detecting unseen planetary 
companions with transit timing variations, ApJ, {\bf 727}, article id.74 (2011)

\bibitem{Vogt10}
Vogt, S. S., Butler, R. P., Rivera, E. J., Haghighipour, N. and Henry G. W.:
The Lick-Carnegie Exoplanet Survey: a 3.1 $M_\oplus$ planets in the habitable zone of
the nearby M3V star Gliese 581, ApJ, {\bf 723}, 954-965 (2010)

\bibitem{Zhou05}
Zhou, J-L., Aarseth, S. J., Lin, D. N. C. and Nagasawa, M.: Origin and ubiquity of short-period Earth-like 
planets: Evidence for the sequential accretion theory of planet formation, ApJ, {\bf 631}, L85-L88 (2005)


\end{thebibliography}
\end{document}